\documentclass[conference]{IEEEtran}

\usepackage{amsmath,graphicx,epsfig,float,subfigure,times,latexsym,verbatim,mathrsfs,amssymb,textcomp,txfonts,color,mathrsfs}
\usepackage[flushleft]{threeparttable}
\usepackage{algorithm}
\usepackage{algorithmic}
\usepackage{url}

\IEEEoverridecommandlockouts

\ifCLASSINFOpdf
\else
 \fi

\begin{document}
%
\title{Wireless Physical-Layer Identification: \\Modeling and Validation}

\author{\IEEEauthorblockN{
\textbf{Wenhao Wang}\IEEEauthorrefmark{1}\IEEEauthorrefmark{3}, \textbf{Zhi Sun}\IEEEauthorrefmark{1}, \textbf{Kui Ren}\IEEEauthorrefmark{2}, \textbf{Bocheng Zhu}\IEEEauthorrefmark{3}, \textbf{Sixu Piao}\IEEEauthorrefmark{2}
\vspace{7pt}
\IEEEauthorblockA{\IEEEauthorrefmark{1}
Department of Electrical Engineering, University at Buffalo, The State University of New York, \\Buffalo, New York 14260, USA, E-mail: \{wenhaowa, zhisun\}@buffalo.edu}
\vspace{7pt}
\IEEEauthorblockA{\IEEEauthorrefmark{2}
Department of  Computer Science and Engineering, University at Buffalo, The State University of New York, \\Buffalo, New York 14260, USA, E-mail: \{kuiren, sixupiao\}@buffalo.edu}
\vspace{7pt}
\IEEEauthorblockA{\IEEEauthorrefmark{3}
School of Electronics Engineering and Computer Science, Peking University, \\Beijing 100871, China, Email: \{wenhaowang, zhubc\}@pku.edu.cn}
}}

\maketitle

\begin{abstract}
The wireless physical-layer identification (WPLI) techniques utilize the unique features of the physical waveforms of wireless signals to identify and classify authorized devices. As the inherent physical layer features are difficult to forge, WPLI is deemed as a promising technique for wireless security solutions. However, as of today it still remains unclear whether existing WPLI techniques can be applied  under real-world requirements and constraints. In this paper, through both theoretical modeling and experiment validation, the reliability and differentiability of WPLI techniques are rigorously evaluated, especially under the constraints of state-of-art wireless devices, real operation environments, as well as wireless protocols and regulations. Specifically, a theoretical model is first established to systematically describe the complete procedure of WPLI. More importantly, the proposed model is then implemented to thoroughly characterize various WPLI techniques that utilize the spectrum features coming from the non-linear RF-front-end, under the influences from different transmitters, receivers, and wireless channels. Subsequently, the limitations of existing WPLI techniques are revealed and evaluated in details using both the developed theoretical model and in-lab experiments. The real-world requirements and constraints are characterized along each step in WPLI, including i) the signal processing at the transmitter (device to be identified), ii) the various physical layer features that originate from circuits, antenna, and environments, iii) the signal propagation in various wireless channels, iv) the signal reception and processing at the receiver (the identifier), and v) the fingerprint extraction and classification at the receiver. 
\end{abstract}

\IEEEpeerreviewmaketitle

\section{Introduction}

In wireless communications, the usage of electromagnetic (EM) waves as signal carrier grants both significant challenges and unique opportunities to users' security and privacy. On the one hand, as the EM waves propagate anywhere within the physics limit through line-of-sight, reflection, diffraction, and refraction paths, it is possible for impostors within the range to participate in the wireless communication. On the other hand, physical waveforms transmitted by any wireless device are inherently stamped with unique features in the physical layer of the communication, which can be utilized to identify impostors and classify authorized users. Such device identification solutions are defined as Wireless Physical-Layer Identification (WPLI) techniques \cite{danev2012physical}. 

While the conventional software-level device identifications (e.g., IP or MAC address) can be easily changed by malwares, the physical layer feature cannot be modified without significant effort. Therefore, the WPLI technique is deemed as a promising wireless security solution, if the physical layer features, or the so called "radio frequency fingerprints (RFFs)", are sufficiently reliable and distinguishable. 
The question is: are RFFs reliable and distinguishable in real and practical applications? Or more specifically, given state-of-art wireless devices and in real operation environments, can WPLI techniques reliably and uniquely identify/classify the authorized users and impostors? The objective of this paper is to answer the question. 

\begin{figure}
\centering
\includegraphics [width=2.5in] {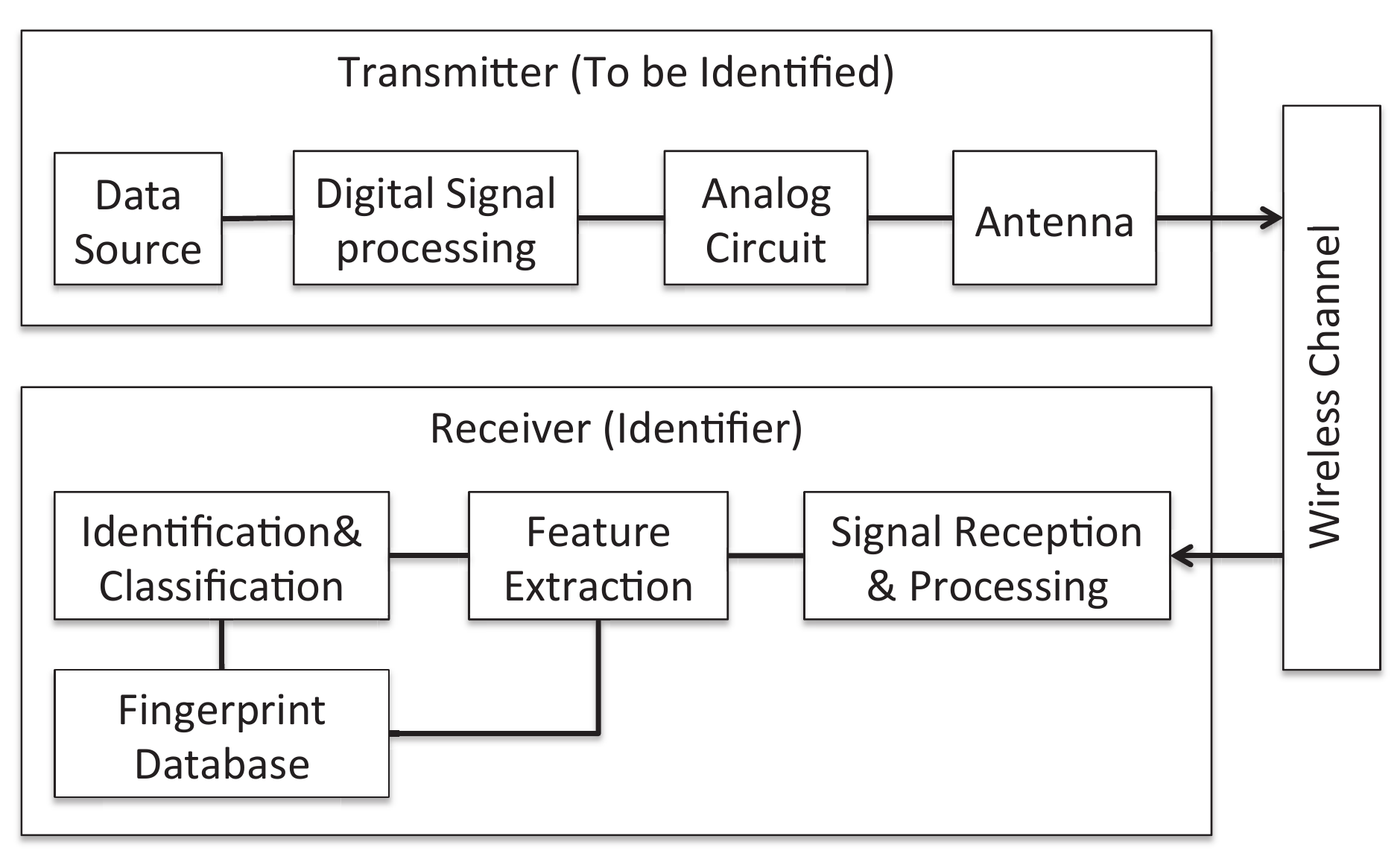}
\vspace{-5pt}
\caption{The logic procedure in wireless physical-layer identification.}
\vspace{-10pt}
\label{fig:system_overview}
\end{figure}

To analyze the reliability and differentiability of RFFs, the origins of various RFFs need to be point out first.
Fig.~\ref{fig:system_overview} illustrate the logic procedure of WPLI during the wireless communication between a transmitter (the device to be identified) and a receiver (the identifier). The RFFs can come from many points along that procedure. At the transmitter side
, the RFFs are rooted in the hardware imperfections of the transmitter device \cite{danev2012physical}, which include clock jitter \cite{zanetti2010physical, jana2010fast}, the DAC sampling error \cite{polak2011identifying}, the mixer or local frequency synthesizer \cite{toonstra1996radio, toonstra1995transient}, the power amplifier non-linearity \cite{polak2011identifying}, \cite{polak2011rf}, \cite{liu2008specific},device antenna \cite{danev2009physical}, modulator sub-circuit (if the analog modulation is used) \cite{brik2008wireless}, among others. The EM waves stamped with the hardware-imperfection-based RFFs are then radiated by the transmitter antenna to the wireless channel, where another type of RFFs are added to the physical waveform, i.e., the unique multipath wireless link characteristics \cite{li2006securing, xiao2007fingerprints}. At the receiver (i.e., the identifier), the RFF-stamped signals are received by the receiver antenna and processed through the analog circuits and digital processing units. Depends on their properties, different RFFs are extracted at (i) different parts of the signals, such as turn-on/off transient \cite{toonstra1995transient, danev2009transient, rehman2012rf}, \cite{hall2004enhancing}; data \cite{zanetti2010physical}, \cite{danev2009physical}, \cite{suski2008using}, \cite{klein2009sensitivity}, \cite{rehman2012analysis}, and clock \cite{zanetti2010physical, jana2010fast}, 
and (ii) different domains, such as time \cite{brik2008wireless}, frequency \cite{danev2009physical}, \cite{danev2009transient}, and wavelet domains \cite{brik2008wireless}. Finally the extracted RFF is matched with the fingerprint database, which complete the physical-layer identification process. 

According to the procedure of WPLI, the factors that can influence the reliability and differentiability of RFFs include: (i) the signal preparing procedure (e.g., modulation and shaping filter) at the transmitter before the RFF is added; (ii) the characteristics of the various hardware imperfections at the transmitter; (iii) the characteristics of the multipath wireless channel and the influence of the operation environments; (iv) the signal reception and processing procedure at the receiver; and (v) the fingerprint extraction and matching strategy at the receiver. Based on the five influential factors, the criteria of reliability and differentiability in WPFL can be specified as: First, the signal generated at the transmitter should provide enough resources (e.g., bandwidth and power) to carry the RFFs to be added. Second, the selected RFFs should be significantly enough for device identification but negligible to communication functionalities. Third, the identification results should not be changed by the effects of EM wave propagations in various wireless channels (e.g., outdoor/indoor or city/rural) or the mobility of users and other objects in the vicinity. Fourth, the RFF-stamped signals can be captured by the off-the-shelf receiver device without additional high-end measuring equipments. Fifth, using the extracted RFFs, the identification/classification algorithm gives an error rate lower than the designed threshold. 


In real and practical applications, it could be challenging for existing WPLI techniques to meet the above five criteria due to various limitations, such as the constraints in receiver devices, the impacts of the complex wireless channels, and the requirements of wireless protocols/regulations. 
First, although the reported error rate of many existing WPLI solutions are impressive (single-digit \% error rate), most existing works use oscilloscopes or spectrum analyzers as the identifier \cite{danev2009physical}, \cite{zanetti2010physical}, which has orders of magnitude higher sampling rate (at least several GHz) than real wireless receivers (e.g., smart phones or wireless sensors). With the much lower sampling rate (usually in the order of MHz), real wireless receivers can lose a large portion of RFFs in higher frequency range and cannot capture the RFFs from the turn-on/off transient of passband signals at all.
In addition, most reported WPLI results are derived in open space with an ideal and fixed wireless channel \cite{danev2009transient}. Many works even just place the receiver right beside the transmitter \cite{danev2009physical}. In fact, practical wireless channel (especially the multipath channel, which is dominant in indoor and metropolitan environments) are highly dynamic and can dramatically distort the transmitted signals as well as the RFF. Consequently, the small difference between devices due to the hardware imperfections will be masked by the strong channel effects. 
Moreover, most existing WPLI solutions require updating RFF database each time when the wireless users or obstructions move or the WPLI system is placed in a new environment. However, mobility is the fundamental requirement of most current wireless applications. It is highly impractical if the identifier has to rebuild the RFF database whenever the users move.
Finally, the close interactions between WPLI and the wireless communications have not been considered in existing works. Actually, while RFFs from severe hardware imperfections or hostile multipath channel effects can deteriorate the communication performance, the communication protocols (e.g., modulation scheme) and wireless regulations (e.g., spectrum mask) can also heavily influence the WPLI.

To address the concern on the reliability and differentiability of RFFs, the modeling and evaluation of the existing WPLI solutions under the influence of the above real-world limitations is of great importance. However, the research in this field is still very limited. There is still no clear answer on whether existing WPLI techniques can fulfill the above five criteria. In \cite{polak2011identifying} and \cite{polak2011rf}, the RF-front-end-based RFF is modeled as nonlinear distortions at the transmitter. Then a hypothesis test-based theoretical model is provided to evaluate the identification accuracy. However, the impacts of receiver device and the wireless channel are not considered. Moreover, while the model provides simple and neat theoretical prediction, it cannot capture the effects of the complex procedures in signal processing procedure of the wireless communication and the fingerprint extraction/match. In \cite{rehman2012analysis}, the influence of the low-end wireless receiver is evaluated through experiments. However, this experiment-based work is limited to one type of devices in one channel, which lacks the universal insights. Similarly, in \cite{danev2009transient}, the stability of the identification accuracy is investigated by experiments in the wireless channel with varied distances and antenna polarizations. The specific-channel experiments also lack the applicability to other wireless environments. All the above evaluation research only focus on specific components in a specific WPLI technique used in a specific environment. To our best knowledge, there is no work that provides a systematic understanding and quantitatively evaluation of the whole WPLI procedure under the aforementioned real-world limitations.

In this paper, through a comprehensive theoretical model and a series of experiments, we rigorously evaluate the reliability and differentiability of WPLI techniques according to the aforementioned five criteria. First, we develop a WPLI model that provides a systematic and mathematic description of the whole WPLI procedure.
More importantly, we then implement the developed model in various WPLI systems that use the spectrum domain RFF due to the non-linear RF front-end. The performance and limitations of that WPLI system is quantitatively evaluated according to our five criteria in different settings of receiver device, wireless channel, and communication protocol. We select the spectrum domain RFF from non-linear RF front-end as the study case because it has relatively low requirement of the receiver hardware while gives very good reported performance . The influence of real-world requirements and constraints are analyzed along each step in the inter-coupled WPLI and wireless communications, including (i) the modulation and filtering of digital signal at the transmitter, (ii) the RFF stamping along the DAC, Mixer, RF front-end amplifier, polarized antennas, and the wireless channel, (iii) the signal propagation along complicated multipath channels with mobile users and obstructions, (iv) the reception, demodulation, sampling of the RFF-stamped signal at the receiver; and (v) the RFF extraction and classification algorithms. 

We conduct in-lab experiments to validate the developed model and our findings. The off-the-shelf wireless devices under investigation include MicaZ wireless sensor nodes\cite{datasheet2006crossbow} and USRP software defined radio platforms \cite{ettus2008universal}, which operate at 2.48 GHz. The influence of multiple factors on WPLI are tested, including the wireless multipath signal propagation, the receiver sampling rate, and fingerprint database updating strategy. 
We use the MicaZ sensors as the transmitter (device to be identified) and USRP as the receiver (identifier), which are deployed in several specially designed indoor multipath fading channels. Due to the reconfigurability of the USRPs, various receiver sampling rates as well as other receiver processing parameters can be changed in real time to highlight the certain influence. 

Through the theoretical modeling and experimental validation, we reveal the limitations of existing WPLI techniques. We find the major challenge comes from the complex and dynamic wireless mobile channel. It is less possible in both theory and real world that the existing WPLI technique can achieve acceptable accuracy in a long distance, non-line-of-sight, fading channel with mobile users and obstructions. Moreover, the more strict physical layer communication protocols (e.g, GSM compared with 802.15) may cause deteriorations of WPLI accuracy, which is due to the regulation of the spectrum mask. In terms of receiver device requirement, higher sampling rate can help increase the identification performance if the noise level is low and the transmitter stamps significant RFFs in wide spectrum. Otherwise, higher sampling rate does not help in reducing error rate due to the accumulation of noise over wider spectrum.

 

The reminder of this paper is organized as follows. The theoretical model is developed to characterize the complete WPLI procedure in Section II. Then, in Section III, the limitations of WPLI are analyzed in details by implementing the theoretical model in various WPLI systems where the RFFs come from the RF-front-end non-linearity. Next, we further validate and discuss our findings by experiments in Section IV. Finally, the paper is concluded in Section V.

\section{Theoretical Model of WPLI}

\begin{figure}[!t]
\centering
\includegraphics [width=3.5in] {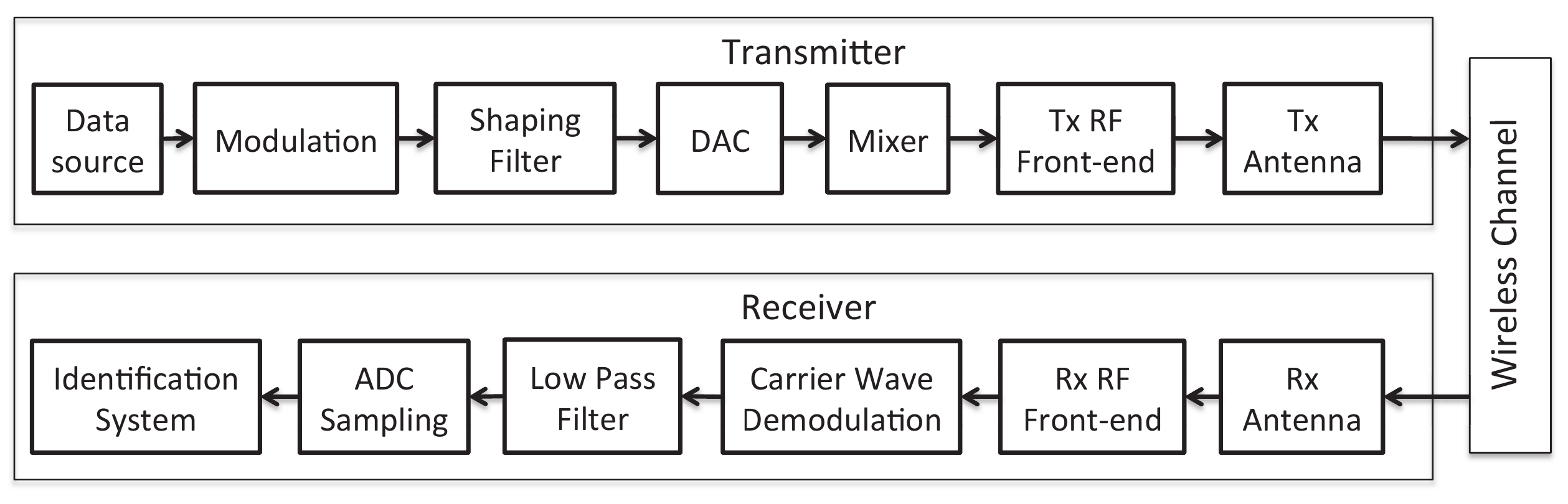}
\caption{WPLI procedure along wireless communication flow block diagram.}
\label{fig_system}
\end{figure}

As the WPLI is closely coupled with wireless communications, the theoretical model of WPLI is also based on the wireless communication procedure, as shown in Fig.~\ref{fig_system}. This section provide general but rigorous mathematical descriptions of the whole WPLI procedure shown in Fig.~\ref{fig_system}, including four major segments: the signal processing and RFF stamping at the transmitter, the signal propagation along wireless channels, the signal reception and feature extraction at the receiver, and the final identification and classification. 

\subsection{Signal Processing and RFF Stamping at Transmitter}

At beginning of the signal processing at the transmitter, the data source send binary bits sequence to the digital modulation block, where the sequential binary bits are first mapped to higher order and complex transmit symbols . We use $m$ to index the sequence of the transmitted symbols so that each symbol is denoted as $x_m=x_m^I+jx_m^Q$, where $x_m^I$ and $x_m^Q$ are the in-phase and quadrature components, respectively. The duration of each symbol is determined by the expected data rate and the order of modulation scheme, which is denoted as $T_{symbol}$. Ideally, $T_{symbol}$ should be a constant in one packet transmission. However, due to the clock-related hardware imperfection, the period/duration of each symbol can vary. As a result, the first possible RFF due to the hardware imperfection appears, which is regarded as a unique time domain feature and is defined as time interval error (TIE) \cite{zanetti2010physical}. We denote this feature as $\sigma^m_{TIE}$ at the $m^{th}$ symbol. Hence, the real symbol duration of the $m^{th}$ symbol becomes $T_m=T+\sigma^m_{TIE}$.

After that the complex symbols $x_m$ is processed by the shaping filter $h_s(t, T_m)$, which gives the output $b_{mod}(x_m, h_s(t, T_m))$, where the formulation of the function $b_{mod}(\cdot)$ is determined by the specific modulation scheme (we will discuss several examples in Section III). Then the digital signal moves to the digital-to-analog converter (DAC). It should be noted that $b_{mod}(x_m, h_s(t, T_m))$ is in fact a desecrate function as the digital shaping filter $h_s(t)$ only change value when the DAC generate new output.
We consider that the DAC has a generation period $T_g$, i.e., the DAC converts the value of $b_{mod}(x_m, h_s(t, T_m))$ to analog current every $T_g$ second. Then the input digital signal $u[n]$ (indexed by $n$) at the DAC can be expressed as
\begin{gather}
\label{eq: mod_signal}
u[n]=A\sum_m b_{mod}(x_m, h_s(nT_g-mT_m, T_m)),
\end{gather}
where $A$ is the amplitude coefficient. It should be noted that the clock-related hardware imperfection could also influence the DAC generation period $T_g$. However, since $T_g$ is much smaller than $T_m$ (due to the high DAC sampling rate), the TIE in $T_g$ is much less significant and can be neglected. It is also worth mentioning that any specific wireless protocol (e.g., 802.11, 802.15, GSM/GPRS, etc.) defines particular modulation schemes and shaping filter, which prepare the transmitted signal with different bandwidth and power spectrum density (PSD). Those parameters can dramatically influence the significance of RFF and eventually affect the performance of WPLI. Detailed examples are discussed in Section III and IV.


Ideally the DAC converts the baseband signal sequence $u[n]$ to a time-continuous analog signal $u(t)$. However, the real output signal $y_u(t)$ from practical DAC is combed with quantization error and integral nonlinearity (INL) \cite{d2010modeling}, which can be expressed as 
\begin{gather}
\label{eq: ADC}
y_u(t)= \sum_{n=-\infty}^{\infty} (u(n)+\Delta_n)g(\frac{t-nT_g}{T_g})+ \Delta_{INL},\notag
\\g(\theta)=
\begin{cases}
1 ,~0\leq \theta <1
\\  0 ,~elsewhere
\end{cases}
\end{gather}
where $\Delta_n$ is quantization noise, for M-bit DAC quantization with input signal dynamic range $[-U, ~U]$, the maximum quantization error $\delta_{\Delta}=2^{-M}U$; 
$\Delta_{INL}$ is the integral nonlinearity that is also regarded as a unique feature rooted in hardware imperfection in \cite{polak2011identifying}. 

After the DAC, the analog baseband signal is moved to passband by the mixer. Ideally the complex passband signal $z(t)$ consists the in-phase component $i(t) \cdot \cos(\omega_c t)$ and the quadrature component $jq(t) \cdot \sin (\omega_c t)$, where $\omega_c$ is the carrier angle frequency and $i(t)=Re\{y_u(t) \},~q(t)=Im\{y_u(t) \}$. However, in practical, the phase offset quadrature error occurs in this procedure due to the imperfection of transmitter's frequency synthesizers (local oscillators), which modifies $z(t)$ to
 \begin{gather}
\label{eq: Mixer_2}
z(t)=\frac{1}{2}( Re\{y_u(t) \}\cdot e^{j\zeta/2}+ Im\{y_u(t) \} \cdot e^{-j\zeta/2})\cdot e^{j\omega_c t}
 \\+\frac{1}{2}(Re\{y_u(t) \}\cdot e^{j\zeta/2}-Im\{y_u(t) \}\cdot e^{-j\zeta/2})\cdot e^{-j\omega_c t} \notag
\end{gather}
where $\zeta$ is the quadrature error. In \cite{toonstra1995transient} and \cite{toonstra1996radio}, the uniqueness of the mixer quadrature error is utilized as RFFs.

Finally, the passband signals go through the TX front-end amplifier and filter to gain enough power for radiation. In this procedure, another important RFF is stamped, i.e., the  non-linearity of the front-end amplifier \cite{gharaibeh2011nonlinear}, which is widely used in many recent WPLI techniques \cite{gharaibeh2011nonlinear, polak2011identifying, polak2011rf, liu2008specific}. The output signal can be formulated as
\begin{gather}
\label{eq: RF_FE_T}
w(t)=h_{PA}(z(t), \mathbf{\tilde{a}_{tx}})\otimes h_{BP}(t)\notag
\\H_{BP}(f)=
\begin{cases}
1, &|f|< W_c
\\  0 , &|f|> W_c
\end{cases}
\end{gather}
where $h_{PA}(\cdot, \mathbf{\tilde{a}_{tx}})$ is nonlinear function of power amplifier at the transmitter (detailed examples are given in Section III); $\mathbf{\tilde{a}_{tx}}$ is complex complex power-series, $\otimes$ stands for convolution computing; and $h_{BP}(t)$ is the bandpass filter function that is nearly ideal within the carrier bandwidth $W_c$. 


\subsection{Signal Propagation between TX and RX Antennas along Wireless Channel}
We consider the antennas as part of the wireless channel since both the antenna polarization and the multipath channel can be used as RFFs in WPLI. With the input signal $w(t)$ given in \eqref{eq: RF_FE_T}, the intensity of the radiated EM waves by a polarized antenna is given by
\begin{gather}
\vec{A}_{tx}(t)=F^h_{tx} [ w(t) ]\cdot e^{-j \phi_{tx}^h} \cdot \vec{\rho}_h+F^v_{tx} [ w(t) ]\cdot e^{-j \phi_{tx}^v} \cdot \vec{\rho}_v
\label{eq: Tx_Antenna}
\end{gather}
where $\vec{\rho}_h$ and $\vec{\rho}_v$ are the direction unit vectors in the horizontal and vertical directions, respectively; $F^h_{tx}(\cdot)$ and $F^v_{tx}(\cdot)$ are the TX antenna projection functions that project the input signal $w(t)$ to the horizontal and vertical polarization, respectively; $\phi_{tx}^h$ and $\phi_{tx}$ are the polarization phases at both directions. In some specific wireless system, such as RFID using the spiral coil antenna, the imperfection of antenna hardware can also contribute to the RFF in WPLI \cite{danev2009physical}. Moreover, even for the simpler antennas that are widely used in contemporary wireless devices (e.g., dipole antennas and patch antennas), although the RFF does not come from antenna imperfection, the random direction of antenna can generate random polarization. Then the antenna polarization can either influence the RFFs due to other hardwares \cite{danev2009transient} or act as a new RFF (coupled with the multipath effect) \cite{li2006securing, xiao2007fingerprints}.

The radiated EM waves then propagate through the multipath wireless channel and finally reach the RX antenna at the receiver. As the RX antenna can be also polarized, the intensity of the EM waves at the RX antenna is also presented in both horizontal and vertical direction: 
\begin{gather}
\vec{A}_{rx}(t)=\sum_{i=1}^{N_{p}} \vec{A}_{tx} (t) \cdot (h_i^h\cdot \vec{\rho}_h + h_i^v \cdot \vec{\rho}_v)  \otimes \delta(t-\tau_i)
\label{eq: Channel}
\end{gather}
where we consider there are $N_p$ paths (including line-of-sight, reflected, diffracted, and refracted paths) connecting the transmitter and receiver; $h_i^h$ and $h_i^V$ are the path loss of the $i^{th}$path with horizontal polarized waves and vertical polarized waves, respectively; $\tau_i$ is the propagation delay of the $i^{th}$path. 

It is obvious that the complex multipath wireless channel can arbitrarily change the amplitude, the delay, the phase of the EM waves at each single path. Consequently, the wireless channel can literally mask most, if not all, RFFs stamped at the transmitter. Hence, the channel effects on WPLI need special attention in the performance evaluation. In \cite{danev2009transient, danev2012physical}, it has been proved that the WPLI accuracy decreases as the distance between the transmitter and receiver increases. It should be noted that the channels considered in \cite{danev2009transient, danev2012physical} are still friendly, which is open space without significant multipath and obstructions. The practical wireless channel, especially in real indoor or metropolitan environments, could be much more hostile in keeping the RFFs from the transmitter. We will do case studies in various multipath channels in the following sections.

It should be also noted that another branch of WPLI systems uses the multipath channel characteristics to authenticate the transmitter (mainly it's location) \cite{li2006securing, xiao2007fingerprints}. The obvious drawback is that such system only works when the transmitter and receiver are fixed and there should be no mobile obstructions in the environment that can change the channel characteristics. However, those conditions are not feasible in most current wireless applications. Hence, we do not further discuss the evaluation of the WPLI systems in this branch.

\subsection{Signal Reception and Processing at Receiver}

At the receiver, the RX antenna captures the EM waves (given in \eqref{eq: Channel}) and convert both the horizontal and vertical polarized components to the received signal:
\begin{gather}
r(t)={F^h_{rx}}^{-1} [ \vec{A}_{rx}(t) \cdot \vec{\rho}_h \cdot e^{j \phi_{rx}^h} ] + {F^v_{rx}}^{-1} [ \vec{A}_{rx}(t) \cdot \vec{\rho}_v \cdot e^{j \phi_{rx}^v} ];
\label{eq: Rx_Antenna}
\end{gather}
where $F^h_{rx}(\cdot)^{-1}$ and $F^v_{rx}(\cdot)^{-1}$ are the RX antenna functions that capture the horizontal and vertical EM wave components, respectively; $\phi_{rx}^h$ and $\phi_{rx}$ are the polarization phases of the RX antenna.

Similar as the transmitter, the received signal also need to pass through the RF front-end at the receiver:
\begin{gather}
 f(t)=h_{PA}(r(t)\otimes h_{BP}(t), \mathbf{\tilde{a}_{rx}})
\label{eq: Rec_signal_FE_T}
\end{gather}
where $h_{PA}(\cdot, \mathbf{\tilde{a}_{rx}})$ is nonlinear function of power amplifier at the receiver; and $h_{BP}(t)$ is the bandpass filter at the receiver, all similar to the case at the transmitter. 

Then the output signal from the RX front-end is convert to baseband by the receiver mixer with quadrature errors (i.e., carrier demodulation). The demodulation signal then goes through the low pass filter $h_{LP}(t)$ to eliminate the higher frequency components, which gives the results of
\begin{gather}
x(t)=\frac{1}{2}( Re\{f(t) \} \cdot e^{j\zeta/2}+ Im\{f(t) \} \cdot e^{-j\zeta/2})\cdot e^{j\omega_c t}\otimes h_{LP}(t)
 \\+\frac{1}{2}(Re\{f(t) \}\cdot  e^{j\zeta/2}-Im\{f(t) \}\cdot  e^{-j\zeta/2})\cdot e^{-j\omega_c t}\otimes h_{LP}(t) \notag
\label{eq: Rec_signal_base_T}
\end{gather}
Note that the low pass filter at the receiver always have a strict cutoff to control the noise level. The gains of the pass-band and stop-band can significantly affect the extracted RFFs, especially the spectrum-domain RFFs. The low pass filter function $h_{LP}(t)$ can be modeled as 
\begin{align}
H_{LP}(f)=
\begin{cases}
A_p , &|f|< W
\\  A_s , &|f|> W
\end{cases}
\label{eq: Filter_F}
\end{align}
where $A_p$ and $A_s$ are the gains in the pass-band and stop-band, respectively.

Finally, the received baseband signals $x(t)$ are sampled by analog-to-digital converter (ADC) to obtain the digital signal sequences which are sent to the identification units to extract the fingerprints. The ADC output is expressed as 
\begin{gather}
 x[n]= x(nT_s)+\Delta_n
\label{eq: ADC_T}
\end{gather}
where $\Delta_n$ is random quantization noise, for M-bit ADC quantization and input dynamic range $[-V, ~V]$, the maximum quantization error is $ \delta_{\Delta}=2^{-M}V$. As discussed in Section I, most existing WPLI solutions utilize the high-end measurement equipment (e.g., oscilloscope and spectrum analyzer) as the receiver. The key difference between the real wireless receivers lies in the sampling rate. While the oscilloscope and spectrum analyzer can directly sample the passband (or the baseband) signal at sampling rate in the order of GHz, most practical wireless device can only sample the baseband signal at the MHz rate. Hence, in the following sections, we will quantitatively analyze the important influence of the receiver's sampling rate.

It should be noted that the above signal reception and processing procedures are also subject to the influence of the hardware imperfections at the receiver's antenna, front end, mixer, and ADC. However, those hardware imperfections are known to the receiver (or the identifier). Hence, the deviation or errors introduced by those hardware imperfections at the receiver can be adjusted and corrected when RFF is extracted.


\subsection{RFF Extraction, Identification, and Classification}

At the identification unit of the receiver, the RFFs are extracted from the sampled digital signals\footnote{The exception is the RFFs residing in the passband turn-on/off transient, which is digital sampled and extracted right after the RX front-end. Our model can be easily adjust to describe such case by deleting the signal processing of mixer.} ($x[n]$ in \eqref{eq: ADC_T}). The extracted feature can be expressed as 
\begin{gather}
\mathbf{S}= Feature(\{x[n], n \in \mathbf{N}\})
\label{eq: feature}
\end{gather}
where $\mathbf{N}$ is the set of all sampled digital signals in this round of identification; $Feature(\cdot)$ is the feature extraction function. Depending on the properties of the RFF and the selected extraction strategy, the feature extraction function $Feature(\cdot)$ usually consists of domain change (e.g., Fourier transform, Wavelet transform, Hilbert transform, among others) and dimensionality reduction (e.g., linear discriminant analysis (LDA), principal component analysis (PCA), or simply picking us a certain parameter such as modulation errors and phase errors ).

By now, the RFF $\mathbf{S}$ is obtained. The next step is to match the RFF with the reference fingerprint, $\mathbf{S}_R$ to calculate the distance matching scores and make the classification or identification decisions. The performance of the classification/identification can be evaluated by the error probabilities, which can be theoretically calculated through hypothesis testing models. 

\subsubsection{Classification}
For $N$-users classification scenario, we adopt $N$-hypothesis testing techniques to apply to $N$ genuine users with $M_N$ reference fingerprints in the database for each user \cite{levine1992training}. The hypothesis $\mathcal{H}_N$ is that the obtained signal is from the genuine user $\#N$. Then the classification probability can be expressed by $P(\mathcal{H}_i|\mathcal{H}_j),~i,~j=1,~...,~N,~ i \neq j$, which is the probability that the RFF from genuine user $\#j$ is classified as the genuine user $\#i$. Such probability can be derived based on the feature distance between the extracted testing feature vector $\mathbf{S}$ with the reference feature vector $\mathbf{S}_R$:
\begin{align}
D_i(\mathbf{S})= Distance(\mathbf{S} ,\mathbf{S}_R )
\label{eq: Distance}
\end{align}

The testing feature vector $\mathbf{S}$ is matched with all the reference fingerprints and assigned to the identity with the smallest distance score. Then the classification error probability can be expressed as
\begin{gather}
P(\mathcal{H}_i|\mathcal{H}_j)= P(D_i=\min (\{D\})|\mathcal{H}_j);
\label{eq: CL_Prob}
\end{gather}
where the formulation of the probability $P(D_i=\min (\{D\})|\mathcal{H}_j)$ depends on the distribution of $D_i(\mathbf{S})$, which is discussed in detail in Section III.
Then the average classification error rate can be derived:
\begin{align}
P_e=\sum_{i=1}^{N} \sum_{j=1,j\neq i}^{N} P(\mathcal{H}_j|\mathcal{H}_i) P(\mathcal{H}_i) 
 \label{eq: CL_Pe}
\end{align}

\subsubsection{Identification}
Unlike the $N$-users classification, in the identification procedure, the distance scores between each authorized class is not the key metric. The whole network can be considered as a two-user hypothesis model \cite{danev2009transient, polak2011identifying, polak2011rf}. $\mathcal{H}_1$ is the hypothesis that the incoming testing fingerprint is from the genuine users while $\mathcal{H}_0$ indicates that the testing fingerprint is from an unknown imposter. In another word, all the $N$ genuine users that have reference fingerprints stored in database can be treated as a whole class. 
A decision threshold $ \lambda$ is set up based on the calculated feature distance in (\ref{eq: LDA_D}). If the minimal distance score is larger than the threshold, the testing feature is identified as an imposter's fingerprint, i.e., $\mathcal{H}_0$, otherwise it is judged as from a genuine user $\mathcal{H}_1$, i.e.,
\begin{gather}
\min (\{D\})\underset{\mathcal{H}_1}{\overset{\mathcal{H}_0}{\gtrless}}  \lambda.
\label{eq:dicision_rule}
\end{gather}

To evaluate the accuracy of identification performance, several probability metrics can be applied \cite{danev2009physical, danev2009transient}, including false accept rate (FAR) $P(\mathcal{H}_1|\mathcal{H}_0)$, false reject rate (FRR) $P(\mathcal{H}_0|\mathcal{H}_1)$, genuine accept rate (GAR) $P(\mathcal{H}_1|\mathcal{H}_1)$, and genuine reject rate (GRR) $P(\mathcal{H}_0|\mathcal{H}_0)$. Based on the decision rule given in \eqref{eq:dicision_rule}, the above probabilities can be calculated in the same way as the classification case according to \eqref{eq: Distance} and \eqref{eq: CL_Prob}. We don't further elaborate the similar formulas here. To generate a single probability metric in identification system, many work put the above error rates together in the Receiver Operating Characteristic (ROC) chart where the FRRs (or GARs) are plotted as a function of different FAR levels. The operating point in ROC, where FAR and FRR are equal, is referred to as the equal error rate (EER).

\section{Real-world Constrains and Requirements: Analysis \\by Implementing the WPLI Model}
In this section, we implement the developed WPLI model in a promising and widely adopted WPLI technique \cite{suski2008using, danev2009transient, danev2009physical, zanetti2010physical}, i.e., the WPLI using frequency domain RFF that originates from the non-linearity of the TX front-end amplifier. The real-world requirements and constraints of the key procedures in WPLI are analyzed. 
It should be noted that, in this section, the RFF only come from the non-linear RF front-end. While other hardware imperfections still exist, they are considered as the hardware noises on top of ideal circuits.

\subsection{Modulation and Shaping Filter: Constraints of Communication Protocols and Regulation}

Preparing the transmitted signal through modulation and shaping filter are the first steps in the WPLI procedure in Fig.~\ref{fig_system}. In fact, the modulation and shaping filter are determined by the specific communication protocol to achieve the expected data rate and to keep the signal power staying within the assigned spectrum band (not interfere other wireless services). The RFF has to be stamped on the resulted signal. Hence, the regulated modulation and shaping filter become the first constraints on WPLI. As mentioned in \eqref{eq: mod_signal}, the function $b_{mod}(\cdot)$ is determined by the specific modulation scheme and shaping filter. Here we implement two widely used modulation schemes: M-PSK and M-QAM. The modulation functions $b_{mod}(\cdot)$ in \eqref{eq: mod_signal} for M-PSK can be further expressed as
\begin{gather}
\label{eq: modulation}
b_{PSK}(x_m, h_s(t))= exp\{j \frac{\pi}{M}x_m \}h_s(t), 
\end{gather}
where $M$ is the modulation order. For M-QAM,
\begin{gather}
\label{eq: modulation}
b_{QAM}(x_m, h_s(t))=x_m h_s(t), 
\end{gather}

Similarly, we implement two widely used shaping filters, including the half-sine (HS) and root-raised-cosine (RRC) filters. The HS shaping is utilized in IEEE 802.15.4 protocols and the RRC is used in GSM cellular systems. The shaping filter function $h_s(t, T_m)$ in \eqref{eq: mod_signal} can be further expressed as
\begin{gather}
\label{eq: shaping}
h_{s,HS}(t, T_m)=sin(\pi\frac{t}{2T_m}) 
\\h_{s,RRC}(t, T_m)=\textstyle 4\beta \frac{\cos((1+\beta) \pi t/T_m) + \sin((1-\beta)\pi t/T_m)/(4\beta t/T_m)}{\pi \sqrt{T_m} (1-(4\beta t/T_m)^2) }
\\0<t<2T_m, \notag
\end{gather}
where $T_m$ is the duration of the $m^{th}$ symbol and $\beta$ is the rolloff factor of RRC shaping filter. After the implementing the specific modulation scheme and shaping filter, the output passband signal $z(t)$ after the processing of the DAC and mixer along WPLI procedure in Fig.~\ref{fig_system} can be analytically characterized. The power spectrum density (PSD) of $z(t)$ using different modulation schemes and shaping filters can be numerically derived. We leave the conclusion of the influence of modulation scheme and shaping filter on RFFs to the next subsection where the RFF (i.e., the front-end non-linearity in this WPLI system) is added to the passband signal.

\subsection{RF Front-end Non-linearity: Significance of Selected RFF}
According to Fig.~\ref{fig_system}, the passband signal $z(t)$ is amplified to gain enough radiation power at the RF front-end, where the RFF considered in this section is added to the signal. The RFF is due to the nonlinear distortion and regrowth of the spectrum at the transmitter RF front-end amplifier, which can be characterized by a complex power-series behavioral model \cite{gard2005impact}. The spectrum of output signal of the RF front-end can be modeled as
\begin{align}
S_{FE}(f)=&  \mathbf{H}( \mathbf{ z},  \mathbf{\tilde{a}_{tx}})=\sum_{n=0}^{\frac{(N-1)}{2}} \sum_{m=0}^{\frac{(N-1)}{2}} \frac{\tilde{a}_{2n+1} \tilde{a}^*_{2m+1}}{2^{2(n+m)}} \notag
\\ & \times  \begin{pmatrix} 2n+1 \\ n+1\end{pmatrix}  \begin{pmatrix} 2m+1 \\ m+1\end{pmatrix} \tilde{S}_{(2n+1)(2m+1)}(f);
\label{eq: RF_FE} 
\end{align}
where $S_{FE}(f)$ is the output signal spectrum after the front-end; $\mathbf{\tilde{a}_{tx}}=\{\tilde{a}_n\}$ is the set of the unique coefficients of the non-linear system; $\tilde{S}_{(2n+1)(2m+1)}(f) $ is the $N^{th}$ odd order autocorrelation spectrum of $z(t)$; $z(t)$ is the input passband signal; and $*$ denotes the conjugate signal. The function $\tilde{S}_{(2n+1)(2m+1)}(f)$ is given by
\begin{align}
\tilde{S}_{(2n+1)(2m+1)}(f) =\int_{-\infty}^\infty\Re_{\tilde{z}_{2n+1}\tilde{z}_{2m+1}}(\tau) e^{-j\omega\tau} d\tau,\notag
\end{align}
where
\begin{align}
& \Re_{\tilde{z}_{2n+1}\tilde{z}_{2m+1}}(\tau) = \lim_{ T \to \infty} \int_{-T}^T \tilde{z}_1^{n+1} (\tilde{z}_{1}^*)^n \tilde{z}_1^{m} (\tilde{z}_{2}^*)^{m+1} dt,\notag
\\& \tilde{z}_1=z(t), ~ \tilde{z}_2=z(t+ \tau), ...\notag
\end{align}
Each non-linear RF front-end has a unique set $\mathbf{\tilde{a}_{tx}}$ that constitutes the foundation of the RFF. In the theoretical modeling in this paper, we derive the coefficient set $\mathbf{\tilde{a}_{tx}}$ by fitting the measured spectrum of the output signal of real wireless devices (MicaZ sensor nodes \cite{datasheet2006crossbow} and USRP software defined radio platform \cite{ettus2008universal}). Note that the antennas in MicaZ sensors and USRPs are simple vertical polarized antennas. Hence, they can be considered as ideal single polarized antennas without obvious imperfection. Hence, the antenna output $\vec{A}_{tx}(t)$ in \eqref{eq: Tx_Antenna} becomes a linear function of the RF front-end output $w(t)$. This fact justify why we can directly use the measured spectrum of the output signal of the MicaZ sensors and USRPs to derive the unique coefficient set $\mathbf{\tilde{a}_{tx}}$.

\begin{figure}[!t]
  \centering
  \subfigure[]{
    \includegraphics [width=1.6in] {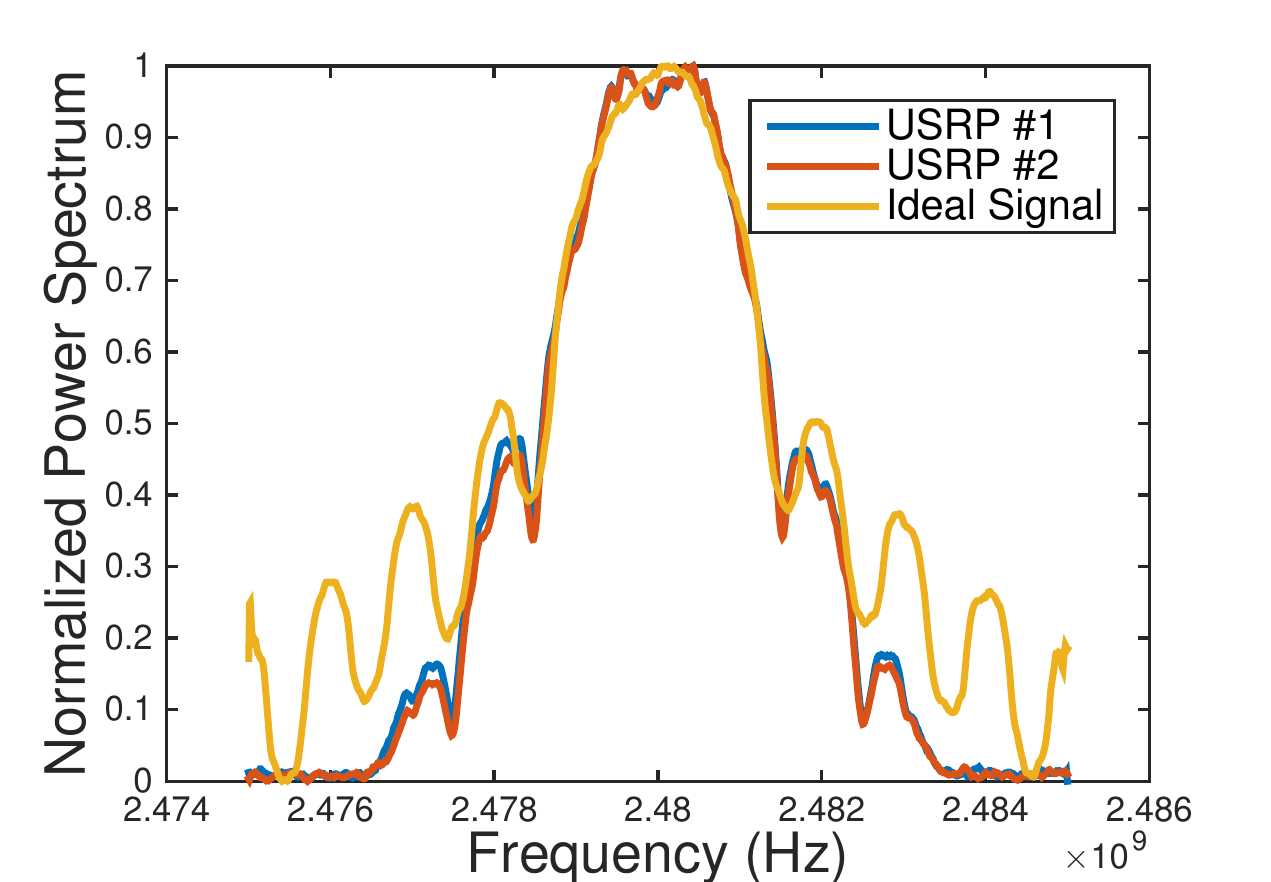}
    \label{Spectrum_USRP_OQPSK_HS}}
  \subfigure[]{
    \includegraphics [width=1.6in] {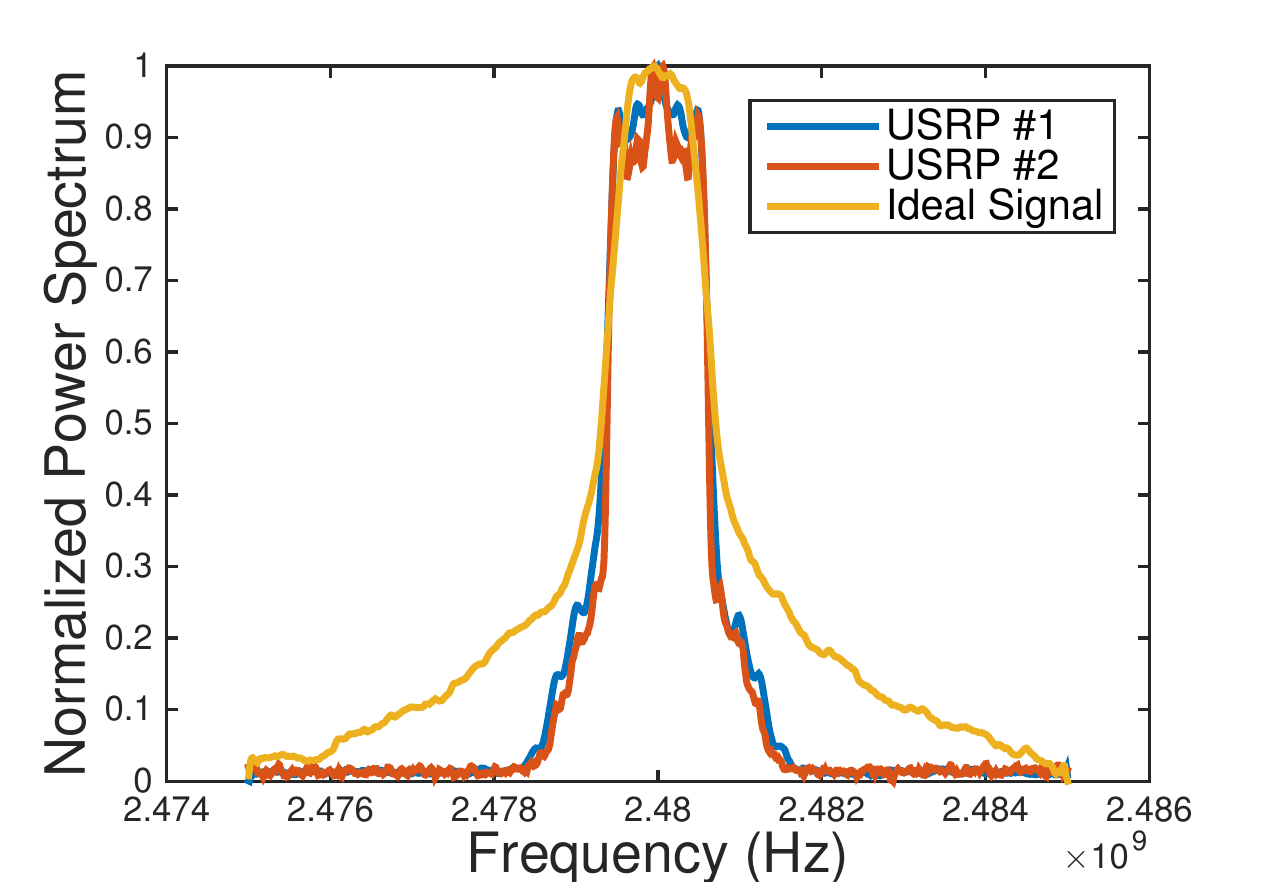}
    \label{Spectrum_USRP_QPSK_RRC}}
    \subfigure[]{
    \includegraphics [width=1.6in] {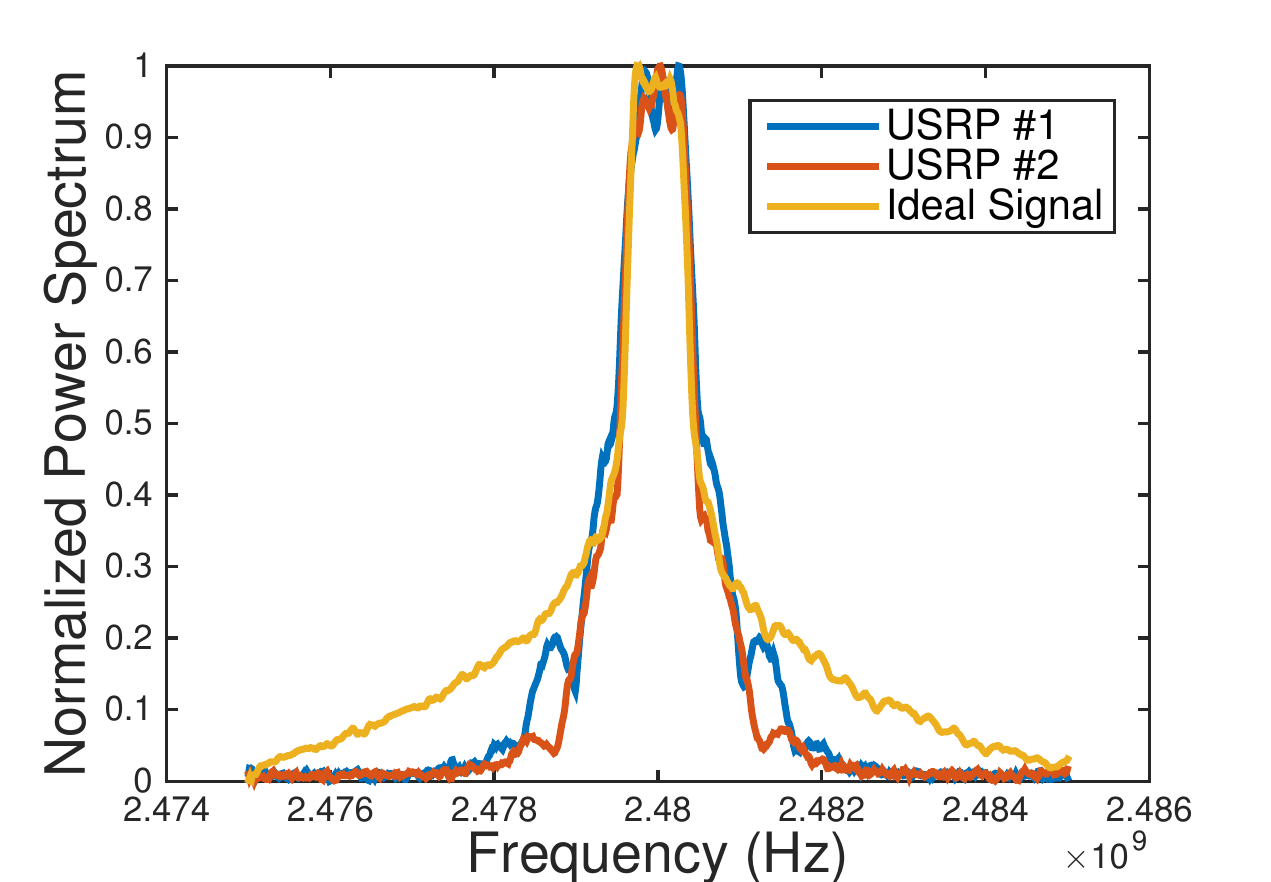}
    \label{Spectrum_USRP_8PSK_RRC}}
     \subfigure[]{
    \includegraphics [width=1.6in] {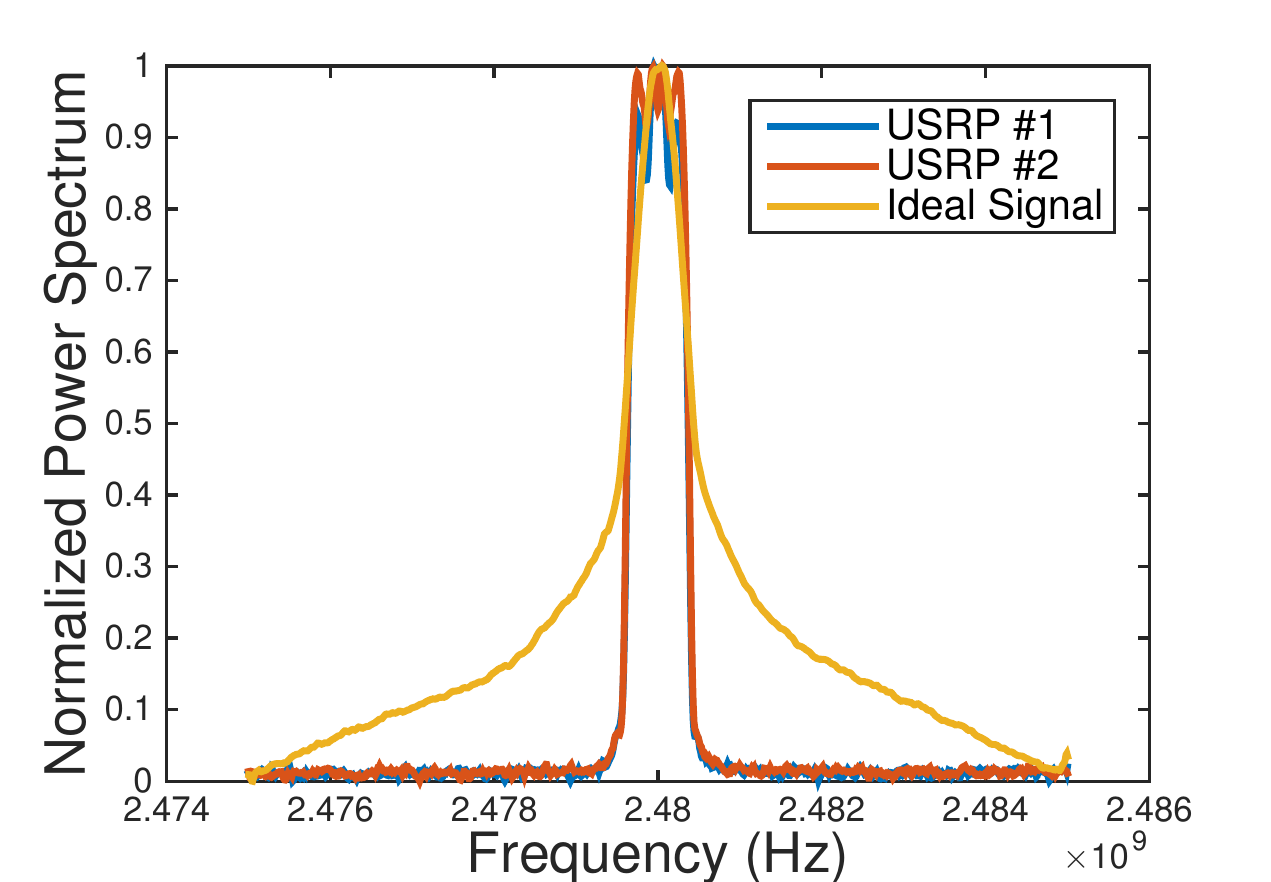}
    \label{Spectrum_USRP_QAM_RRC}}
    \subfigure[]{
     \includegraphics [width=2.5in] {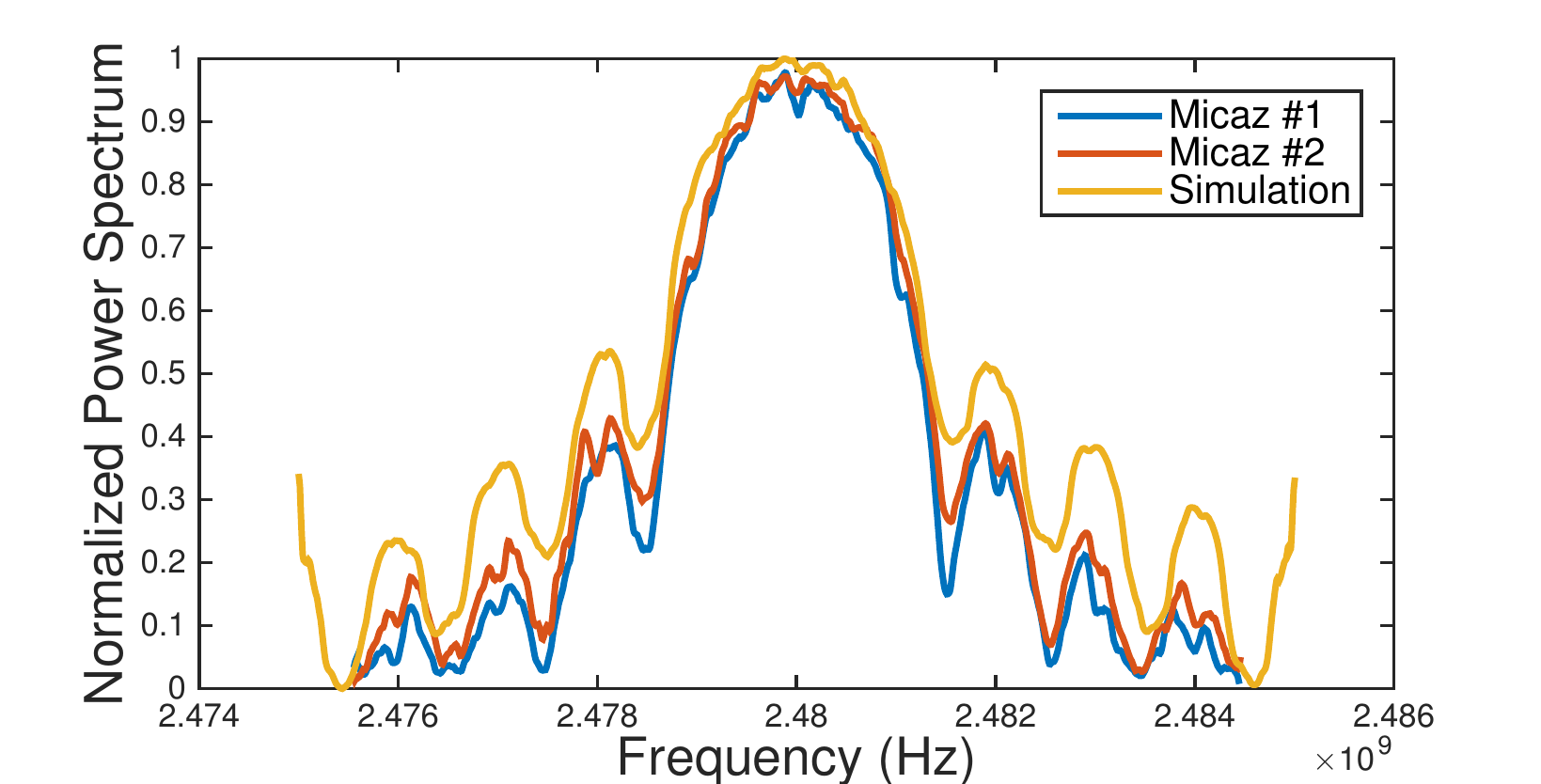}
    \label{Spectrum_sensor}}
  \caption{PSDs of signals (a) using OQPSK and Half-sine Shaping (USRP). (b) using OQPSK and RRC Shaping (USRP). (c) using 8PSK and RRC Shaping (USRP). (d) using 16QAM and RRC Shaping (USRP). (e) using OQPSK and Half-sine Shaping (MicaZ).}
\end{figure}

Now we can numerically check the significance of the selected RFF as well as the influence of modulation scheme and shaping filter. Fig.~\ref{Spectrum_USRP_OQPSK_HS} - Fig.~\ref{Spectrum_USRP_QAM_RRC} show the PSDs of the transmitted wireless signal at the TX side with and without the RFF (front-end non-linearity), using different modulation schemes and shaping filters. The PSD without RFF is derived by our model with the ideal RF front-end. To derive the PSD with RFF, two non-linear system coefficient sets ($\mathbf{\tilde{a}_{tx}}$) are derived based on the measurements from two USRP transmitters, which have the carrier frequency $f_c=2.48GHz$ (IEEE 802.15.4) and the fixed baseband data rate $R_b=2Mbps$. 

Fig.~\ref{Spectrum_USRP_OQPSK_HS} shows the PSDs using OQPSK modulation and half-sine shaping filter. The PSDs with RFF are significantly different from the PSD without RFF. Moreover, the RFF-stamped PSDs of different devices also show obvious uniqueness, which confirms the effectiveness of the selected RFF (at least at the TX side before wireless channel). More importantly, we observe that RFF become more significant as the signal spectrum moves away from the carrier frequency. In another word, RFF is mainly added to the higher frequency segment of the baseband signal. 

In Fig.~\ref{Spectrum_USRP_QPSK_RRC}, the shaping filter is changed from half-sine to RRC. We can see that the half-sine shaping filter can result in a spectrum with more side lobes and larger bandwidth, while the RRC shaping filter strictly restrains the bandwidth and eliminates the leakage and harmonics of the side lobes. From the perspective of the communication regulation, RRC shaping filter definitely performs better in terms of eliminating the interference to other wireless service in the adjacent frequency band. However, since the RFF considered in this paper shows more uniqueness in the spectrum side lobes, the RRC also reduce the effectiveness of the RFF.

Similar effects are shown in different modulation schemes. Fig.~\ref{Spectrum_USRP_QPSK_RRC} and Fig.~\ref{Spectrum_USRP_QAM_RRC} compare the PSDs using 8PSK and 16QAM modulation schemes. As the modulation order increases, the signal power is more restricted in the mail lobes and the RFF uniqueness is only observable in the small bandwidth of the main lobes. 


Besides the USRPs, we also take the non-linear system coefficient sets ($\mathbf{\tilde{a}_{tx}}$) from the measurements of two MicaZ sensors (OQPSK modulation and half-sine shaping filter), which show more obvious nonlinear distortions. In Fig.~\ref{Spectrum_sensor}, the PSDs of the two sensors are shown companying with the ideal signal without fingerprinting distortions. The results confirm that the RFFs due to non-linear front-end amplifier are more significant in the side lobes of the signal spectrum.

\subsection{Wireless Multipath Channel: Impact of Practical Environments}

As discussed in Section II-B, the complex multipath channel can dramatically change the effectiveness of RFFs from the transmitter. While the ray-tracing model given in \eqref{eq: Channel} is very accurate, it lacks the theoretical insights and cannot be further abstracted. Hence, in this subsection, we utilize the widely used statistical multipath channel models, including the AWGN channel, Rayleigh channel, Rician channel, and Nakagami channel, to analyze the channel impacts. 
The Rayleigh model is frequently used to model multipath fading with no direct line-of-sight path. The Rician model is often used when there is one strong direct line-of-sight component and many random weaker components. The Nakagami-m model gives the best fit to land mobile and indoor-mobile multipath propagation. As the parameter $m$ increases to infinite, the Nakagami-m channel converges to a AWGN channel.
We consider the signal bandwidth is smaller than the channel coherent bandwidth \cite{simon2005digital}. As a result, the channel effects on spectrum fingerprints can be considered as a flat spectrum amplitude fading, which consists of a large scale path loss $\alpha_{pl}$ and small scale multipath fading $\alpha_{ch}$ \cite{akyildiz2010wireless}. Then the intensity of the EM waves at the RX antenna can be simplified as 
\begin{gather}
\vec{A}_{rx}(t)=\vec{A}_{tx} (t) \cdot \alpha_{pl} \cdot \alpha_{ch} + N_{AWGN}
\label{eq: statistical_channel}
\end{gather}
where $N_{AWGN}$ is the channel background noise; the large scale path loss $\alpha_{pl}$ in logarithm scale at distance $d$ is given by
\begin{gather}
\alpha_{pl}[dB]=\alpha_{pl}(d_0)-10\eta \cdot log(\frac{d}{d_0})
\label{eq: path_loss}
\end{gather}
where $\alpha_{pl}(d_0)$ is the path loss at the reference distance $d_0$, $\eta$ is the path loss exponent, $X_\sigma$ is a normal random variable.

For the small scale channel effects, while the AWGN channel simply has the $\alpha_{ch}=1$, the probability density function (PDF) of $\alpha_{ch}$ in Rayleigh, Rican, and Nakagami-m channel are \cite{simon2005digital}:
\begin{gather}
\label{eq: multipath}
p_{Ray}(\alpha)=\frac{2\alpha}{\Omega} exp(-\frac{\alpha^2}{\Omega})
 \\p_{Nak}(\alpha)=\frac{2m^m\alpha^{2m-1}}{\Omega^m \Gamma(m)} exp(-\frac{m \alpha^2}{\Omega})\notag
 \\p_{Ric}(\alpha)=\frac{2(1+n^2)e^{-n^2}\alpha}{\Omega} exp \left [-\frac{(1+n^2)\alpha^2}{\Omega} \right ] I_0 \left ( 2n\alpha \sqrt{\frac{1+n^2}{\Omega}} \right)\notag
\end{gather}
where $\alpha$ is channel fading amplitude with mean-square $\Omega=\bar{\alpha^2}$; $m$ is the Nakagami-m fading parameters which ranges from $\frac{1}{2}$ to $\infty$; where $n$ is the Nakagami-n fading parameter is related to the Rician $K$ factor by $K=n^2$; $I_0(\cdot)$ is the zeroth-order modified Bessel function of the first kind.

\begin{figure}[!t]
  \centering
  \subfigure[]{
    \includegraphics [width=1.6in] {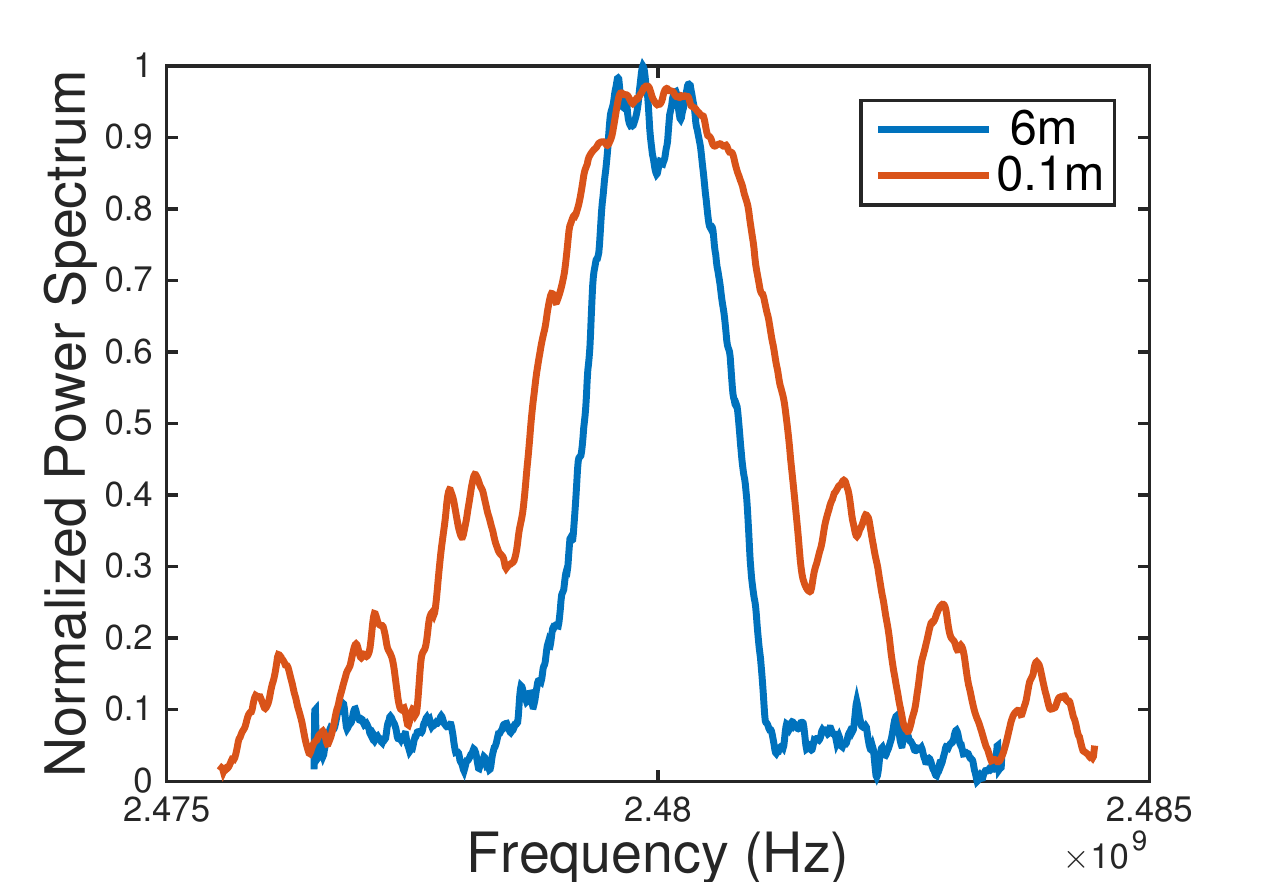}
    \label{fig_spectrum_6m}}
  \subfigure[]{
    \includegraphics [width=1.6in] {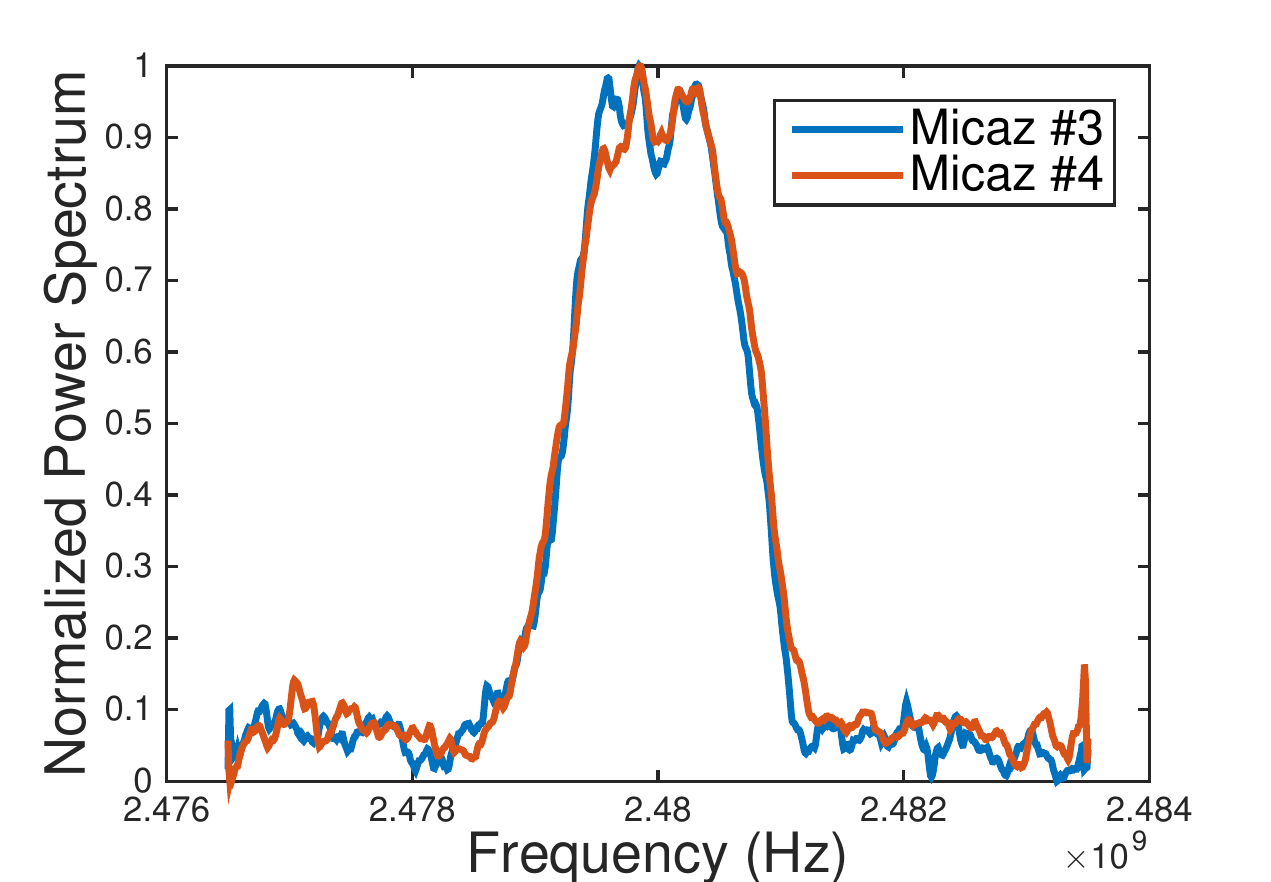}
    \label{fig_spectrum_6m_2}}
  \caption{ (a) PSDs of RFF-stamped signals at different distance. (b) PSDs of RFF-stamped signals from two devices at 6 m.}
\end{figure}

By applying the above channel effects on the PSDs of the radiated signal at the transmitter, Fig.~\ref{fig_spectrum_6m} show how the signal PSD changes along a Rayleigh multipath channel. The red curve is the signal PSD at 0.1m away from the transmitter while the blue curve is the PSD 6m away from the transmitter. A very significant distortion can be observed due to the wireless channel effect.   
Fig.~\ref{fig_spectrum_6m_2} shows the PSDs of the two MicaZ sensors 6m away from the transmitters. Compared with the PSD of the same sensors in Fig.~\ref{Spectrum_sensor} when the distance is 0m, we find that the side lobes of the PSDs (where the most identity information resides) are mostly ruined by channel effects. That is the reason why most WPLI systems need to update the identifiers' fingerprint database and threshold whenever the user moves or the channel condition changes. It also explains why the WPLI performance are poor over the multipath channel.

\subsection{Sampling and FFT: Influence of Receiver Device}
As modeled in Section II-C, the receiver captures/processes the wireless signal and finally derives the sampled digital signals $\{x[n]\}$. Since the WPLI technique considered in this section requires the frequency domain feature, the fast Fourier transform (FFT) is applied to the sampled digital signals, which is actually part of the feature extraction modeled in \eqref{eq: feature}. The output of the $N_{FFT}$-points FFT transformation is: 
\begin{gather}
 X[k]=\sum_{n=0}^{N_{FFT}-1} x[n] \cdot e^{-2\pi i k n/N_{FFT}} 
\label{eq: FFT}
\end{gather}
where $f_s= 1/T_s$ is the ADC sampling rate. The ADC sampling rate $f_s$ as well as the bandwidth of the LPF in \eqref{eq: Filter_F} determine the bandwidth of the FFT spectrum. The FFT point number $N_{FFT}$ controls the frequency domain resolution. According to the FFT properties, $N_{FFT} \geq L$, where $L$ is the length of the sampled signal. In most WPLI systems, $N_{FFT}$ is fixed to the length of the tested data (e.g., the signal preamble \cite{suski2008using}). 

\begin{figure}[!t]
\centering
\includegraphics [width=3.5in] {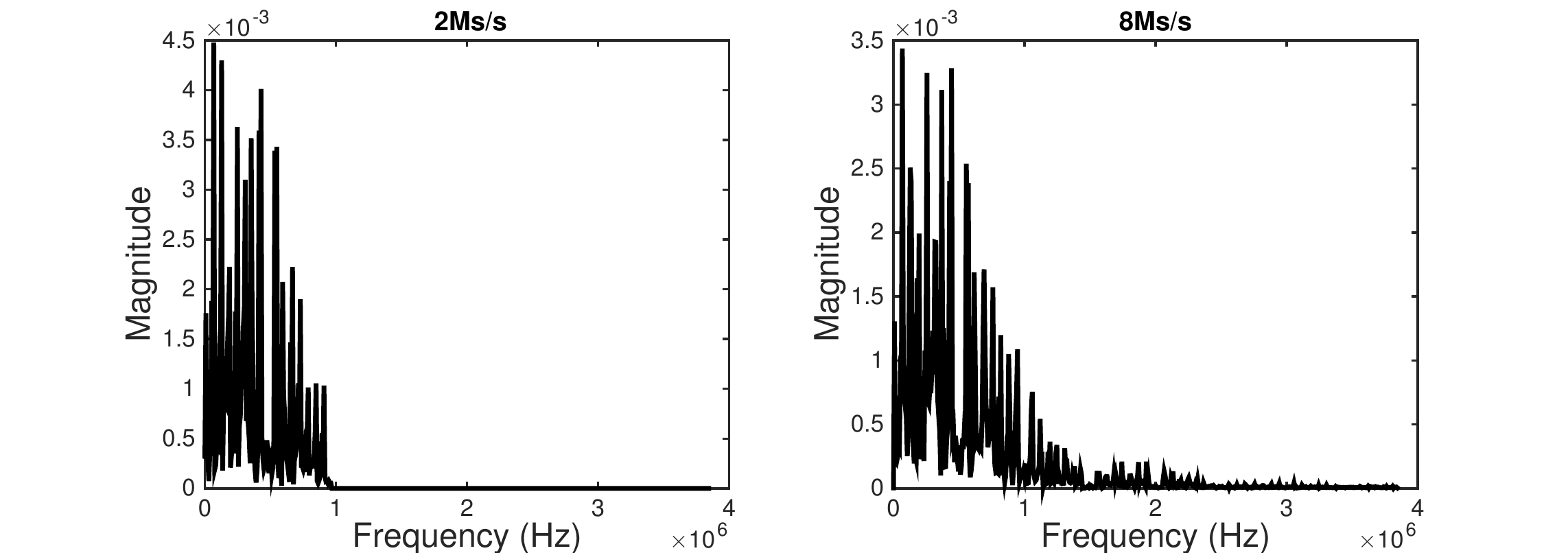}
\caption{FFT PSDs with different sampling rate.}
\label{fig: PSD_fs}
\end{figure}

\begin{figure}[!t]
\centering
\includegraphics [width=3.5in] {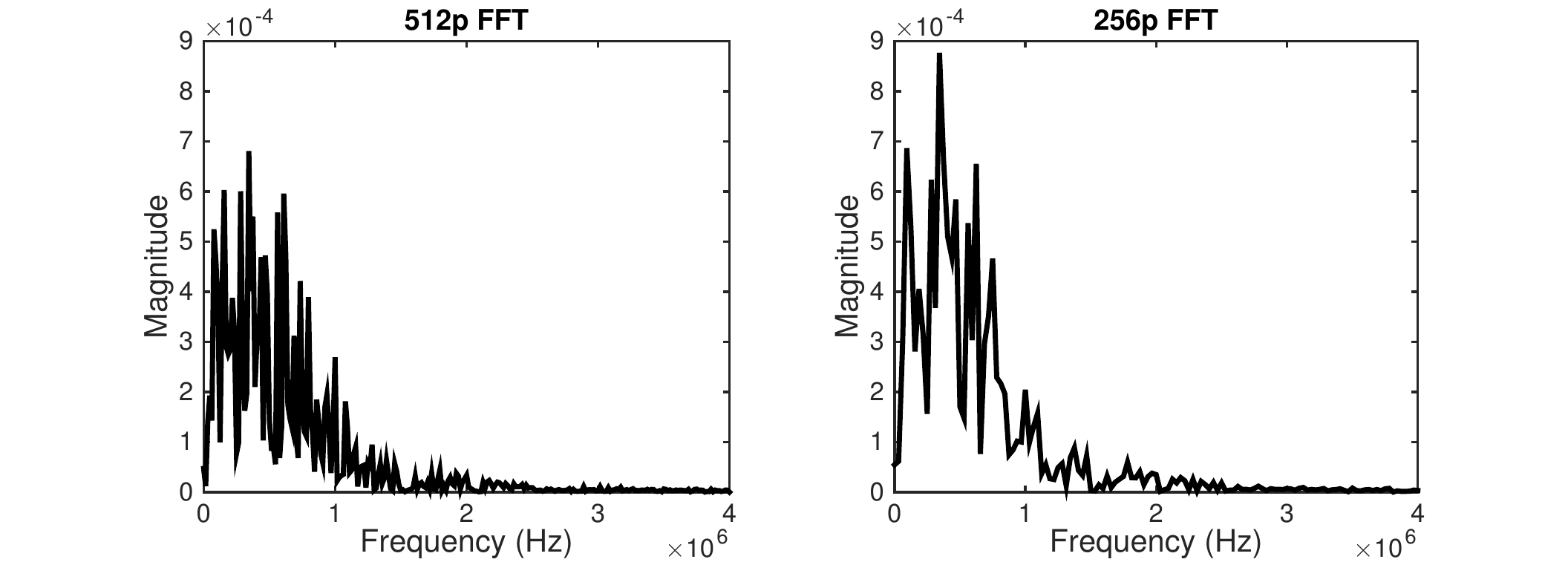}
\caption{FFT PSDs (8Ms/s) with different number of FFT points.}
\label{fig: PSD_FFT}
\end{figure}

The effects of different receiver sampling rates are illustrated in Fig.~\ref{fig: PSD_fs}, where the FFT spectrum from one of the MicaZ sensors in Fig.~\ref{fig_spectrum_6m_2} are plotted with two different sampling rates. We observe that the higher sampling rate keeps more RFF identification information that mainly resides in the side lobes of the signal PSD. However, it should be also noted that the high sampling rate may also increase the noise band. If the SNR become very low in a long distance fading channel, the high sampling rate may not be a favorable option. Existing work keeps SNR high when the high-end measurement equipments are used. 
In the Fig. \ref{fig: PSD_FFT}, the FFT PSDs of the same signal with different FFT points are shown. It is clear that the higher FFT point number increases the frequency domain resolution. However, we don't find clear relationships between the higher FFT resolution with the better identification performance. Part of the reason is that we always keep $N_{FFT}$ larger than the length of sampled signal in time domain. 

\subsection{Error Rate in Classification \& Identification: Final Metrics}

Although the above analysis gives qualitative explanation on the real-world constraints and requirements of WPLI systems, the quantitative evaluation needs the error rates in classification or identification as the final metrics, which is theoretically derived in this subsection. Before calculating the error rates, the RFF needs to be extracted and compared with the reference RFFs.

\subsubsection{Fingerprint Extraction}
As discussed at the beginning of this section, the only RFF considered in this WPLI system is the TX RF front-end non-linearity. The other hardware imperfections are considered as additional noises on top of ideal circuit while the multipath channel effects are considered as random process defined by the statistical channel models. Hence, we can rewrite the FFT result of the sampled digital signal $X[k]$ as a function of the TX baseband signal $u(t)$:
\begin{gather}
\mathbf{X}=\alpha_{pl} \cdot \alpha_{ch} \cdot \mathbf{\tilde{U}} \cdot \mathbf{H_{tx}} \cdot \mathbf{H^T_{rx}}+\mathbf{N};
\label{eq: Rec_siganl_V}
\end{gather}
where $\mathbf{X}$ is the vector $\{X[k], k=1,2,...\}$; the channel effect is represented by the path loss $\alpha_{pl}$, the fading $\alpha_{ch}$, and the background noise (part of $\mathbf{N}$), as modeled in \eqref{eq: statistical_channel}; the RF front-end non-linearity at transmitter is characterized by $\mathbf{H_{tx}}$, which is determined by the a unique coefficient set $\mathbf{\tilde{a}_{tx}}=\{a_n\}$ of the non-linear system at the TX RF front-end described in \eqref{eq: RF_FE}, in particular:
\begin{gather}
 \mathbf{H_{tx}}= \begin{bmatrix} |\tilde{a}_1|^2 & ... & \frac{\tilde{a}_{2n+1} \tilde{a}^*_{2m+1}}{2^{2(n+m)}} \scriptstyle  \begin{pmatrix} 2n+1 \\ n+1\end{pmatrix}  \begin{pmatrix} 2m+1 \\ m+1\end{pmatrix}    \end{bmatrix}^T.  \notag
\label{eq: Rec_siganl_V}
\end{gather}
Similarly, $\mathbf{H_{rx}}$ in \eqref{eq: Rec_siganl_V} describes the non-linearity at the receiver front-end but is considered as a known vector by the identifier. The receiver can cancel the effect of the non-linearity at the receiver front-end before the identification or classification. The influences of all the other hardware imperfections (clock circuit, DAC, and mixer) contribute to the other part of the addictive noise $\mathbf{N}$. In \eqref{eq: Rec_siganl_V}, the TX baseband signal input is represented by a diagonal matrix $\mathbf{\tilde{U}}$:
\begin{gather}
\mathbf{\tilde{U}}=Diag \begin{Bmatrix} \tilde{U}_{11} & ... & \tilde{U}_{2n+1,2m+1}  \end{Bmatrix} \notag;
\end{gather}
where $\tilde{U}_{2n+1,2m+1}$ has the same formulation as the passband autocorrelation spectrum $\tilde{S}_{(2n+1)(2m+1)}$ given in \eqref{eq: RF_FE}; the only difference is that the passband input $z(t)$ is replaced with the baseband input $u(t)$.


After canceling the known influence of the RX front-end non-linearity, the sampled digital signal $\mathbf{X}$ become the fingerprint at the receiver $\mathbf{S}$:
\begin{gather}
\mathbf{S}=\mathbf{X} \cdot \mathbf{H^T_{rx}}^{-1}= \alpha_{pl} \cdot \alpha_{ch} \cdot \mathbf{\tilde{U}} \cdot \mathbf{H_{tx}} +\mathbf{\tilde{N}} 
\label{eq: fingerprint}
\end{gather}
We approximately modeled $\mathbf{\tilde{N}}$ as a Gaussian noise since many independent sources jointly contribute to it.

The reference fingerprints of the authorized users in the database at the identifier $\mathbf{S}_R$ can be expressed by
\begin{gather}
\mathbf{S}_R=\alpha_{pl,R}\cdot \alpha_{ch,R} \cdot \mathbf{\tilde{U}} \cdot \mathbf{H_{tx,R}};
\label{eq: Ref_fingerprint}
\end{gather}
where $\alpha_{pl,R}$ and $\alpha_{ch,R}$ usually do not take effect since the reference fingerprints are obtained when the transmitter is very close to the receiver; all reference and device to be identified use the same transmitted signal packet so that they share the same known $\mathbf{\tilde{U}}$; and $\mathbf{H_{tx,R}}$ is the source of the fingerprints of the reference.


\subsubsection{Fingerprint Matching}
There are multiple ways to calculate the distance between the testing fingerprint $\mathbf{S}$ with the reference fingerprint $\mathbf{S}_R$. The most straightforward way is to calculate the norm of the difference of the two vectors, i.e., $D(\mathbf{S})=||\mathbf{S}-\mathbf{S}_R||$. However, this method treats the difference of each frequency point equally and adds them together. In contrast, as we show in Fig.~\ref{Spectrum_USRP_OQPSK_HS}-Fig.~\ref{Spectrum_sensor}, the uniqueness of the RFFs shows more significant difference at the higher frequency range of the baseband signal. Hence, a more efficient way is to assign higher weight to the frequency points where the uniqueness of the RFFs is more obvious. To this end, the Linear Discriminant Analyses (LDA) strategy is introduced \cite{bishop2006pattern, danev2009transient}, which assigns different weights to different frequency points according to the information from fingerprint database.

Consistency with the previous discussion, the covered bandwidth of the extracted fingerprint vector is determined by the receiver sampling rate, i.e., $BW=f_s/2$. The length of the fingerprint vector is determined by the FFT points i.e. $L_N=N_{FFT}$. The prerequisite of effective LDA is that the receiver sampling rate is high enough to keep the majority of the fingerprint information and the FFT point number is larger than the data/preamble length. Under the LDA scheme, the feature distance $D_i(\mathbf{S})$ between the incoming feature vector $\mathbf{S}$ and the $i^{th}$ reference feature vector $\mathbf{S}_R^i$ is given by \cite{bishop2006pattern}:
\begin{align}
D_i(\mathbf{S})=\frac{ \Big\| \mathbf{W}_{LDA}^T \cdot (\mathbf{S} -\mathbf{S}_R^i) \Big\|}{\sigma\Big(\mathbf{W}_{LDA}^T \cdot \mathbf{S}_R^i\Big)};
\label{eq: LDA_D}
\end{align}
where $\mathbf{W}_{LDA}$ is the LDA feature matrix that is derived based on the fingerprint database of the authorized users;  $(\cdot)^T$ indicate the matrix transpose; $\sigma(\cdot)$ is the standard deviation function of a vector. We do not elaborate the procedures in training the LDA feature matrix. Detailed formulas can be found in \cite{bishop2006pattern}.

\subsubsection{Classification Error Rate}

In device classification, the testing feature vector $\mathbf{S}$ is matched to all the reference fingerprints $\{\mathbf{S}_R^i, i=1,2,...\}$ and assigned to the identity with the smallest distance score. Then the classification decision probability $P(\mathcal{H}_i|\mathcal{H}_j)$ can be calculated according to \eqref{eq: CL_Prob}. Without loss of generality, we consider a two-user scenario to derive the classification probabilities. 

The two class classification scenario is between two genuine user, $\mathcal{H}_1$ and $\mathcal{H}_2$. The feature distance from the incoming vector to the two reference vectors are $D_1$ and $D_2$. Then the classification devision rule becomes $D_{\Delta}\overset{def}=D_1 -D_2 \underset{\mathcal{H}_1}{\overset{\mathcal{H}_2}{\gtrless}} 0$. Then the average classification error rate can be derived:
\begin{align}
P_e=P(D_{\Delta}>0|\mathcal{H}_1) P(\mathcal{H}_1) + P(D_{\Delta}<0|\mathcal{H}_2) P(\mathcal{H}_2);
 \label{eq: LDA_Pe}
\end{align}
where we can safely assume $P(\mathcal{H}_1)=P(\mathcal{H}_2)=0.5$; and
\begin{gather}
 \label{eq: LDA_Prob_1}
 P(D_{\Delta}>0|\mathcal{H}_1) =\int_{ \frac{ \sigma_{\mathbf{G}_R^2}  \mathbf{S}_R^1  -  \sigma_{\mathbf{G}_R^1} \mathbf{S}_R^2    } {\sigma_{\mathbf{G}_R^2} -\sigma_{\mathbf{G}_R^1}} } ^{\mathbf{S}_R^2} \frac{ \| W  \| (\sigma_{\mathbf{G}_R^2}-\sigma_{\mathbf{G}_R^1})}{\sigma_{\mathbf{G}_R^1}\sigma_{\mathbf{G}_R^2}} p(\mathbf{S}|\mathcal{H}_1)d\mathbf{S}.
\end{gather}
Note that $P(D_{\Delta}<0|\mathcal{H}_2)$ can be derived in the same way. 

In order to compute $ p(\mathbf{S}|\mathcal{H}_i)$, we define $\mathbf{L}\overset{def}=\alpha_{pl}  \alpha_{ch} \cdot \mathbf{\tilde{U}} \cdot \mathbf{H_{tx}} $ so that $\mathbf{S}=\mathbf{L}+\mathbf{\tilde{N}}$ according to \eqref{eq: fingerprint}. Then, the conditional probability density $ p(\mathbf{S}|\mathcal{H}_i)$ can be derived by
\begin{gather}
  p(\mathbf{S}|\mathcal{H}_i)=\int_{-\infty}^{+\infty} p_L(\mathbf{x}|\mathcal{H}_i) p_{\tilde{N}}(\mathbf{S}-\mathbf{x}) d\mathbf{x};
 \label{eq: condi_prob}
\end{gather}
where $p_{\tilde{N}}$ is the probability density of a zero-mean Gaussian variable with a predetermined variance; $p_L(\mathbf{x}|\mathcal{H}_i)$ is determined by the statistical channel models given in \eqref{eq: multipath}:
\begin{align}
 \label{eq: LDA_Prob_ch}
&p_L^{Ray}(\mathbf{L}|\mathcal{H}_i)=\frac{2 \| (\alpha_{pl}^{-1}\mathbf{\tilde{U}}^{-1}\mathbf{L}\mathbf{H^i_{tx}}^{-1}) \|}{\Omega} 
 \cdot e^{-\frac{(\| \alpha_{pl}^{-1}\mathbf{\tilde{U}}^{-1}\mathbf{L}\mathbf{H^i_{tx}}^{-1}\| )^2}{\Omega}}
\\
&p_L^{Nak}(\mathbf{L}|\mathcal{H}_i)=\frac{2m^m(\| \alpha_{pl}^{-1}\mathbf{\tilde{U}}^{-1}\mathbf{L}\mathbf{H^i_{tx}}^{-1}\|)^{2m-1}}{\Omega^m \Gamma(m)} 
 \cdot e^{-\frac{m (\|\alpha_{pl}^{-1}\mathbf{\tilde{U}}^{-1}\mathbf{L}\mathbf{H^i_{tx}}^{-1}\|)^2}{\Omega}} \notag
\\
&p_L^{Ric}(\mathbf{L}|\mathcal{H}_i)=\frac{2(1\!+\!n^2)e^{-n^2}(\| \alpha_{pl}^{-1}\mathbf{\tilde{U}}^{-1}\mathbf{L}\mathbf{H^2_{tx}}^{-1}\| )}{\Omega} \!\cdot\! e^ {-\frac{(1+n^2)(\| \alpha_{pl}^{-1}\mathbf{\tilde{U}}^{-1}\mathbf{L}\mathbf{H^i_{tx}}^{\!\!-1}\| )^2}{\Omega}}\notag
\\
 &\quad\quad\quad\quad\quad\quad\quad\quad\quad\cdot I_0 \left ( 2n(\| \alpha_{pl}^{-1}\mathbf{\tilde{U}}^{-1}\mathbf{L}\mathbf{H^i_{tx}}^{-1}\| ) \sqrt{\frac{1+n^2}{\Omega}} \right) \notag
\end{align}
Note that the AWGN channel does not have multipath fading. Hence, $\mathbf{S}$ becomes a Gaussian variable with the same variance but different mean values, whose condition probability can be easily derived.

\begin{figure}[!t]
  \centering
  \subfigure[]{
    \includegraphics [width=2.5in] {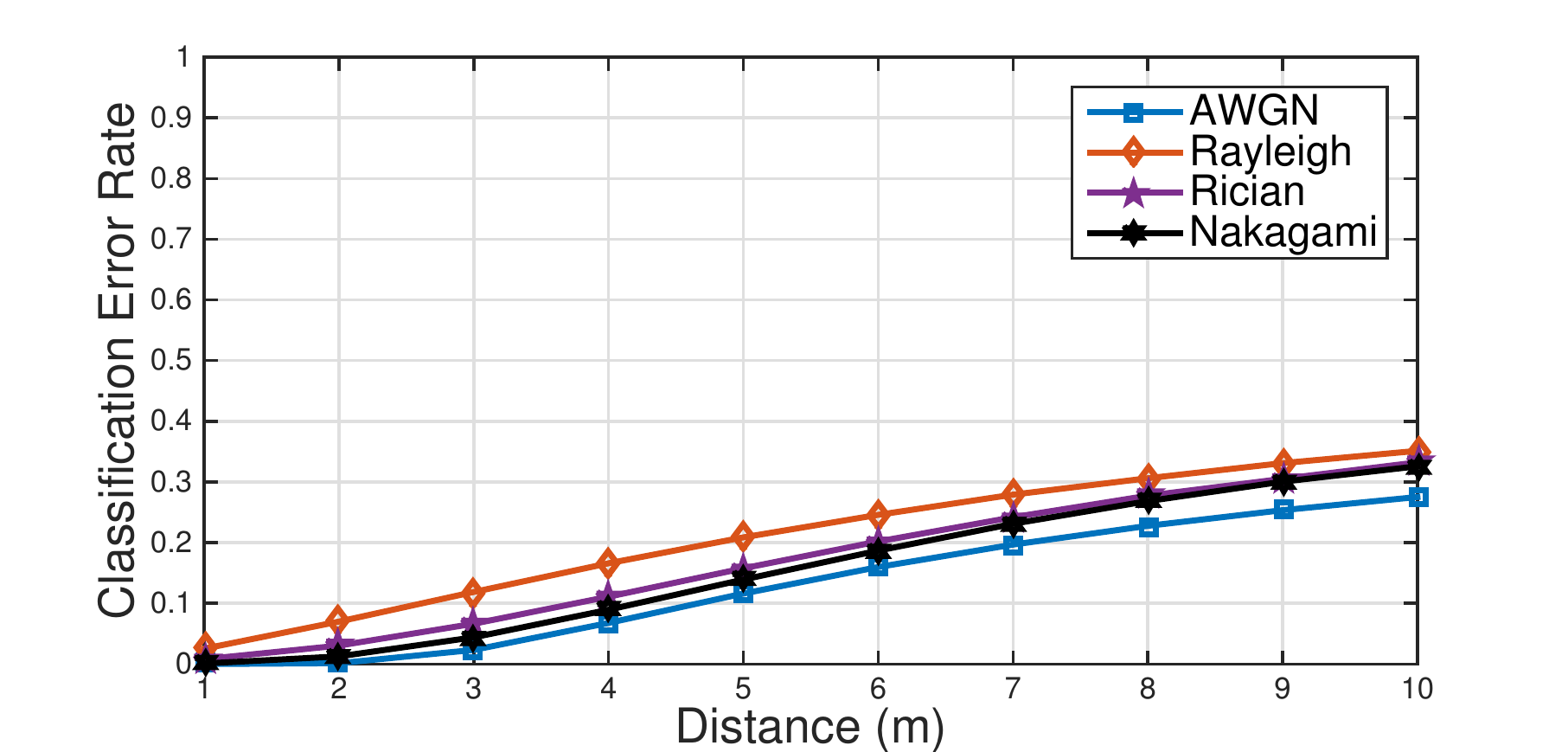}
    \label{fig: Ch_class_updated}}
  \subfigure[]{
    \includegraphics [width=2.5in] {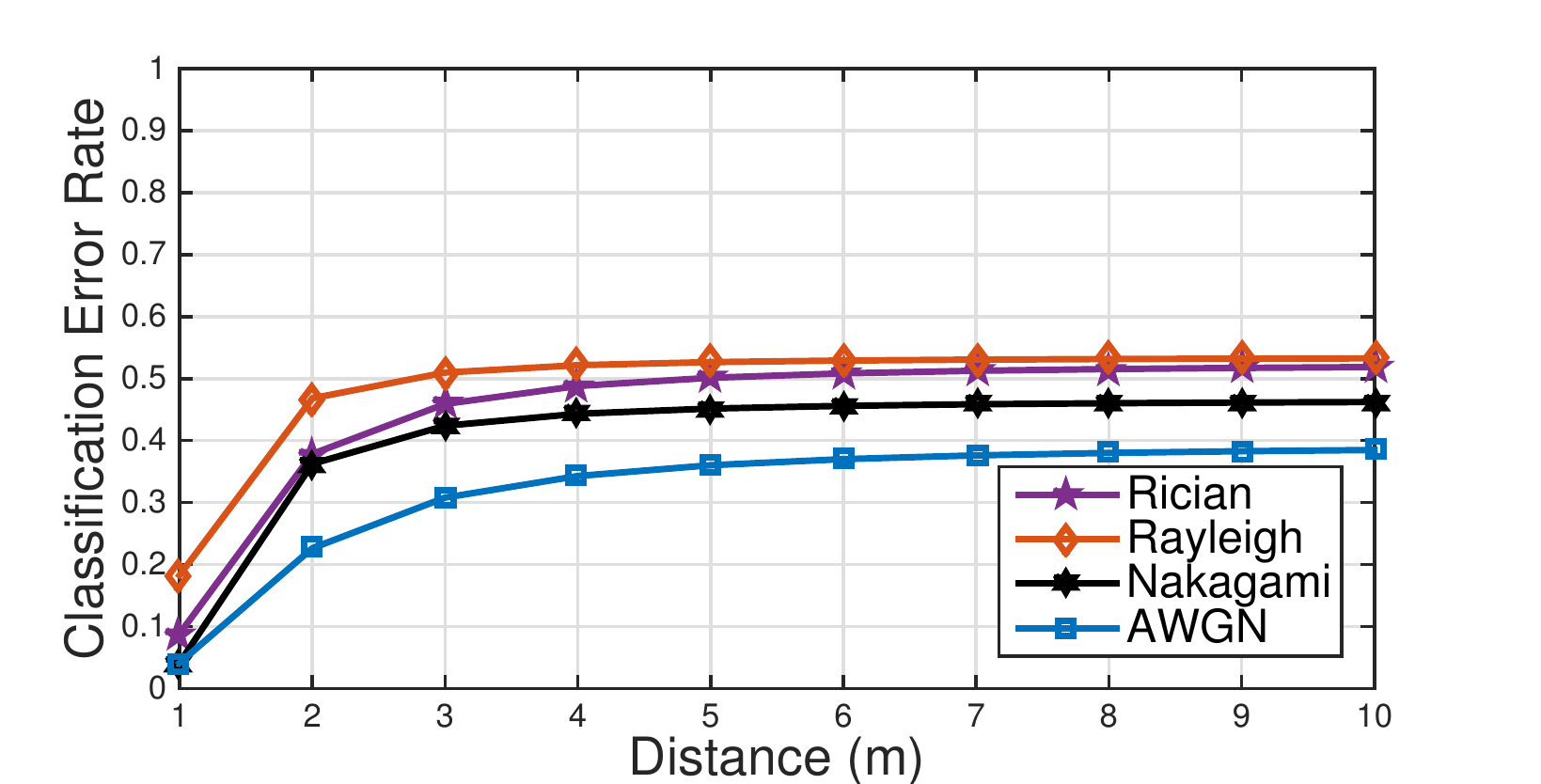}
    \label{fig: Ch_class_fixed}}
  \caption{Classification error rates over different channels. (a) with updated database. (b) with fixed database.}
  \vspace{-10pt}
  \label{fig: Ch_class_2}
\end{figure}

Based on \eqref{eq: LDA_Pe}-\eqref{eq: LDA_Prob_ch}, we can numerically evaluate the classification error rate in different channels (AWGN, Rayleigh, Nakagami $m=3$, Rician $K=4$) and with different receiver sampling rate (2M-8MHz). When different channel/distance is considered, we need to first determine the strategy on how to build the fingerprint database. On the one hand, in most existing work \cite{danev2009transient}, the fingerprint database is updated with the new channel characteristics whenever the transmitter or receiver change location. On the other hand, we believe the frequent database update is not feasible in many wireless applications. Hence, we evaluate both strategies. Fig.~\ref{fig: Ch_class_updated} shows the classification error rate of different channel as a function of distance. The fingerprint database is updated at each point along the curves. Even with this infeasible strategy, we observe that the error rate increases dramatically if the transmission distance increases or the multipath channel is involved. Due to the lack of line-of-sight path, the Rayleigh channel gives the worse performance while the more friendly AWGN channel has the best performance. In Fig.~\ref{fig: Ch_class_fixed}, we fix the reference fingerprint (measured at the very close distance, 0.1 m) and plot error curves at different distances over different channels. We observe very poor classification performance. The error rate approach 50\% (i.e., loss of classification capability) within 10 m in the Rayleigh channel.

\begin{figure}[!t]
\centering
\includegraphics [width=2.3in] {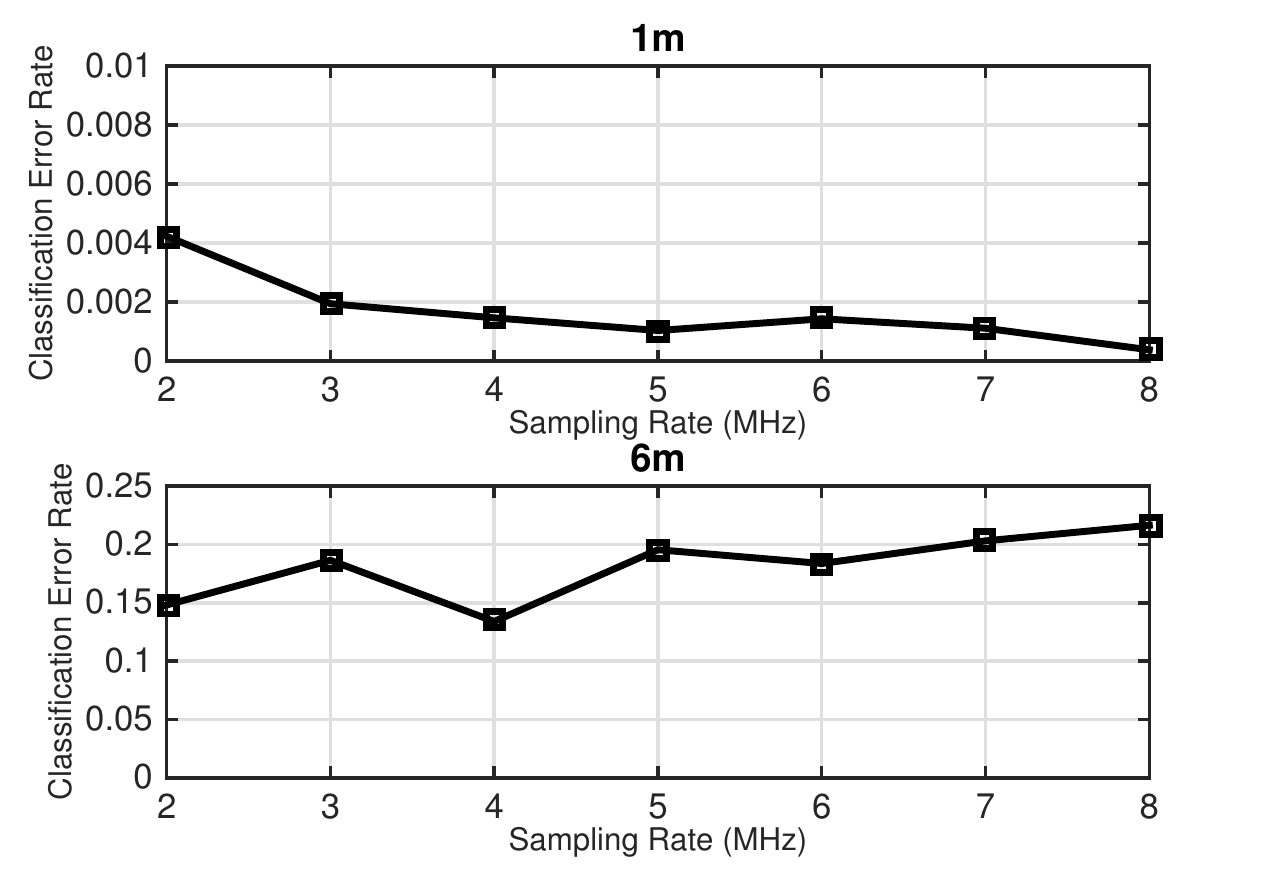}
\caption{Classification error rates with different receiver sampling rates.}
  \vspace{-20pt}
\label{fig: Sampling_CL}
\end{figure}

Besides the channel effect, we also evaluate the influence of the receiver sampling rate on the classification error rates. In Fig.~\ref{fig: Sampling_CL}, the numerical results of average classification rates due to different sampling rate are shown. To avoid the multipath effects and also to show the trade off in sampling rates, we calculate the numerical results at 1m and 6m under AWGN channel. At the close distance, where the SNR level of the channel is high (25dB), the classification error rate decreases with the increasing of sampling rate. This is because more  side-lobe information of spectrum are covered by the fingerprinting sampling bandwidth. At 6m distance, the classification performance become worse in higher sampling rate situation. The reason for this phenomena is that most side-lobe information are ruined by increased noise band when the channel SNR level is low (15dB). Hence the choice of the sampling rate is the trade off between the different application scenarios due to the analyses in the previous section.

\begin{figure*}[t]
\begin{minipage}[t]{0.3\linewidth}
\includegraphics [width=2.5in] {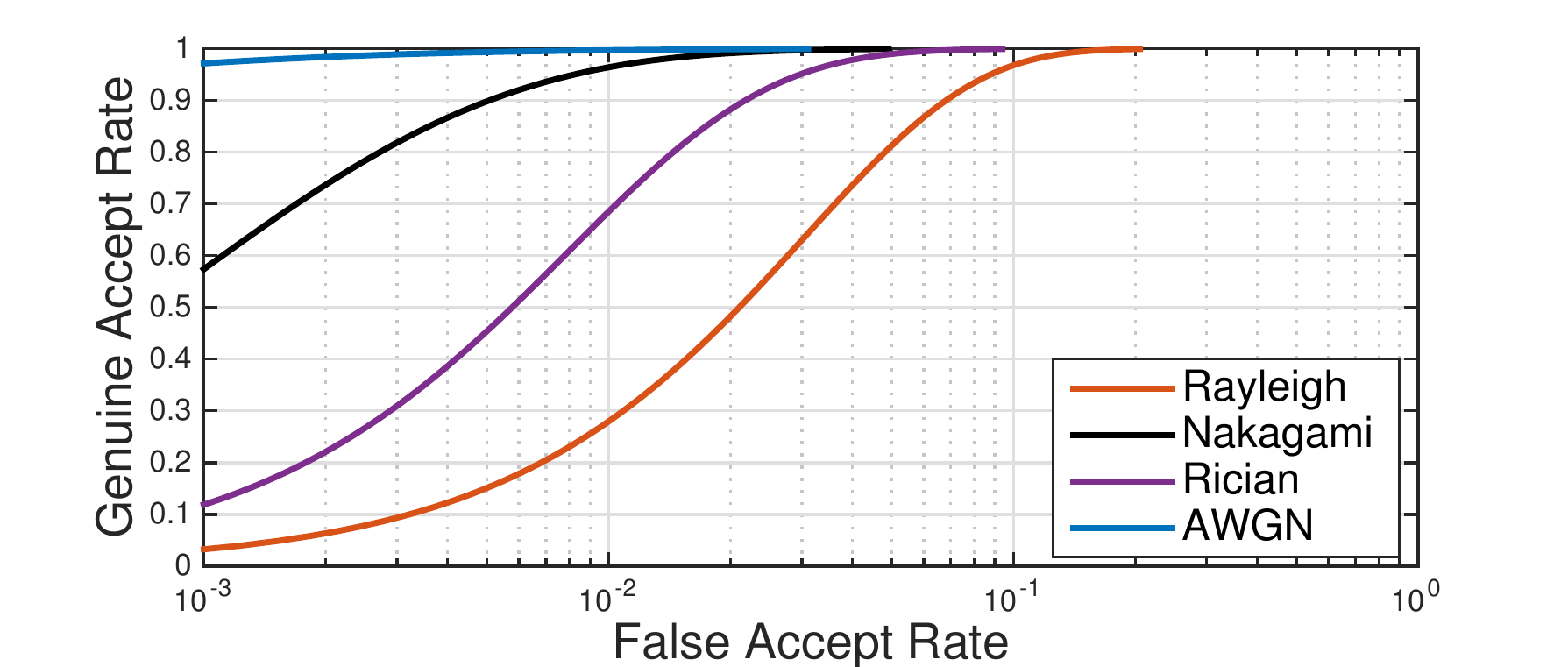}
\caption{Theoretical ROC curves over different channels.}
\label{fig: ROC_1}
\end{minipage}
\quad
\begin{minipage}[t]{0.3\linewidth}
\includegraphics [width=2.5in] {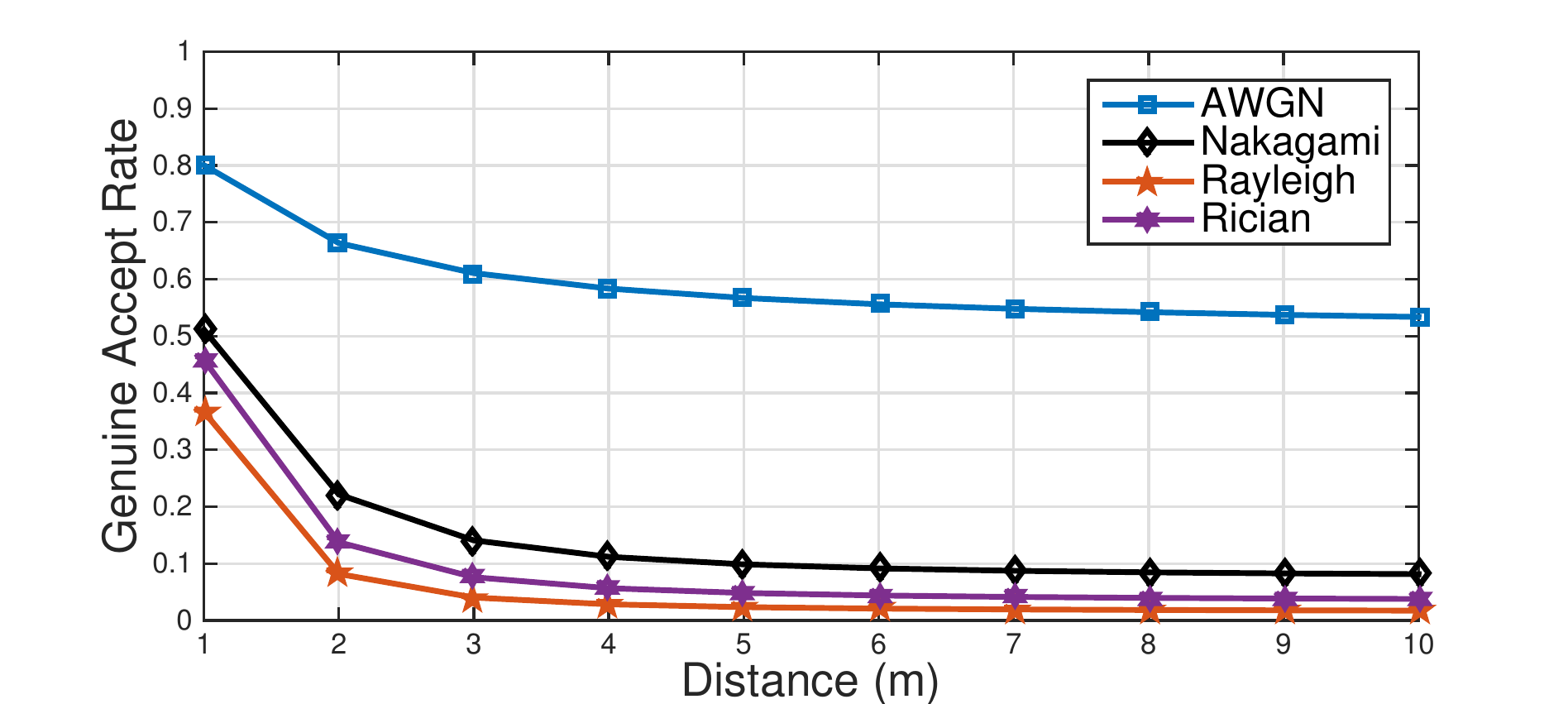}
\caption{Genuine Accept Rate over different channels (with fixed database).}
\label{fig: Ch_fixed}
\end{minipage}
\quad\quad
\begin{minipage}[t]{0.3\linewidth}
\includegraphics [width=2.5in] {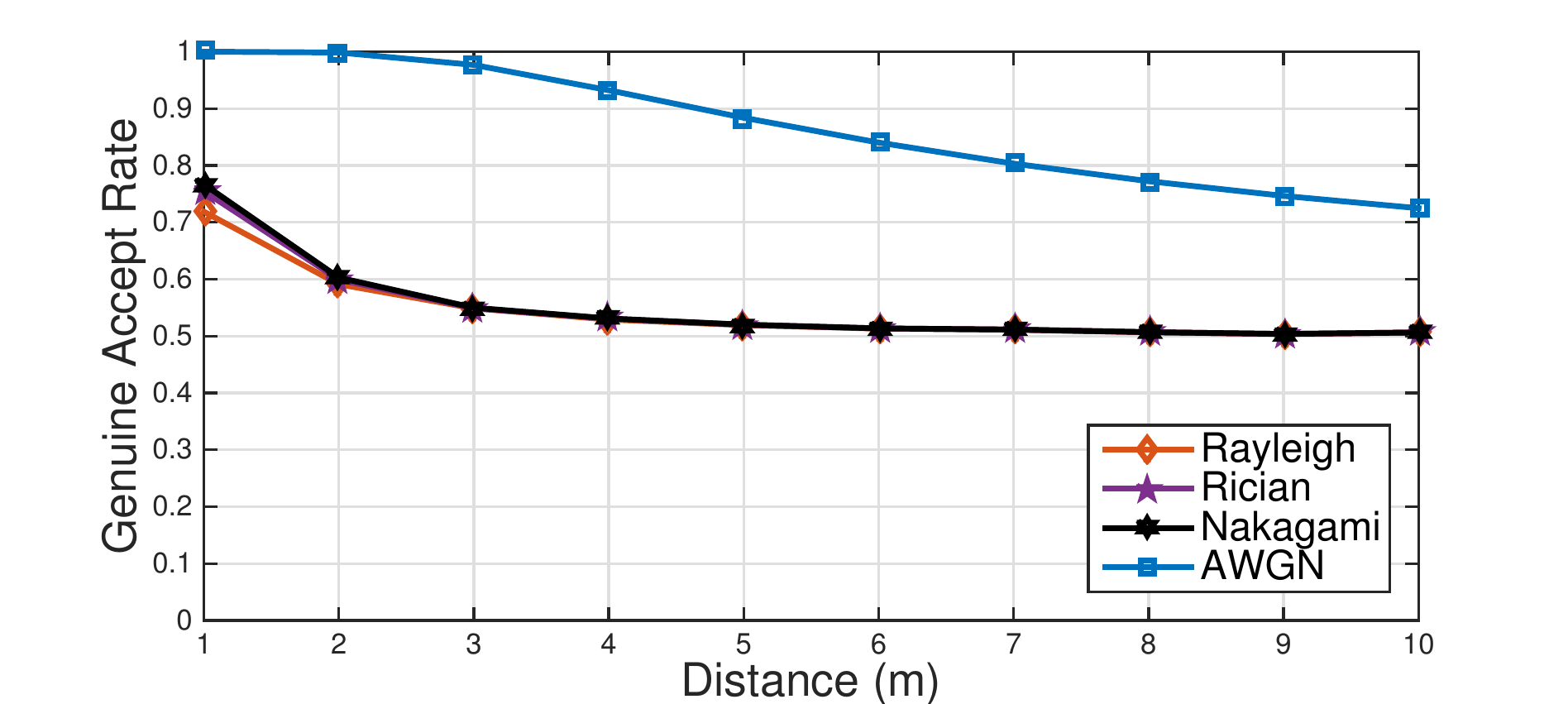}
\caption{Genuine Accept Rate over different channels (with updated database).}
\label{fig: Ch_updated}
\end{minipage}
\end{figure*}

\subsubsection{Identification Error Rate}
Different from the classification, authorized user/imposter detection can be considered as a two-user hypothesis model. Due to the lack of knowledge on imposters' fingerprints, we cannot know which frequency point carries more significant identification information. As a result, the LDA scheme assigns the same weights for all frequency points of the feature vectors. Then the identification simply relies on the direct difference between the incoming feature vector and the reference vector $ \mathbf{d} =\mathbf{S}-\mathbf{S}_{R}^1 $.
As previously defined, $\mathcal{H}_1$ denotes the authorized users while $\mathcal{H}_0$ denotes imposters. According to the decision rule in \eqref{eq:dicision_rule}, the judging threshold $\lambda$ needs to be determined first by finding the EER point in the ROC chart. In most existing work, $\lambda$ is updated whenever the fingerprint database of authorized users is updated. However, we can also change the strategy by fixing the database and threshold after the first update. The difference between the incoming feature and the reference under $\mathcal{H}_1$ can be deduced as
 \begin{align}
 \label{eq: H_1}
 \mathbf{d}|\mathcal{H}_1 & =\mathbf{S}_1-\mathbf{S}_{R}^1 
  \\&= \alpha_{pl,1}\cdot \alpha_{ch,1} \cdot \mathbf{\tilde{U}} \cdot \mathbf{H_{tx}^1} +\mathbf{\tilde{N}}
   -\alpha_{pl,1}\cdot \alpha_{ch,1} \cdot \mathbf{\tilde{U}} \cdot \mathbf{H_{tx}^1}  \notag
 \\&=\mathbf{\tilde{N}}. \notag
\end{align}
Similarly,
\begin{align}
\label{eq: H_0}
 \mathbf{d}|\mathcal{H}_0 &  = \mathbf{S}_0-\mathbf{S}_R  
 \\&= (\alpha_{pl,0}\cdot \alpha_{ch,0} \cdot \mathbf{\tilde{U}} \cdot \mathbf{H_{un}}
  -\alpha_{pl,1}\cdot \alpha_{ch,1} \cdot \mathbf{\tilde{U}} \cdot \mathbf{H_{tx}^1}) +\mathbf{\tilde{N}}  \notag
\end{align}
where $\mathbf{H_{un}}$ is RF front-end series of the imposter (unknown to identifier). According to \eqref{eq: H_1} and \eqref{eq: H_0}, the identification problem can be simplified as a signal detection problem:  $\mathcal{H}_1$ indicates that the difference vector $\mathbf{d}$ is a zero-mean Gaussian noise; while $\mathcal{H}_0$ indicates that $\mathbf{d}$ is a signal under the influence of the multipath fading channel and the Gaussian noise. Then we can apply the classical energy detection model \cite{digham2003energy} that detects unknown signal over fading channels. In AWGN channels, the identification GRR can be calculated by
\begin{align}
 P_{AWGN}(\mathcal{H}_0|\mathcal{H}_0)=Q_u(\sqrt{2\gamma},\sqrt{\lambda}),
\label{eq: Pd_AWGN}
\end{align}
where $Q_u(..)$ is the generalized Marcum Q-function, $\gamma$ is the SNR of difference vector $\mathbf{d}$ over the Gaussian noise $\mathbf{\tilde{N}}$; and $\lambda$ is the judging threshold. In Rayleigh channels, the identification GRR is
\begin{align}
\label{eq: Pd_Ray} 
   P(\mathcal{H}_0|\mathcal{H}_0)_{Ray}&=e^{-\frac{\lambda}{2}} \sum_{n=0}^{\mu-2} \frac{1}{n!} \left( \frac{\lambda}{2} \right)^2 
\\&\quad\quad\quad\textstyle  +\left( \frac{1+\gamma}{\gamma} \right)^{\mu-1} \left [ e^{ - \frac{\lambda}{2(1+\gamma)} }- e^{ - \frac{\lambda}{2} } e^{-\frac{\lambda}{2}} \sum_{n=0}^{\mu-2} \frac{1}{n!} \frac{\lambda\gamma}{2(1+\gamma)} \right ], \notag
\end{align}
where $\mu$ is the time bandwidth product.
And in Nakagami-m channels, identification GRR is:
\begin{gather}
P(\mathcal{H}_0|\mathcal{H}_0)_{Nak} =\alpha \left [ G_1  + \beta  \sum_{n=1}^{\mu-1} \frac{(\lambda/2)^n}{2(n!)} ~_1F_1 \scriptstyle \left(m; n+1; \frac{\lambda}{2} \frac{\gamma}{m+\gamma} \right) \right ] \notag
\end{gather}
where $m$ is the Nakagami-m parameter;$~_1F_1(...)$ is the confluent hypergeometric function; $\alpha$ and $G_1$ are parts of the equations which can be find in the specific form in \cite{digham2003energy}.  Finally, in Rician channels, GRR is given by
\begin{align}
P(\mathcal{H}_0|\mathcal{H}_0)_{Ric}=Q  \left( \sqrt{ \frac{2K\gamma}{K+1+\gamma}}, \sqrt{ \frac{\lambda(K+1)}{K+1+\gamma}} \right),
\label{eq: Pd_Ric}
\end{align}
where $K$ is the Rician factor; $Q(..)=Q_1(..)$ is the  first-order Marcum Q-function.

For all the four types of channels (AWGN, Rayleigh, Nakagami, and Rician), the FRR remains the same, which is
\begin{align}
P(\mathcal{H}_0|\mathcal{H}_1)=\frac{\Gamma(\mu,\frac{\lambda}{2})}{\Gamma(\mu)}.
\label{eq: Pf}
\end{align}
where $\Gamma(..)$ is the incomplete gamma function. 

It should be noted that the GRRs and FRRs given in \eqref{eq: Pd_AWGN}-\eqref{eq: Pf} are all functions of the SNR $\gamma$. Here the SNR is not the ratio of the total signal over noise but the ratio of feature difference over noise. Since the feature of imposters is unknown, higher receiver sampling rate can help increase $\gamma$ by involving more possible significant device differences in a wider spectrum, if the overall noise level is low. However, if the noise level is high, higher receiver sampling rate may reduce $\gamma$ and harm the identification performance.

Based on the derived model, we calculate the theoretical error rate to evaluate the influence of the wireless channel and receiver sampling rate. Similar to the settings in the classifications, we derive the authorized user's fingerprint and judging threshold at a very close distance (0.1m). Fig.~\ref{fig: ROC_1} shows the ROC curves of different channel types at the distance of 0.1 m. Then we choose the EER point (0.1\%) on the AWGN curve as the judging threshold.
In Fig.~\ref{fig: Ch_fixed}, we fix the threshold and reference fingerprints. The identification GARs at different distance over different channels are plotted. We observe that if the reference fingerprint and threshold are fixed, the identification performance dramatically deteriorates as the transmission distance increases or the multipath channel model is applied. In contrast, Fig.~\ref{fig: Ch_updated} shows the identification GARs when the reference database and judging threshold get updates at each plot point. Obvious performance improvements are observed but the influence of longer transmission distance and the multipath channel is still significant. 

\begin{figure}[!t]
\centering
\includegraphics [width=2.5in] {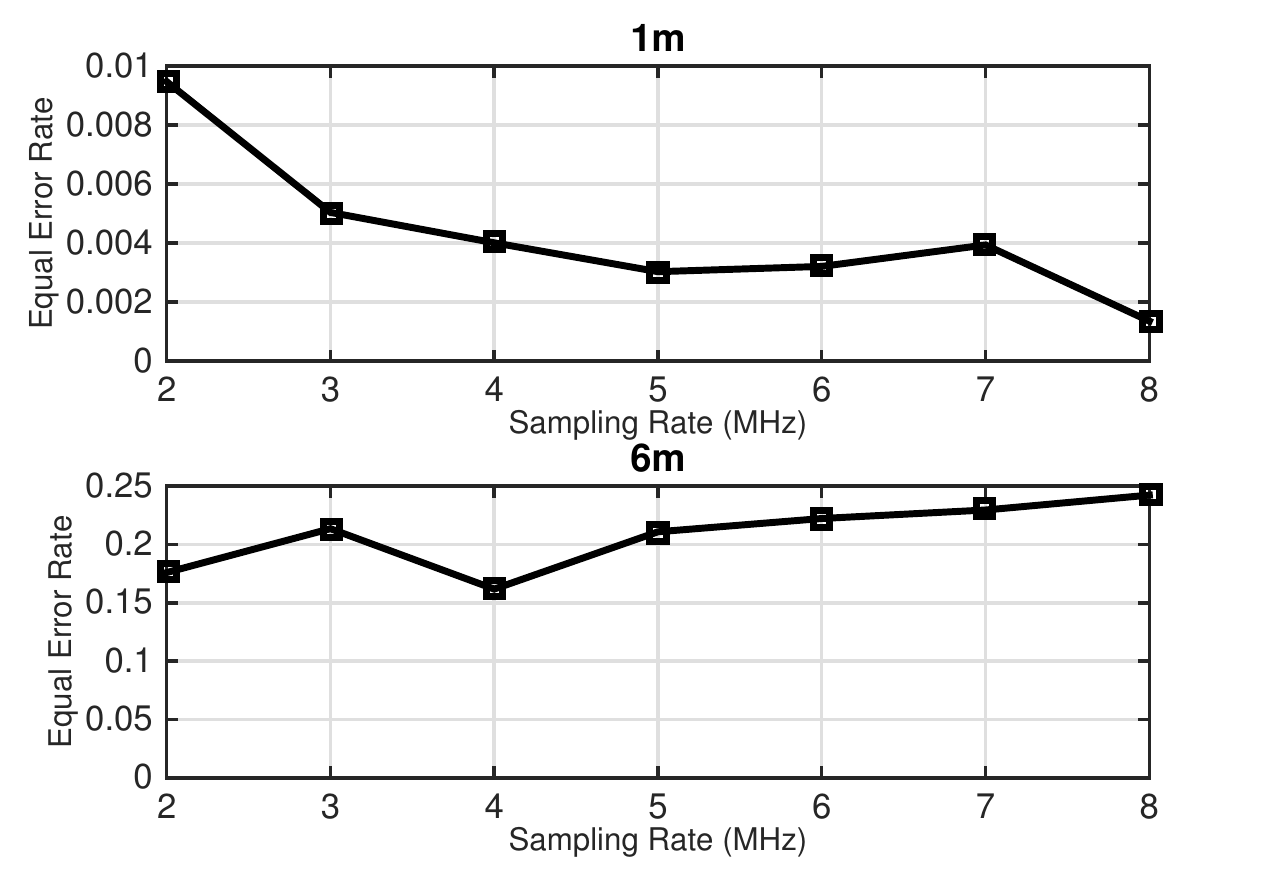}
\caption{Equal Error Rates with different receiver sampling rates.}
\label{fig: Sampling_ID}
\end{figure}

In Fig.~\ref{fig: Sampling_ID},  the numerical results of equal error rate due to different sampling rate are shown. Similarly, we calculate the numerical results at 1m and 6m under AWGN channel. At the close distance, the equal error rate decreases with the increasing of sampling rate. However, at longer distance, the performance of identification become worse with the increasing of sampling rate. Hence the choice of the sampling rate is still a trade off for identification performance under different application scenarios. 

\section{Experimental Analysis}
In this section, we experimentally validate the real-world constraints of WPLI analyzed in Section III. The experimental setup and scenarios are first described. Then the influence of wireless channels, receiver sampling rates, and FFT points on the identification and classification performance of WPLI are analyzed through a series of experiments.


\subsection{Experimental Setup and Scenarios}

\begin{figure}
  \centering
    \includegraphics[width=3.5in]{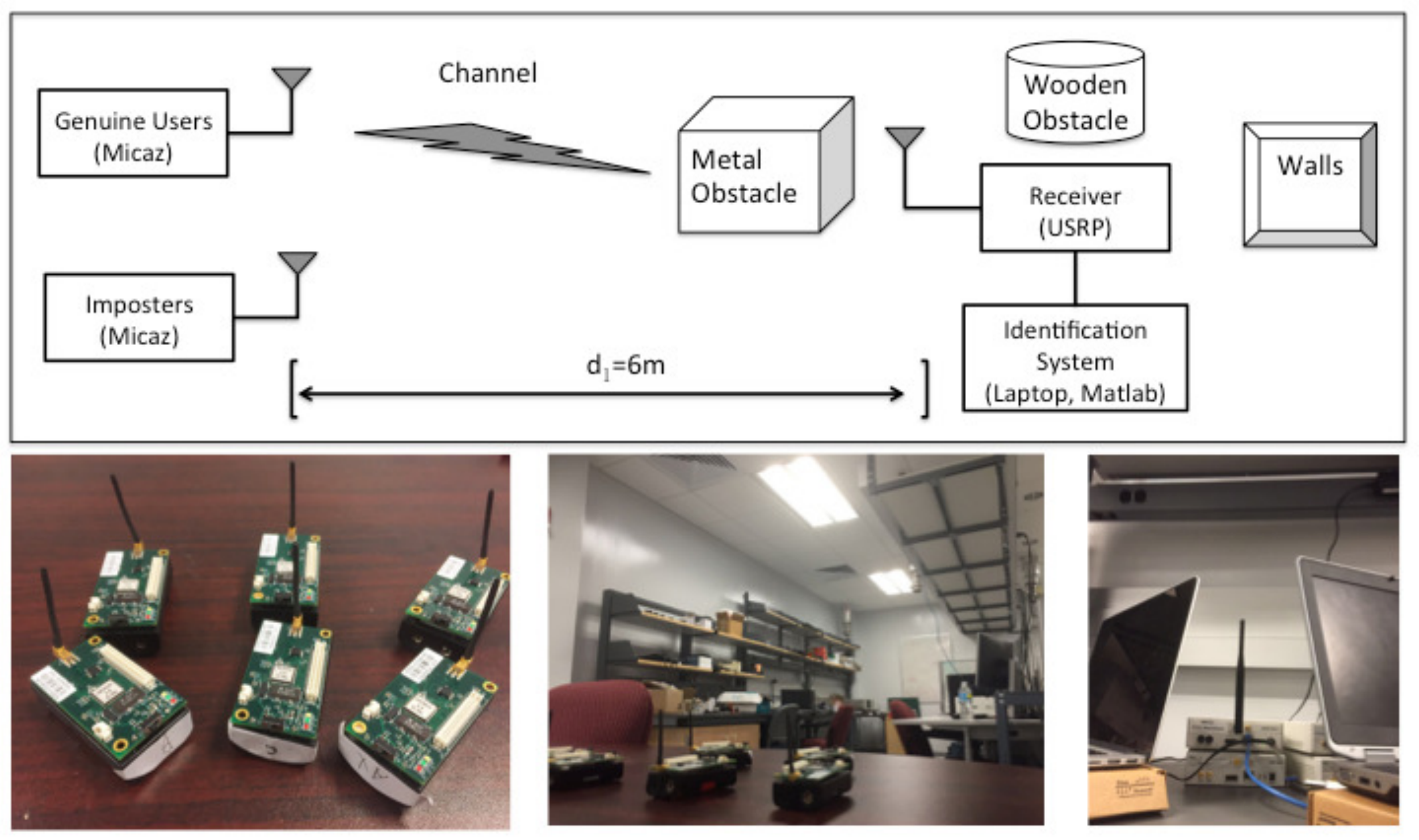}
  \caption{Experimental setup.}
    \vspace{-5pt}
  \label{fig:exp_setup}
\end{figure}

\begin{figure*}[t]
\begin{minipage}[t]{0.3\linewidth}
\centering
\includegraphics [width=2.2in] {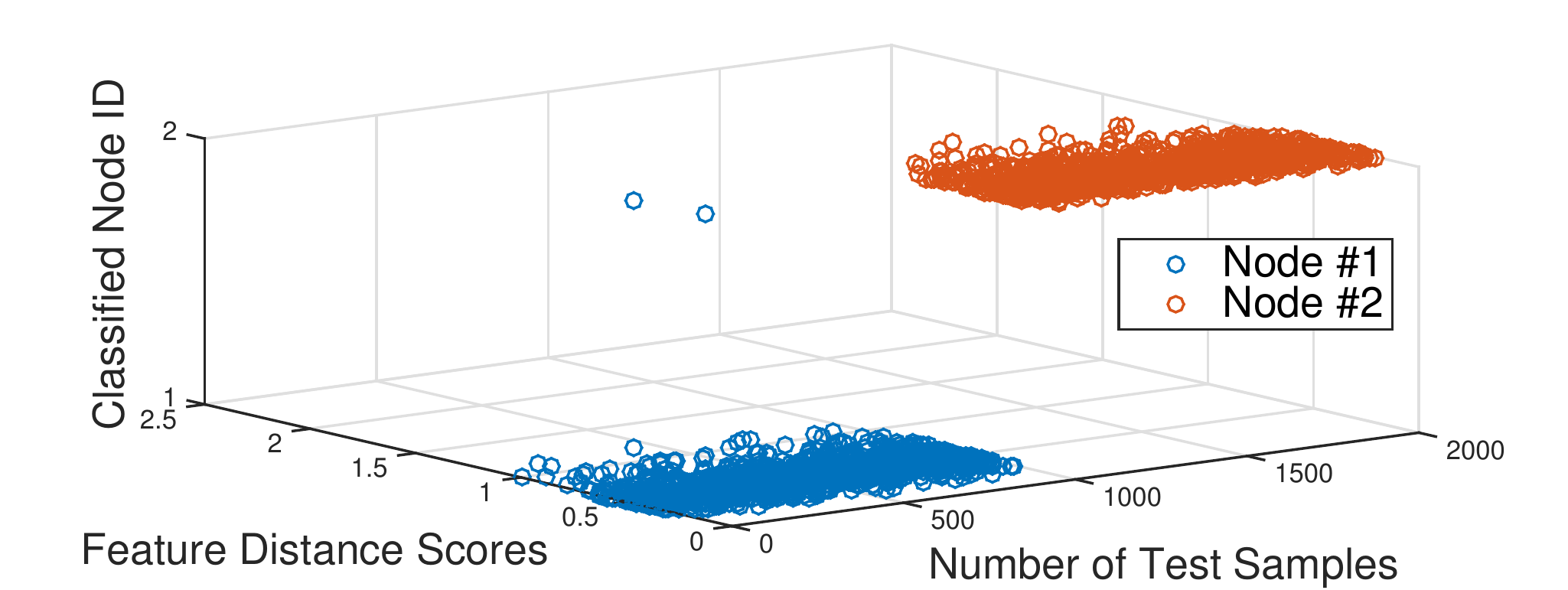}
\caption{Classification results at 0.1m}
\label{fig: CL_0.1m}
\end{minipage}
\quad
\begin{minipage}[t]{0.3\linewidth}
\centering
\includegraphics [width=2.2in] {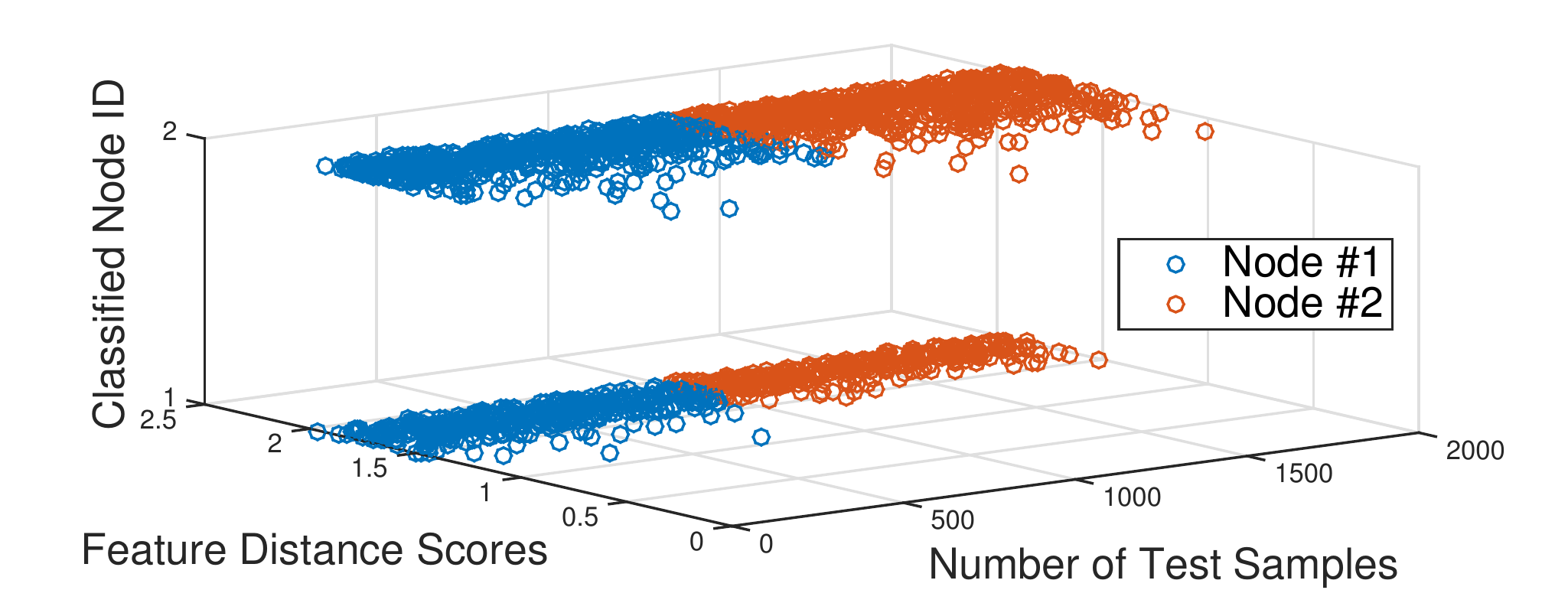}
\caption{Classification results at 6m with fixed database.}
\label{fig: CL_6m_fixed}
\end{minipage}
\quad
\begin{minipage}[t]{0.3\linewidth}
\centering
\includegraphics [width=2.2in] {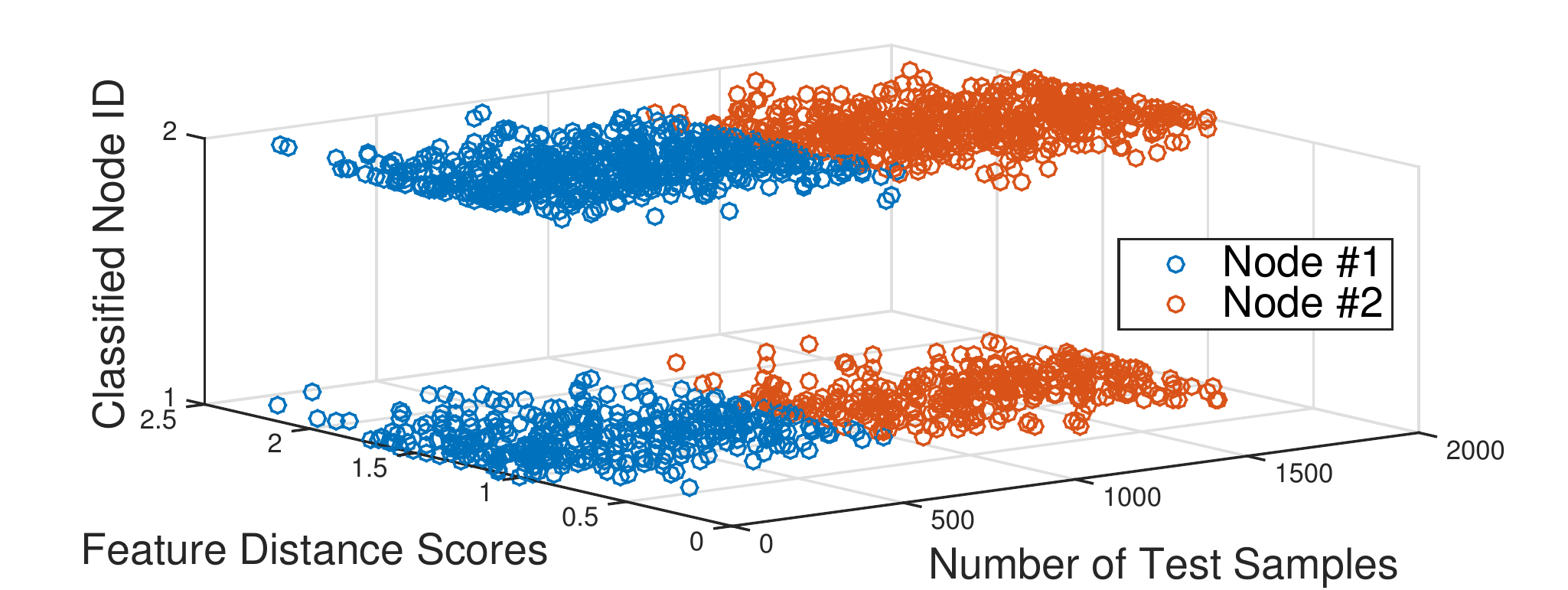}
\caption{Classification results at 6m without path loss effects.}
\label{fig: CL_6m_power}
\end{minipage}
\end{figure*}

\begin{figure*}[t]
\begin{minipage}[t]{0.3\linewidth}
\centering
\includegraphics [width=2.2in] {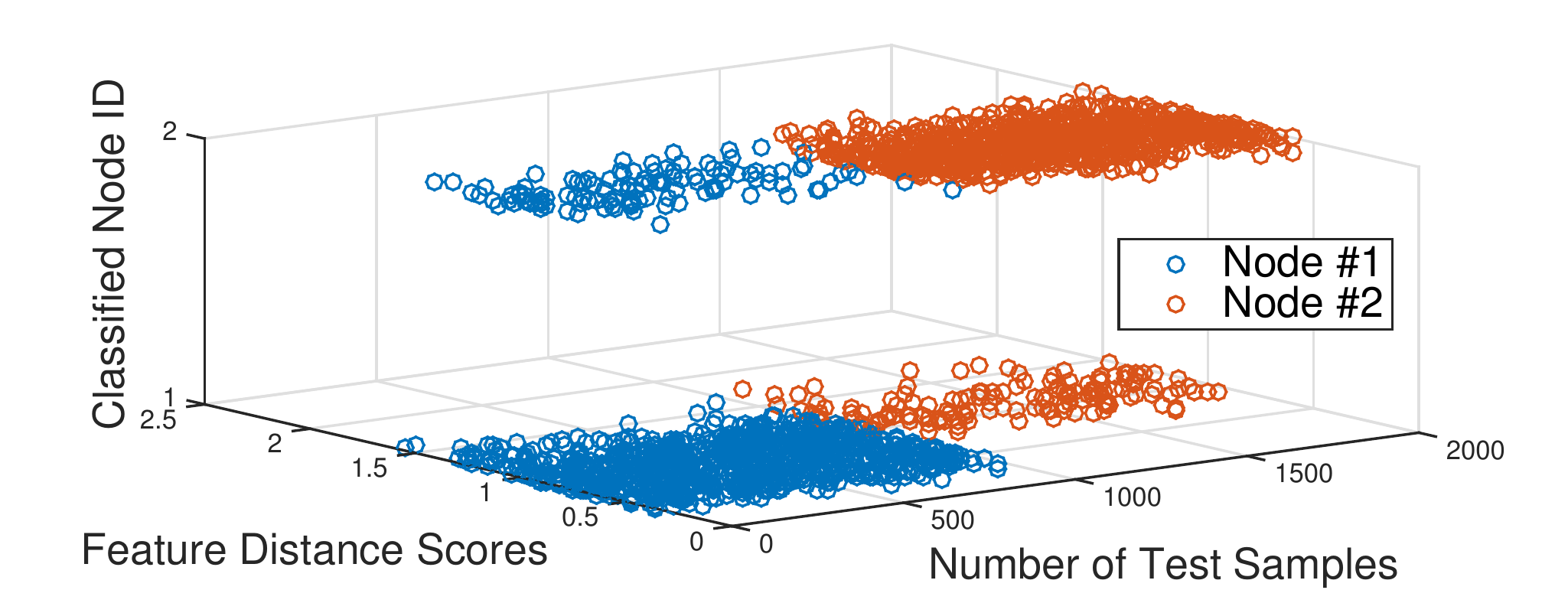}
\caption{Classification results at 6m with updated database}
\label{fig: CL_6m_updated}
\end{minipage}
\quad
\begin{minipage}[t]{0.3\linewidth}
\centering
\includegraphics [width=2.2in] {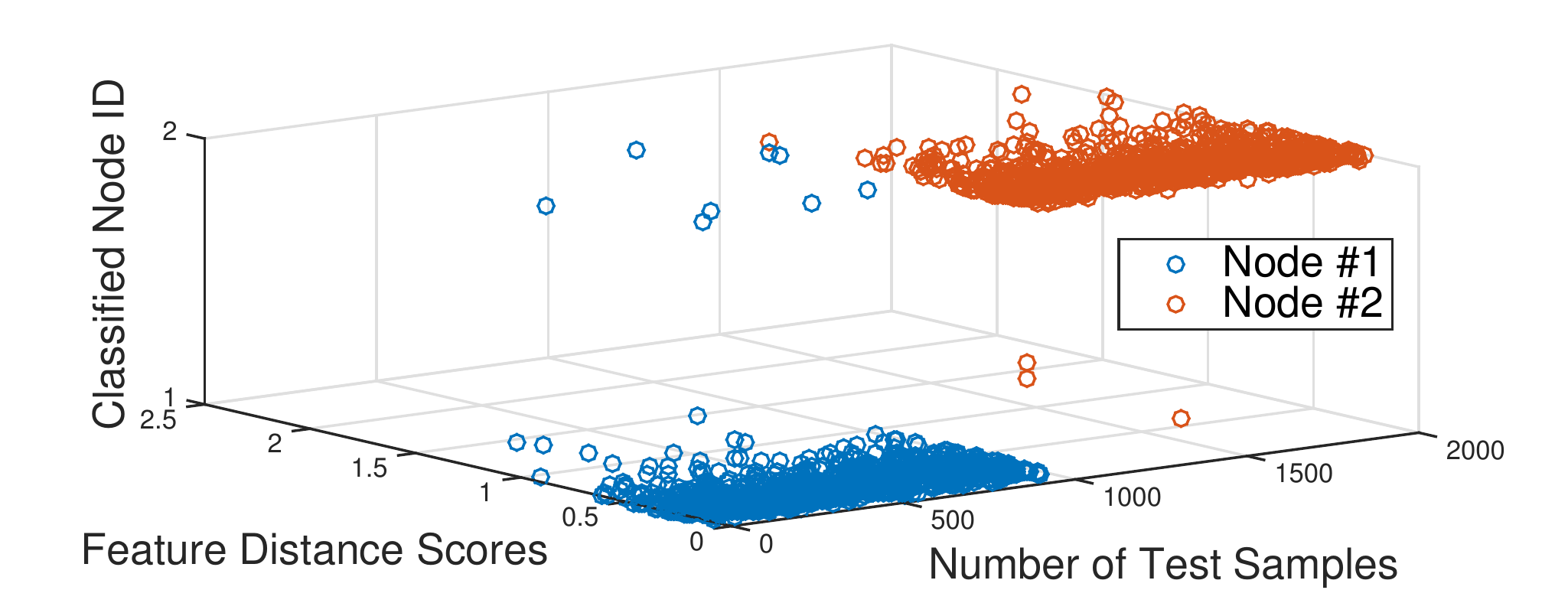}
\caption{Classification results at 1m, 2Ms/s}
\label{fig: CL_2M}
\end{minipage}
\quad
\begin{minipage}[t]{0.3\linewidth}
\centering
\includegraphics [width=2.2in] {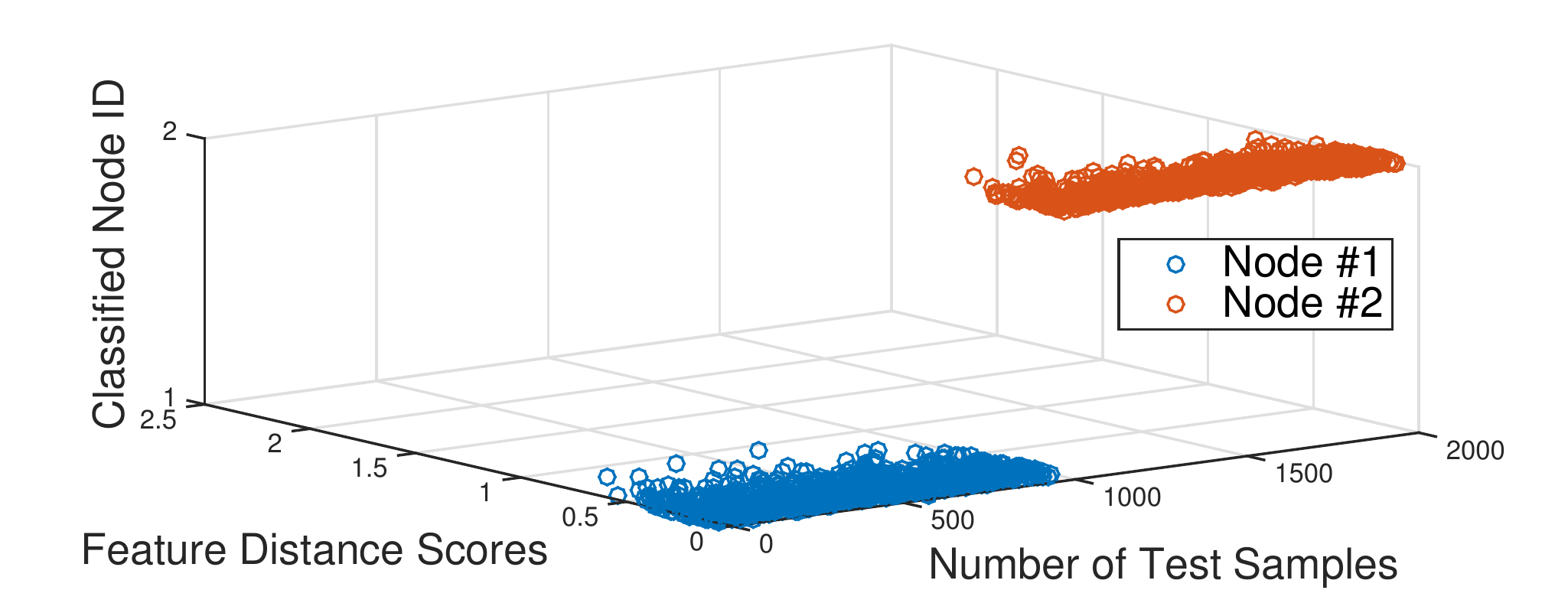}
\caption{Classification results at 1m, 8Ms/s}
\label{fig: CL_8M}
\end{minipage}
\end{figure*}

\begin{figure*}[t]
\begin{minipage}[t]{0.3\linewidth}
\centering
\includegraphics [width=2.2in] {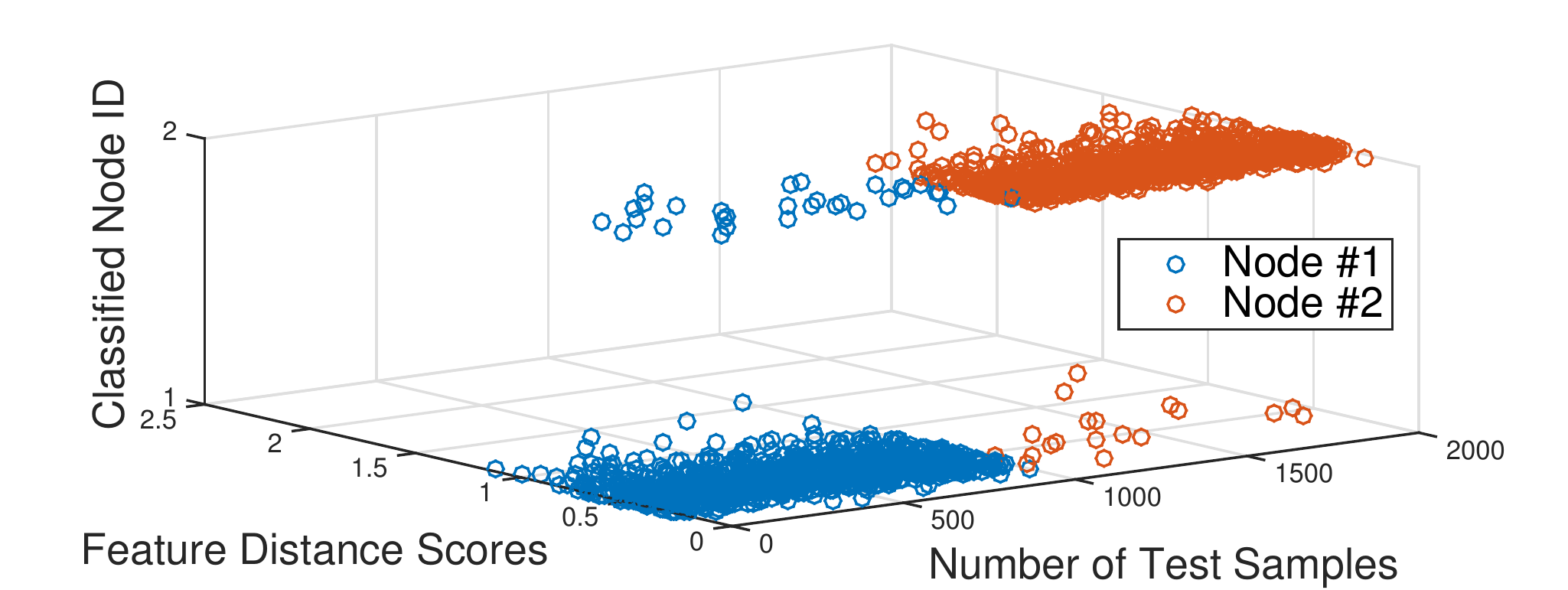}
\caption{Classification results at 3m,  256p FFT}
\label{fig: CL_256}
\end{minipage}
\quad
\begin{minipage}[t]{0.3\linewidth}
\centering
\includegraphics [width=2.2in] {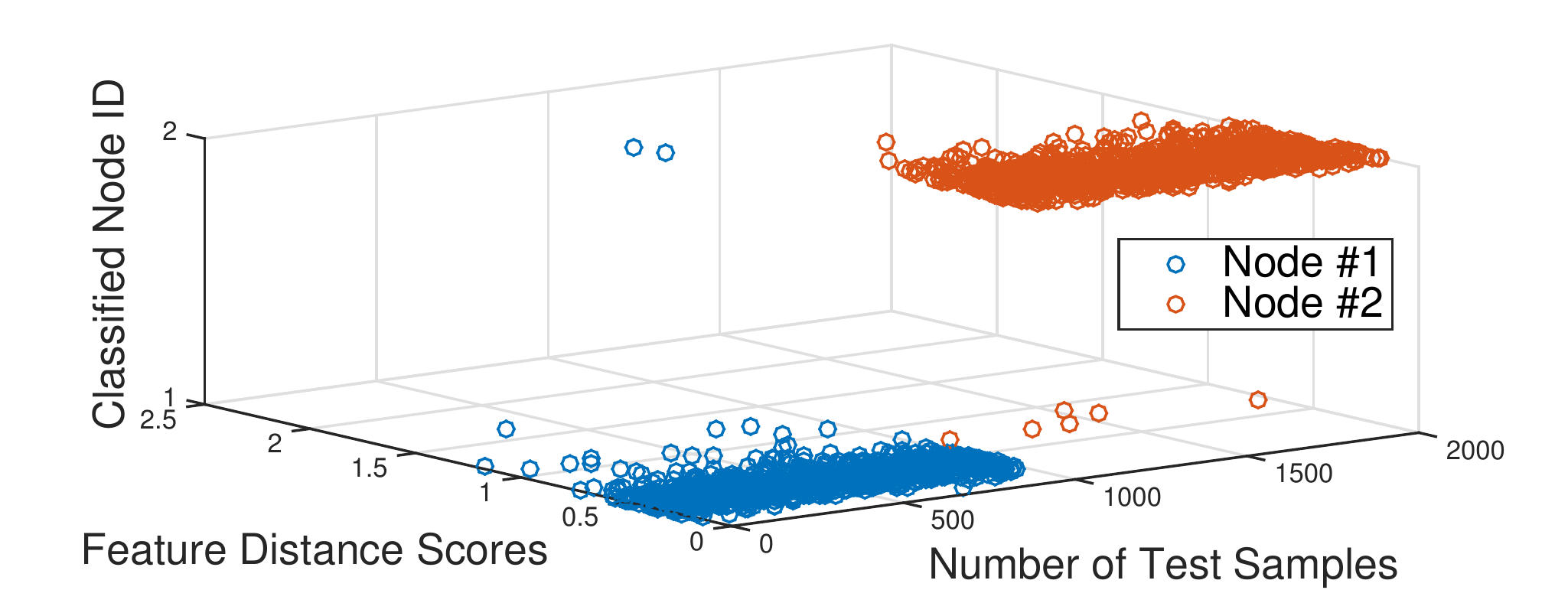}
\caption{Classification results at 3m,  512p FFT}
\label{fig: CL_512}
\end{minipage}
\quad
\begin{minipage}[t]{0.3\linewidth}
\centering
\includegraphics [width=2.2in]  {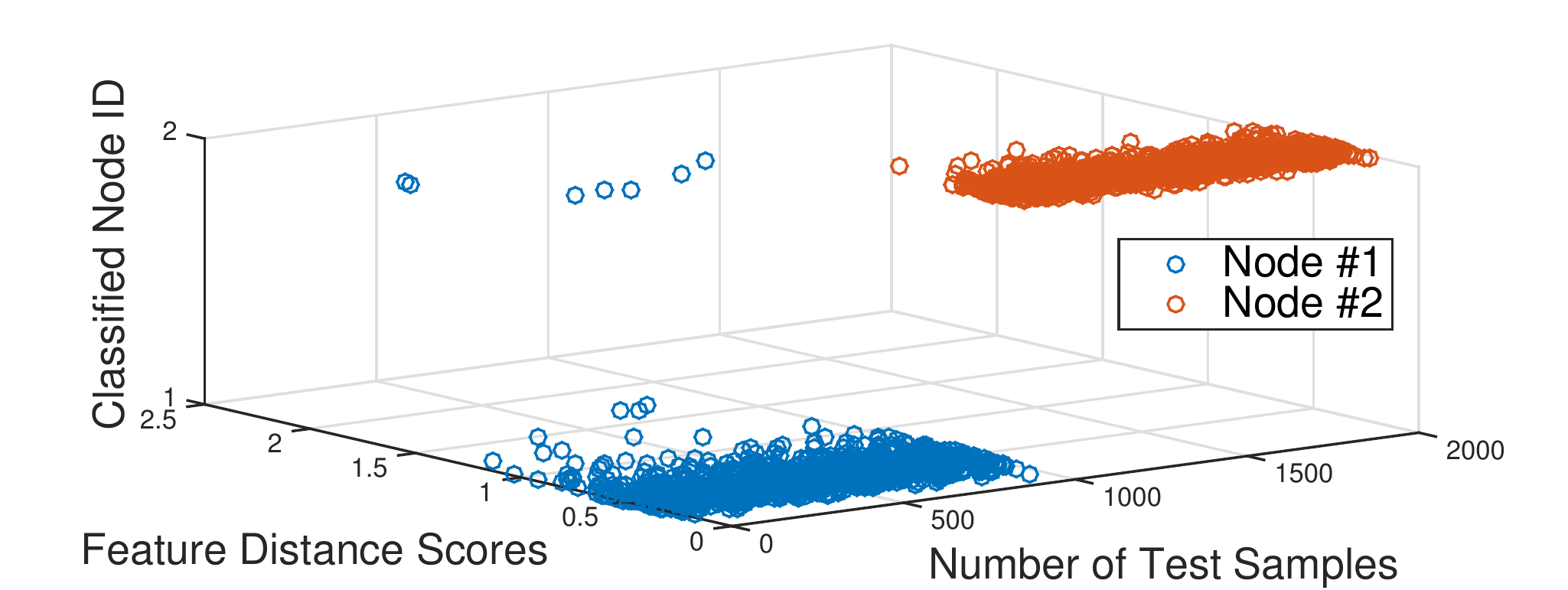}
\caption{Classification results at 3m, 1024p FFT}
\label{fig: CL_1024}
\end{minipage}
\end{figure*}

The experimental setup is illustrated in Fig.~\ref{fig:exp_setup}. The transmitters (device to be identified) are 6 Micaz sensors nodes, which is a IEEE 802.15.4 compliant RF transceiver with the RF frequency band ranging from 2.4 to 2.48 GHz  \cite{datasheet2006crossbow}. The modulation scheme and shaping filter used in IEEE 802.15.4 protocol are fixed, which are O-QPSK with half-sine pulse shaping filter \cite{instruments2007cc2420}. The baseband transmitting symbol rate 1 M symbol per second. The data packet preamble is used to extract the RFF, which has a length of 32 bits all-zero data (128 symbols for either I/Q-phase signal after modulation and spread spectrum procedure).  

We use USRP N210 companied with SBX daughter board \cite{ettus2008universal} as the receiver (identifier). The SBX daughter board is a RF front-end system with a carrier frequency range from 400 MHz to 4.4 GHz, which allows communication with Micaz nodes at the 2.4 GHz band. The USRP N210 is equipped with a dual 14-bit ADC operating at 100 MHz and dual 16-bit DAC operating at 400 MHz. The data finally stored in files are sampled with the setting sampling rate $f_s$ in GNU Radio software platform. The maximum sampling rate is 25 Ms/s due to limitation of direct gigabit Ethernet link \cite{pesovic2012implementation}. It should be noted that in the digital down conversion procedure to achieve the setting sampling rate, the USRP utilizes a half-band filter which is the low pass filter with pass bandwidth $W=f_s/2$ \cite{kodali2013ddc}. 

The receiver is either 1 m, 3 m, or 6 m away from the transmitters to obtain the large scale attenuation effects. Meanwhile, the receiver is placed in the middle of a circle of the metal obstacles (laptops), wooden obstacles (book shelfs), and the concrete wall, as shown in Fig.~\ref{fig:exp_setup}. Hence, no direct line-of-sight wireless link exists and obvious multipath effects can be observed. In the experiments to validate the effects of receiver sampling rates and FFT points, we place the receiver in an in-door open space to build an AWGN-type channel so that the receiver's settings take major effects (other than the multipath channel). 

A laptop with MATLAB 2014b is connected to the USRP receiver. The packet preambles are extracted from data files obtained in USRP and the LDA-based identification and classification algorithms are realized in MATLAB. To extract the preambles from the data files, the variance-based threshold detection algorithm \cite{rasmussen2007implications} is adopted. The LDA dimension $\kappa=5$ is fixed in all experiments.




All the classification and identification decisions are made based on the LDA feature distance between the fingerprints under different experiment conditions. The same as in Section III, the average classification error rates are derived to evaluate the classification performance, while the false accept rate (FAR), genuine accept rate (GAR), and equal error rate (EER) are derived to evaluate the identification performance. 


In order to analyze the influence from wireless channels, we first build fingerprint database and obtain identification threshold at very small distance (0.1 m). The classification/identification performance is also obtained as the best case for comparison. Then we place the transmitter and receiver in the aforementioned multipath fading channel and conduct the experiment in three scenarios. 
In the first scenario, we update the reference database and identification threshold at each new location, which is the common strategy that existing work adopt. 
In the second scenario, we do not update the reference database and check the classification/identification results. The objective of this scenario is to validate whether the existing WPLI systems with a universal fingerprint database and threshold can work with mobile users or in dynamic channels. 
Moreover
In the third scenario, we do not update the reference database but normalize the test samples from different channels and distances. The purpose is to check whether the path loss is the main source of the channel effects and whether the normalization can eliminate the influence from channels.

In order to analyze the influence from the receiver sampling rate, we set two different sampling rates (i.e., 2 Ms/s and 8 Ms/s) in USRP. Fixed spectrum resolution is used when building the fingerprints database and extract fingerprints from incoming signal. 
The 2 Ms/s sampling rate can only cover the main lobes of the spectrum while the 8 Ms/s can recover more side lobe information, as shown in Fig. \ref{fig: PSD_fs}. Whether the sampling rate effectively covers enough bandwidth is determined by both the hardware of the receiver and the modulation/shaping filter of the transmitter. Since the MicaZ sensors have fixed modulation scheme and shaping filter, the receiver sampling rate plays major role in our experiments.
In order to check the effects of FFT point number at the receiver, the same data file is obtained with a single sampling rate in USRP receiver. However, we use the different FFT point number in feature extraction to realize the fingerprints with the same signal bandwidth but different spectrum resolution, as shown in Fig.~\ref{fig: PSD_FFT}.


\subsection{Classification Performance}
In this group of the experiments, the classification performance are tested to capture the effects of channel, database updating strategy, sampling rates, and FFT point number. The sensor node \#1 and \#2 are used as the two genuine users to be classified. 
In Fig.~\ref{fig: CL_0.1m} - Fig.~\ref{fig: CL_1024}, the x-axis is the index of the test samples. The first 1000 are samples from Node \#1 and the next 1000 are from Node \#2. The y-axis is the minimal feature distance of each test sample. The color of each testing sample indicates the original identity of the signal: blue - Node \#1 and red - Node \#2. The z-axis is the classification results: there is a classification error if blue circles appear in upper level or red circles appear in lower level.

\subsubsection{Initialization}
We start the roll in and the classification at the close distance (0.1m, $f_s=4Ms/s$, $N_{FFT}=512$). According to the results shown in Fig.~\ref{fig: CL_0.1m}, the classification performance is exceptional where only two samples are wrongly classified, i.e., the average classification error rate $P_e=0.001$.

\subsubsection{Channel effects}
We place the receiver in the aforementioned non-line-of-sight multipath channel at 6 m away from transmitter. At first, we do not update the reference database. The results are shown in Fig. \ref{fig: CL_6m_fixed} where the classification error rate $P_e=0.4910$, i.e., complete loss of classification capability. 

Then in order to check whether the attenuation over distance (i.e., path loss) is the main impact on the performance, we eliminate the path loss effects by normalizing the test samples and redo the classification. The results are shown in Fig.~\ref{fig: CL_6m_power} where $P_e=0.4830$. The feature distance scores are smaller but the classification performance is as poor as before. Hence, we can conclude that path loss is not the main source of the channel effects and the sample normalization cannot mitigate the impact.

Finally, we adopt the existing approach to update the reference database at the new location before classification. The results are shown in Fig.~\ref{fig: CL_6m_updated} where $P_e=0.1388$. It is obvious that the system resume part of the classification capability if the less practical database updating strategy is used.
However, compared with the results at the close distance, the performance is much worse due to the multipath channel. It should be noted that the real indoor or metropolitan channel can be much worse and hostile than the short-range artificial multipath channel we build in our experiments.

\begin{figure*}[t]
\begin{minipage}[t]{0.3\linewidth}
\centering
\includegraphics [width=2.2in] {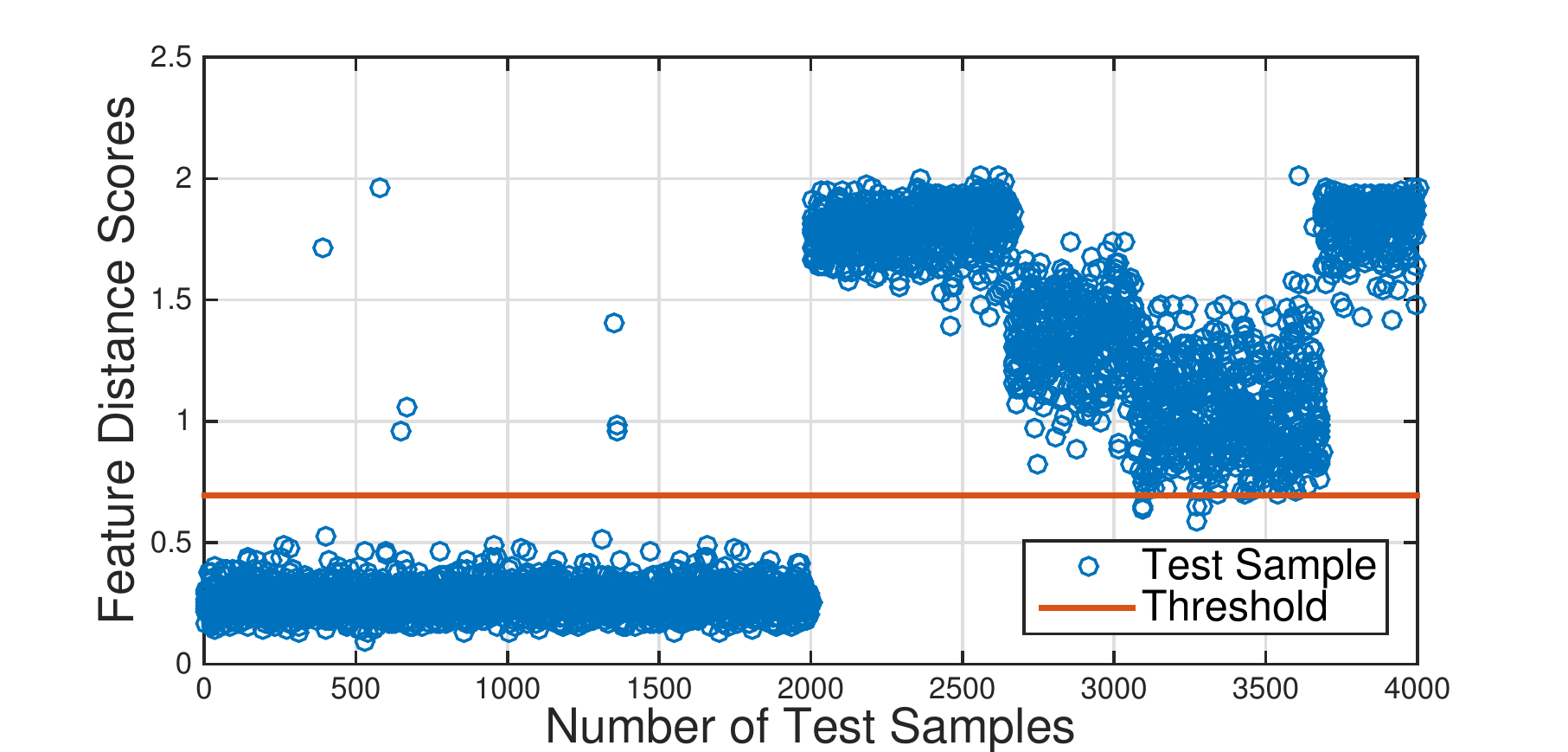}
\caption{Identification results at 0.1m.}
\label{fig: ID_0.1m}
\end{minipage}
\quad
\begin{minipage}[t]{0.3\linewidth}
\centering
\includegraphics [width=2.2in] {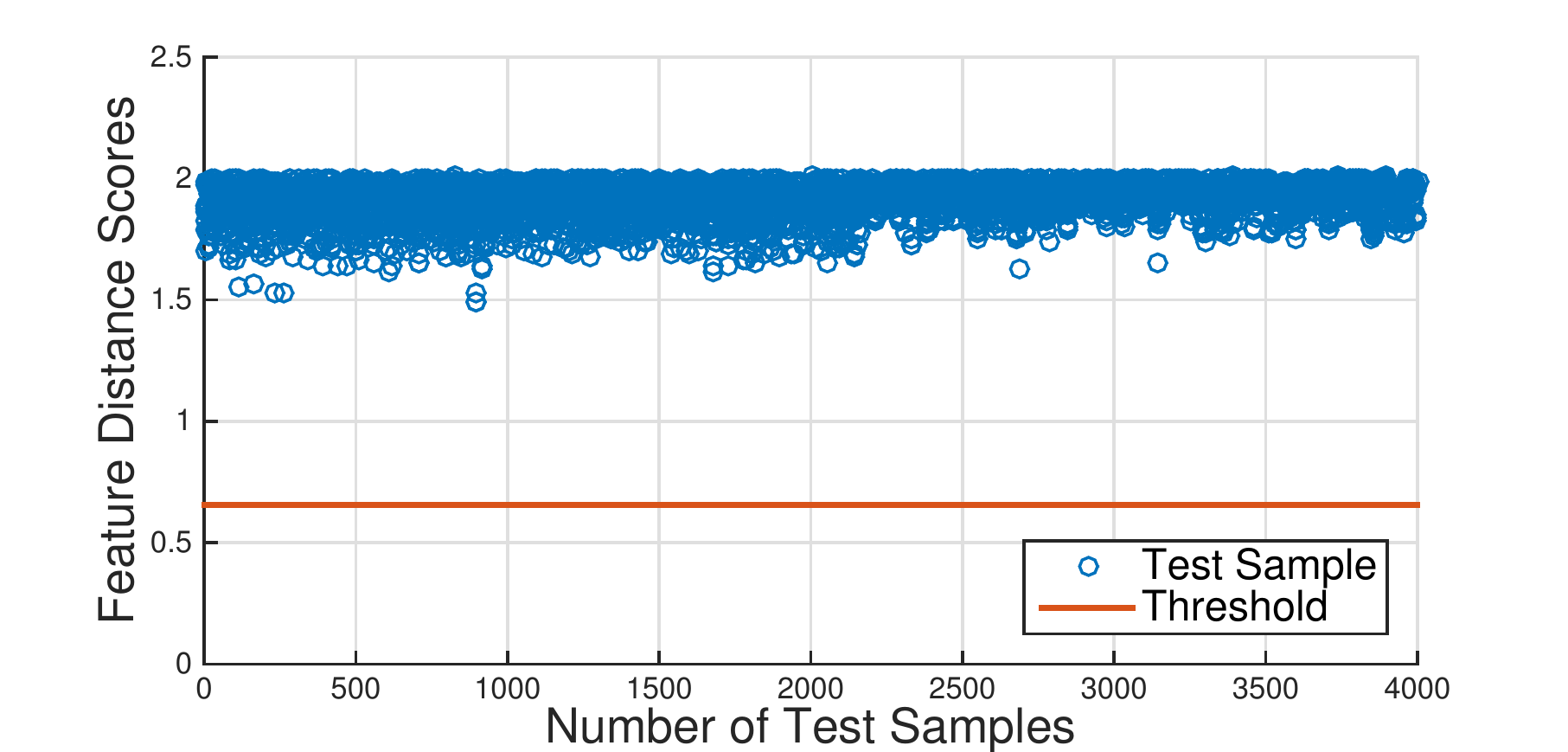}
\caption{Identification results at 6m, fixed database}
\label{fig: ID_6m_fixed}
\end{minipage}
\quad
\begin{minipage}[t]{0.3\linewidth}
\centering
\includegraphics [width=2.2in]  {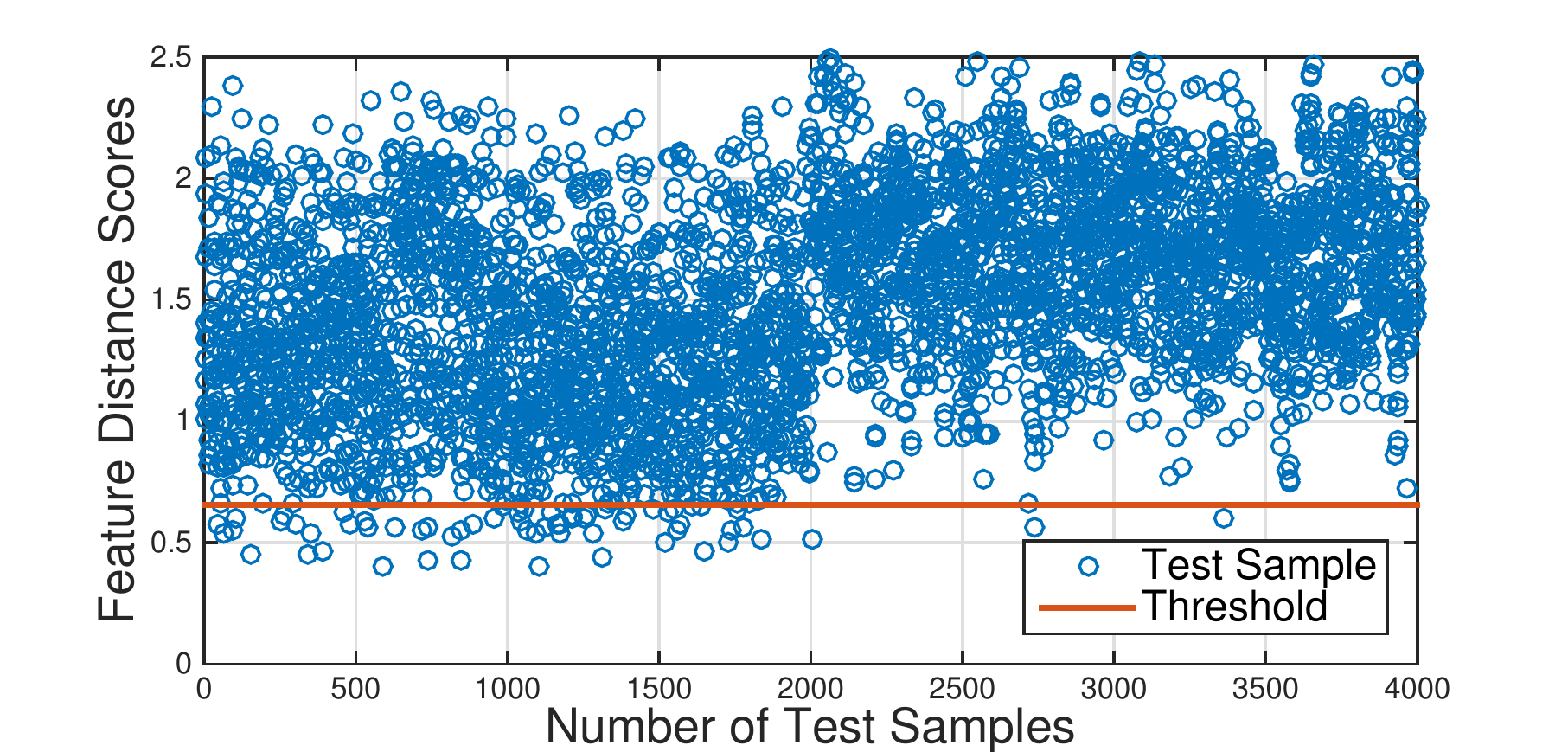}
\caption{Identification results at 6m, without path loss.}
\label{fig: ID_6m_power}
\end{minipage}
\end{figure*}

\begin{figure*}[t]
\begin{minipage}[t]{0.3\linewidth}
\centering
\includegraphics [width=2.2in] {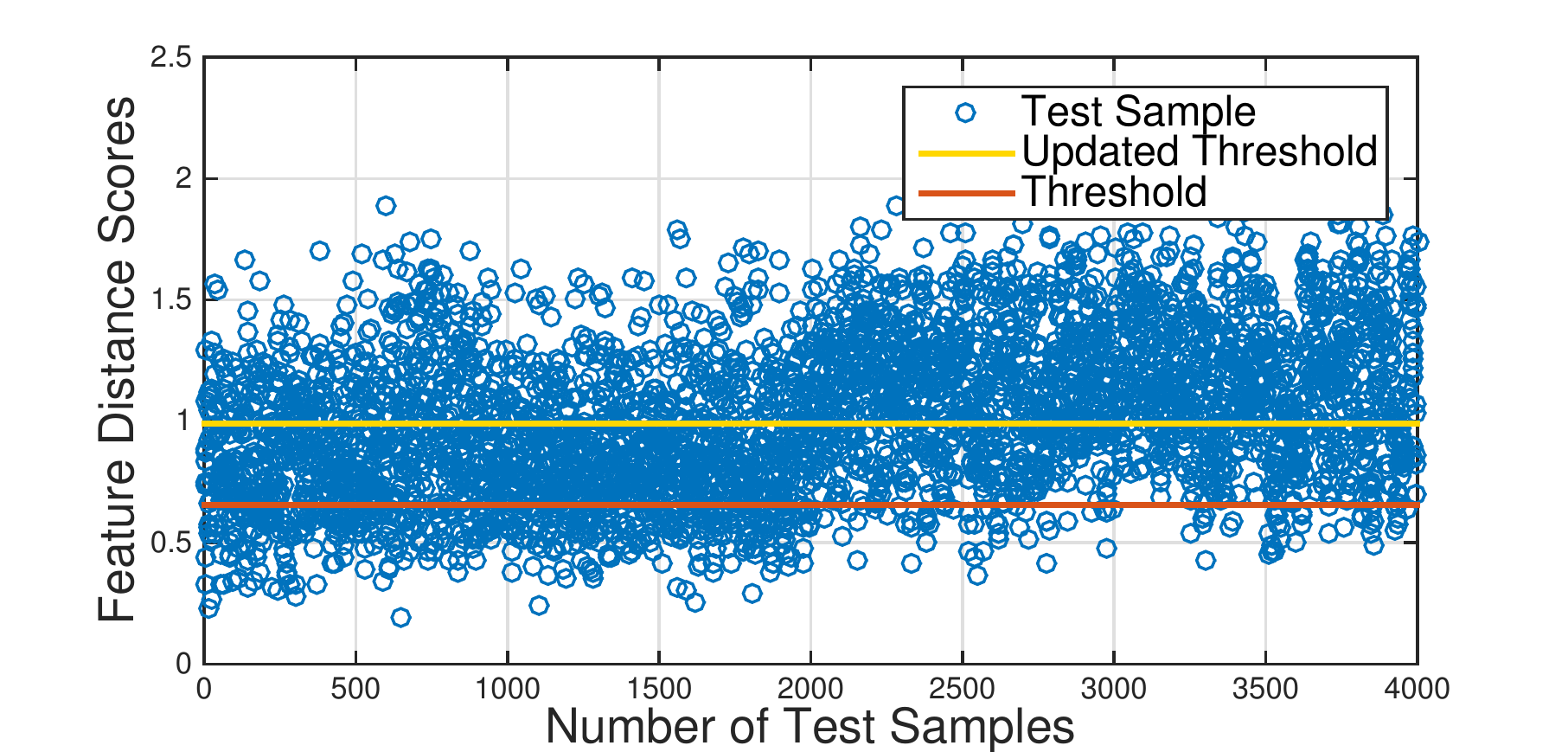}
\caption{Identification results at 6m, updated database.}
\label{fig: ID_6m_update}
\end{minipage}
\quad
\begin{minipage}[t]{0.3\linewidth}
\centering
\includegraphics [width=2.2in] {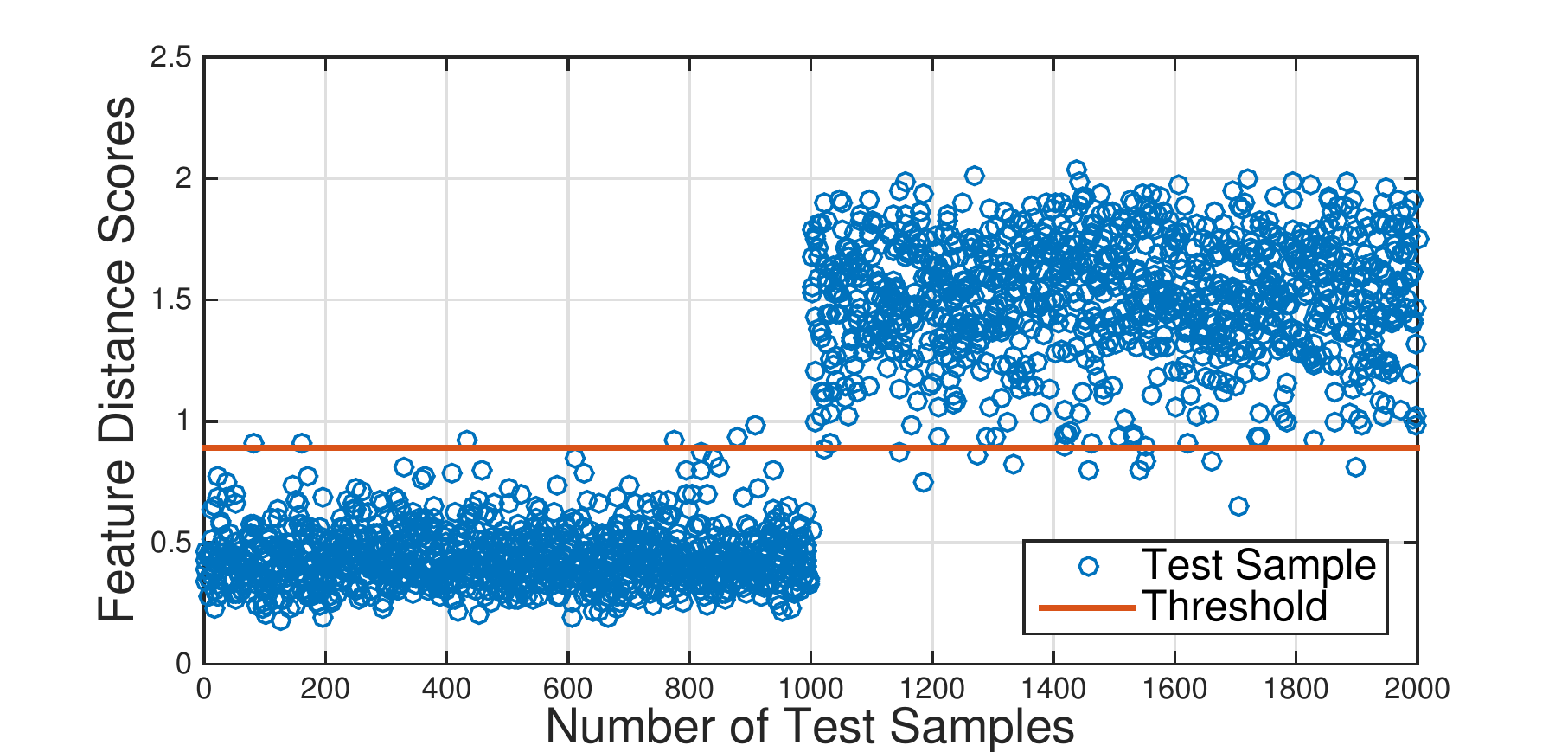}
\caption{Identification results at 1m,2Ms/s.}
\label{fig: ID_1m_2M}
\end{minipage}
\quad
\begin{minipage}[t]{0.3\linewidth}
\centering
\includegraphics [width=2.2in]  {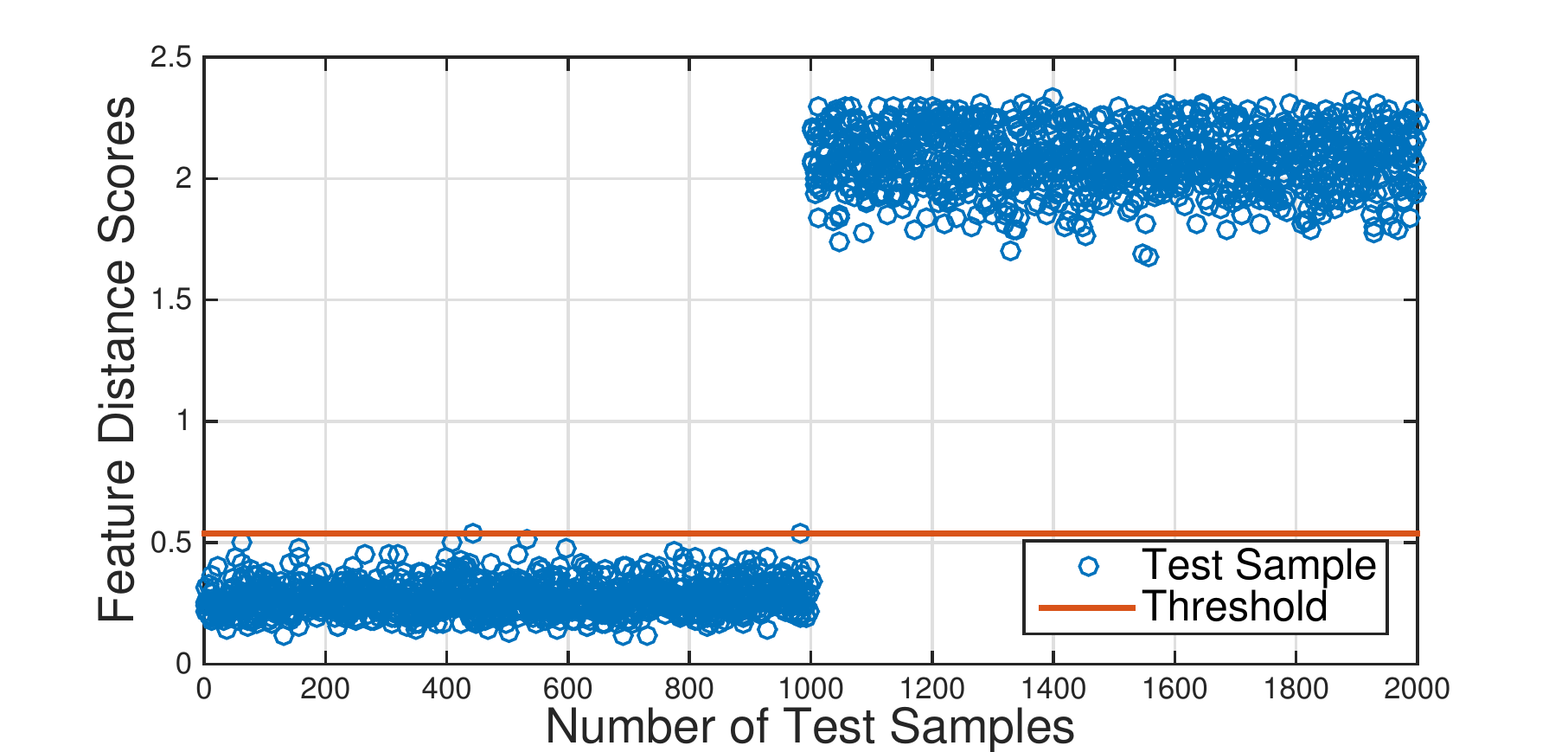}
\caption{Identification results at 1m, 8Ms/s.}
\label{fig: ID_1m_8M}
\end{minipage}
\end{figure*}

\begin{figure*}[t]
\begin{minipage}[t]{0.3\linewidth}
\centering
\includegraphics [width=2.2in]  {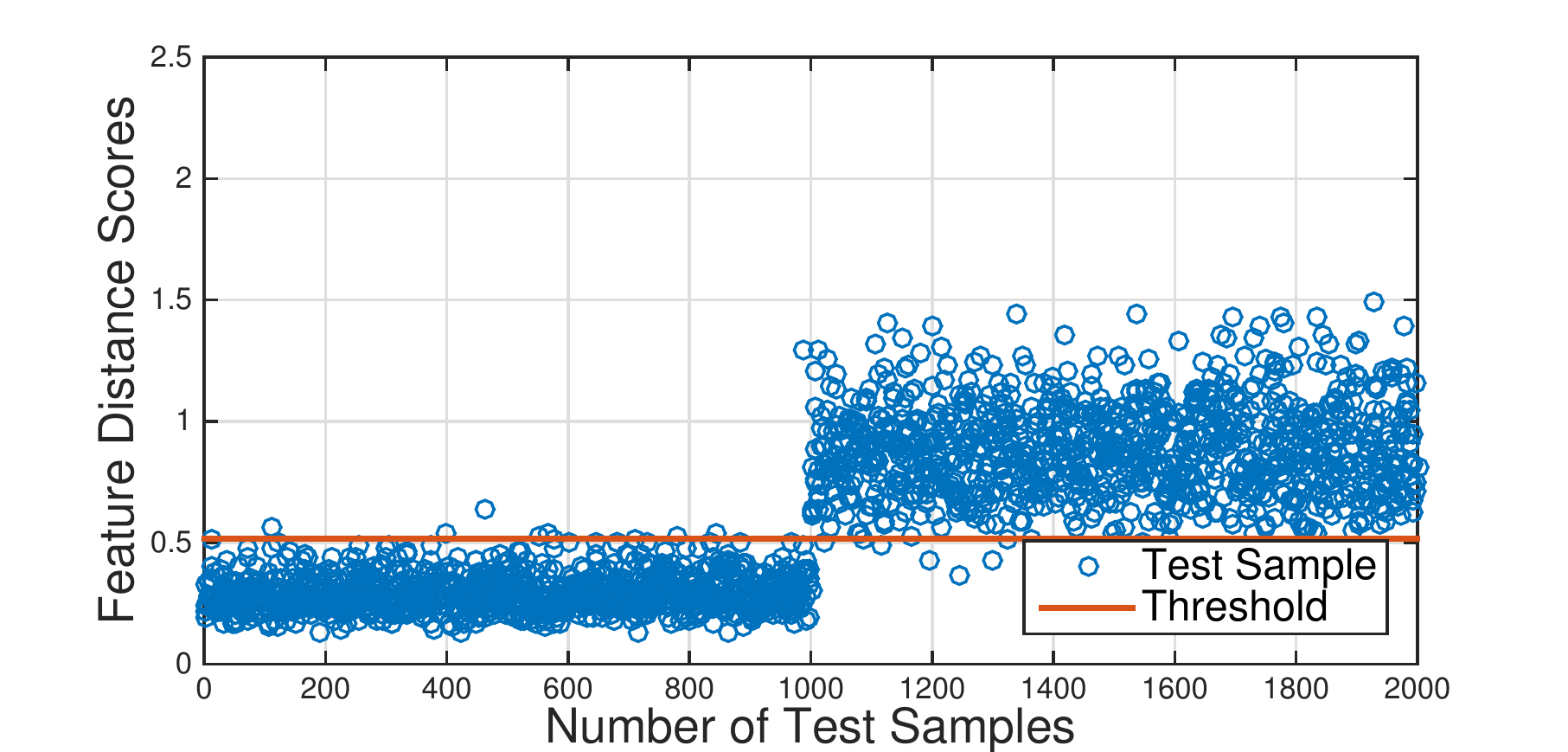}
\caption{Identification results at 3m, 256p FFT.}
\label{fig: ID_256}
\end{minipage}
\quad
\begin{minipage}[t]{0.3\linewidth}
\centering
\includegraphics [width=2.2in] {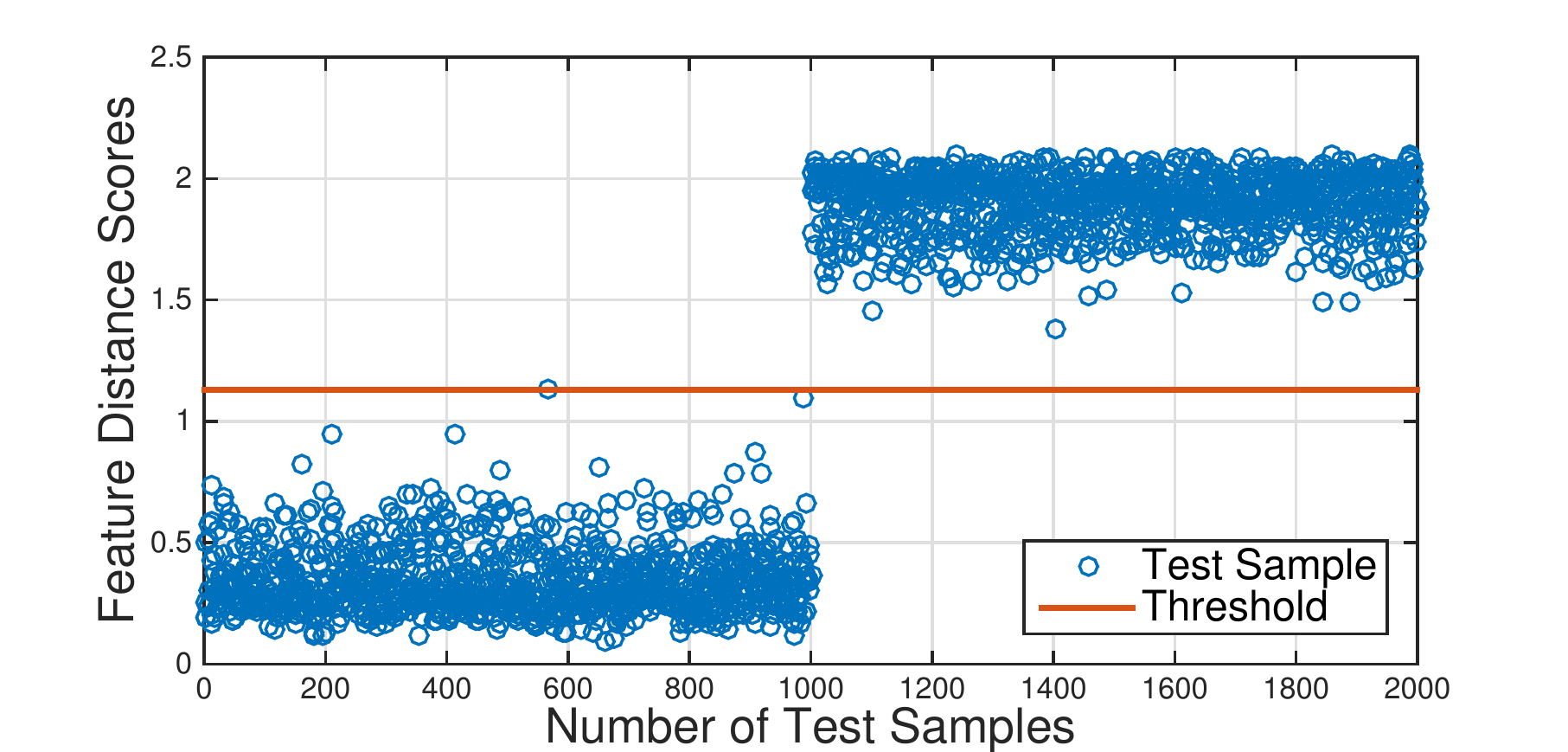}
\caption{Identification results at 3m, 512p FFT.}
\label{fig: ID_512}
\end{minipage}
\quad
\begin{minipage}[t]{0.3\linewidth}
\centering
\includegraphics [width=2.2in]  {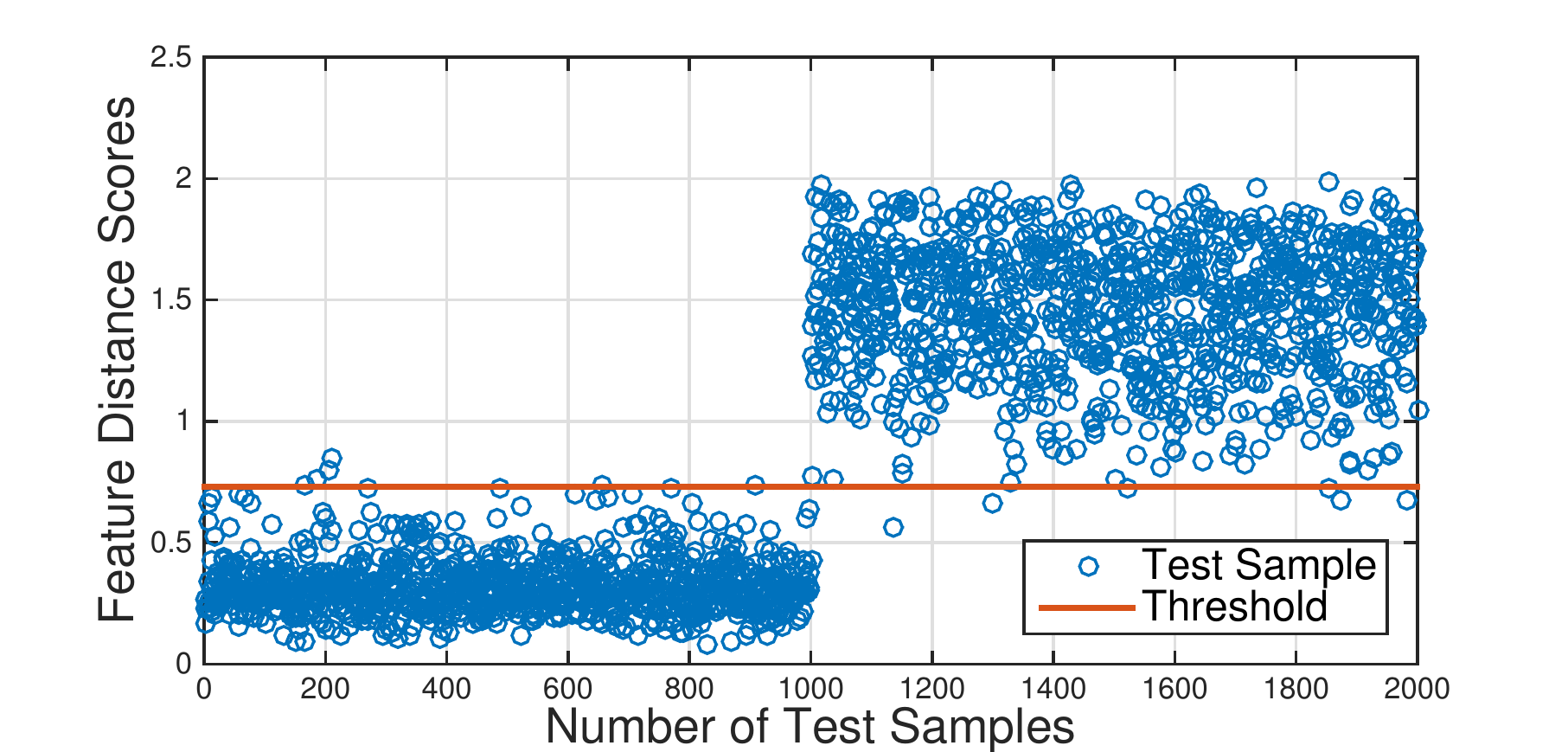}
\caption{Identification results at 3m, 1024p FFT.}
\label{fig: ID_1024}
\end{minipage}
\end{figure*}

\subsubsection{Effects of sampling rate}
Due to property of USRP hardware, the bandwidth of its half-band filter are set to be $W=f_s/2$. Hence the fingerprint bandwidth $BW$ equals to the receiver bandwidth $W$. The classification results can be influenced by both the two factors. We set the sampling rate as either $f_s=2Ms/s$ or $f_s=8Ms/s$. To make sure the receiver effects are not merged by the channel effects, we put the receiver only 1 m away from the transmitter and there is no obstructions in between. To achieve the same spectrum fingerprinting resolution, $N_{FFT}=256$ when $f_s=2Ms/s$ and $N_{FFT}=1024$ when $f_s=8Ms/s$. The classification results are shown in Fig.~\ref{fig: CL_2M} and Fig.~\ref{fig: CL_8M}, where the classification error rates are $P_e=0.0055$ and $P_e=0$, respectively. The higher sampling rate improves the performance by involving higher frequency spectrum where the RFF is still significant.
However, high sampling rate can help only when the distance is short, the SNR is high, and the RFF indeed occupy the higher frequency band. If the transmission distance is high and the SNR is low, the non-significant RFF may be merged by noise. Increasing the sampling rate can introduce more noise if the extra fingerprinting spectrum has low SNR. Hence, the optimal sampling rate need to be determined by jointly consider the channel effects, the origin of the RFF, and the regulation of the modulation scheme and shaping filter.

\subsubsection{Effects of number of FFT points}
To test the effects of FFT point number, the receiver is again placed at a close distance at 3 m and the sampling rate is set as $f_s=8Ms/s$. The classification results using $N_{FFT}=256$, $N_{FFT}=512$, and $N_{FFT}=1024$ are shown in Fig.~\ref{fig: CL_256}, Fig.~\ref{fig: CL_512}, and Fig.~\ref{fig: CL_1024}, where the classification error rates are $P_e=0.0255$, $P_e=0.004$, and $P_e=0.0035$, respectively. There is no clear difference on the performance with different FFT point number. As the frequency point where the RFF is significant is not regularly distributed, the extra FFT points obtained by higher point number may either pick up the significant RFF or just add more noise. Hence, there is no need to use ultra high FFT point number that add additional computation burden.

\subsection{Identification Performance}
In this group of the experiments, the above influence factors are analyzed again but in the identification procedure. The sensor node \#5 is used as the genuine user, while the nodes \#1~\#4 are used as the imposters. 
In Fig.~\ref{fig: ID_0.1m} - Fig.~\ref{fig: ID_1024}, the x-axis is the index of the test samples. The first half of samples (1000 or 2000) are samples from genuine user \#5. The rest of samples are from imposters. The y-axis is the feature distance score of each test sample. The horizontal red line is the judging threshold. The samples below the threshold are identified as genuine user, otherwise are identified as imposters. 

\subsubsection{Initialization}
Similarly to the classification experiments, we start the roll in and the classification at the distance of 0.1 m. After obtaining the feature distance scores of all the testing samples, we choose the point, where the equal error rate $EER=0.0030$ is achieved, as the judeging threshold of the identification system, i.e., $\lambda_{EER}=0.655$. 

\subsubsection{Channel effects}

Again, we place the receiver in the artificial non-line-of-sight multipath channel at 6 m away from transmitter. We fix the reference database and judging threshold. The results are shown in Fig.~\ref{fig: ID_6m_fixed} where $GAR=0$ and $FAR=0$. Similar to the classification case, the system completely lose the identification capability. 

Then we eliminate the path loss effects by normalizing the test samples. The results are shown in Fig.~\ref{fig: ID_6m_power} where $FAR=0.0010$ and $GAR=0.0305$. The identification capability is still very weak. Hence, we can again conclude that the multipath channel impact cannot be mitigated by simply compensating the path loss.

\begin{figure}[!t]
\centering
\includegraphics [width=2.5in] {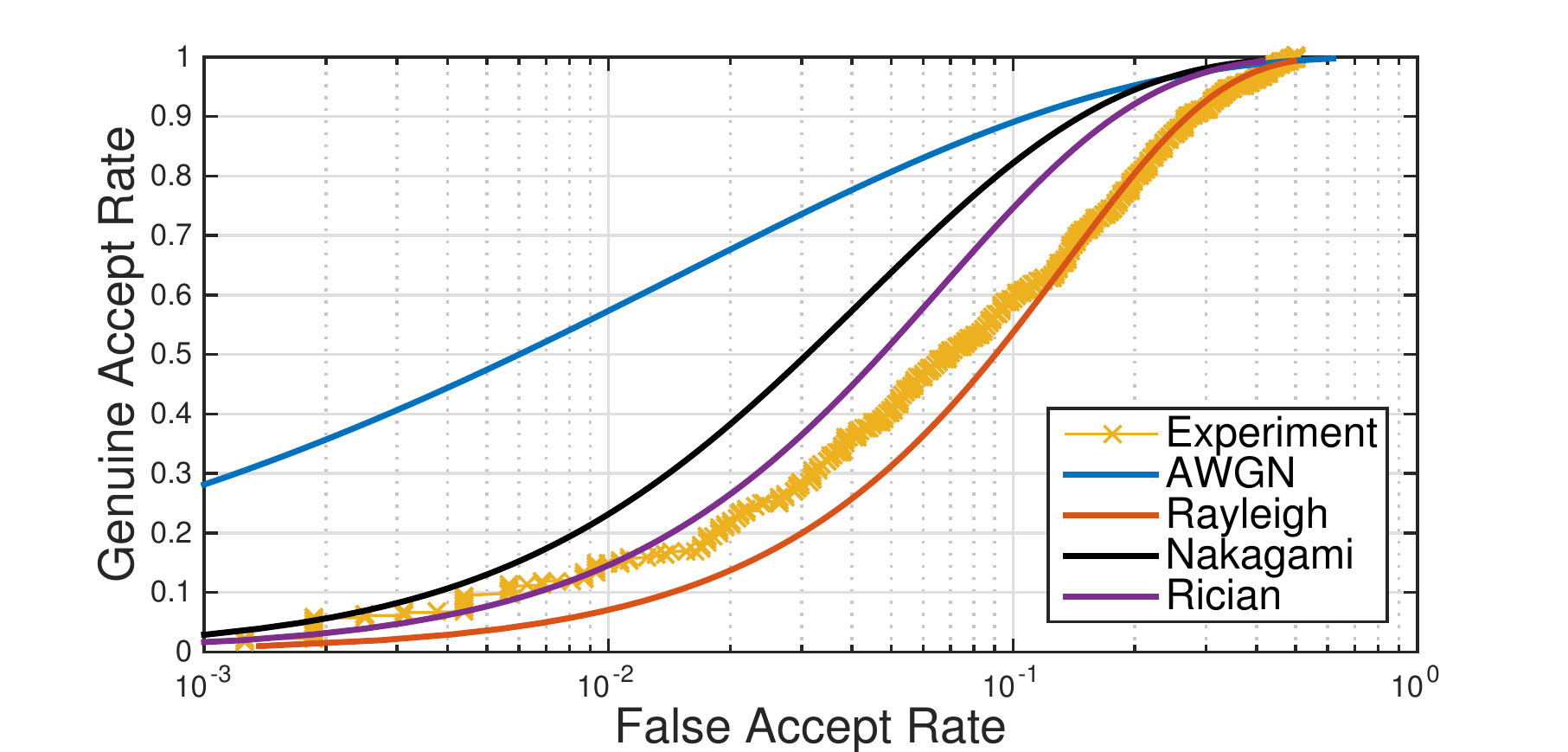}
\caption{Experimental and theoretical ROC curves.}
\label{fig: ROC_2}
\end{figure}

Finally, we update the reference database and reset the threshold at every new location: not practical in most applications but the same as most existing WPLI solutions. The new results are shown in Fig.~\ref{fig: ID_6m_update} where the threshold, $\lambda_{EER}=0.9900$, is reset to achieve the $EER=0.3205$. However, the identification performance is still much worse than the case when distance is short, which is obviously caused by the effects of multipath channel. Fig.~\ref{fig: ROC_2} shows the ROC curve derived in the experiment with the multipath channel and 6 m distance. The theoretical ROC curves under the same SNR level derived in Section III are also plotted for comparison. We can clearly see the experimental curve has the best match with the theoretical Rayleigh ROC curve, which is consistent with expectation in the non-line-of-sight multipath channel we build in our lab.

\subsubsection{Effects of sampling rate}
To focus on the effects of the sampling rate of the receiver, we only use two transmitters, i.e., genuine user - Node \#5 and imposter - Node \#4. 1000 samples are taken for each device that is 1 m away from each other. When lower sampling rate are used, i.e., $f_s=2Ms/s, N_{FFT}=256$, the results are shown in Fig.~\ref{fig: ID_1m_2M} where $EER=0.008$ and $\lambda_{EER}=0.892$. When higher sampling rate are used, i.e., $f_s=8Ms/s,  N_{FFT}=1024$, the results are shown in Fig.~\ref{fig: ID_1m_2M} where $EER=0$ and $\lambda_{EER}=0.537$. Obviously, higher sampling rate improves the identification performance. However, the same comments as in the classification experiment also apply here. The effects of the sampling rate are also determined by the channel, the RFF property, and the modulation/shaping filter.

\subsubsection{Effects of number of FFT points}

Finally, the identification results with different FFT point number are given in Fig.~\ref{fig: ID_256}, Fig.~\ref{fig: ID_512}, and Fig.~\ref{fig: ID_1024}. The data are collected at 3m, with 8 Ms/s sampling rate. The equal error rates and threshold for the each case are $EER=0.01, \lambda_{EER}=0.5150$; $EER=0, \lambda_{EER}=1.129$; and $EER=0.01, \lambda_{EER}=0.5150$. Again, similar conclusion as the classification experiments can be obtained here: extra FFT points do not have clear effects on the identification performance due to the irregularly distributed RFF along the spectrum.

\section{Conclusion}
\label{sec: C}

In this work, we rigorously evaluate the influence of real-world constraints along every single step in the WPLI techniques, using a new theoretical model and in-lab experiments. The theoretical model provide the first systematical description on the whole WPLI procedure, which is applicable in most WPLI that based on digital wireless communications. The model is then implemented to comprehensively characterize various WPLI techniques that utilize the frequency domain RFF from the non-linear RF front-end, with different settings in transmitters, receivers, and wireless channels. From both theoretical deduction and experiment validation, we reveal several key influence factors in real-world applications, including the constraints in wireless regulation, the origins of RFFs, the multipath fading channel and mobile users, and the sophistication of the receiver device. Our results show that existing WPLI techniques is less likely to achieve acceptable accuracy in real-world operation environments with off-the-shelf wireless devices, which motivate the discovery of new sources of more reliable and differentiable RFFs.

\bibliographystyle{IEEEtran}
\bibliography{PHY_modeling.bib}

\end{document}